\documentclass[10pt,openright]{book}

\RequirePackage{textcase}

\usepackage[hyperfootnotes=false]{hyperref}
\hypersetup{colorlinks=true,
	citecolor=blue,
	linkcolor=black,
	urlcolor=RedViolet}
\usepackage[all]{hypcap}

\usepackage{classicthesis}

\usepackage[bottom=3cm]{geometry}                

\usepackage{graphicx}
\usepackage{amssymb}

\usepackage{rotating}

\usepackage{amsmath}
\numberwithin{equation}{section}

\usepackage{epstopdf}
\usepackage{eso-pic}

\usepackage{cancel}
\usepackage{lipsum}
\usepackage{wrapfig}
		
\usepackage[numbers,sort&compress]{natbib}
\usepackage[chapter,notbib,nottoc,notlof]{tocbibind}

\usepackage{bbold}

\newcommand{\BackgroundPic}{
\put(-4,0){
\parbox[b][\paperheight]{\paperwidth}{\centering
\includegraphics[width=\paperwidth,height=\paperheight]{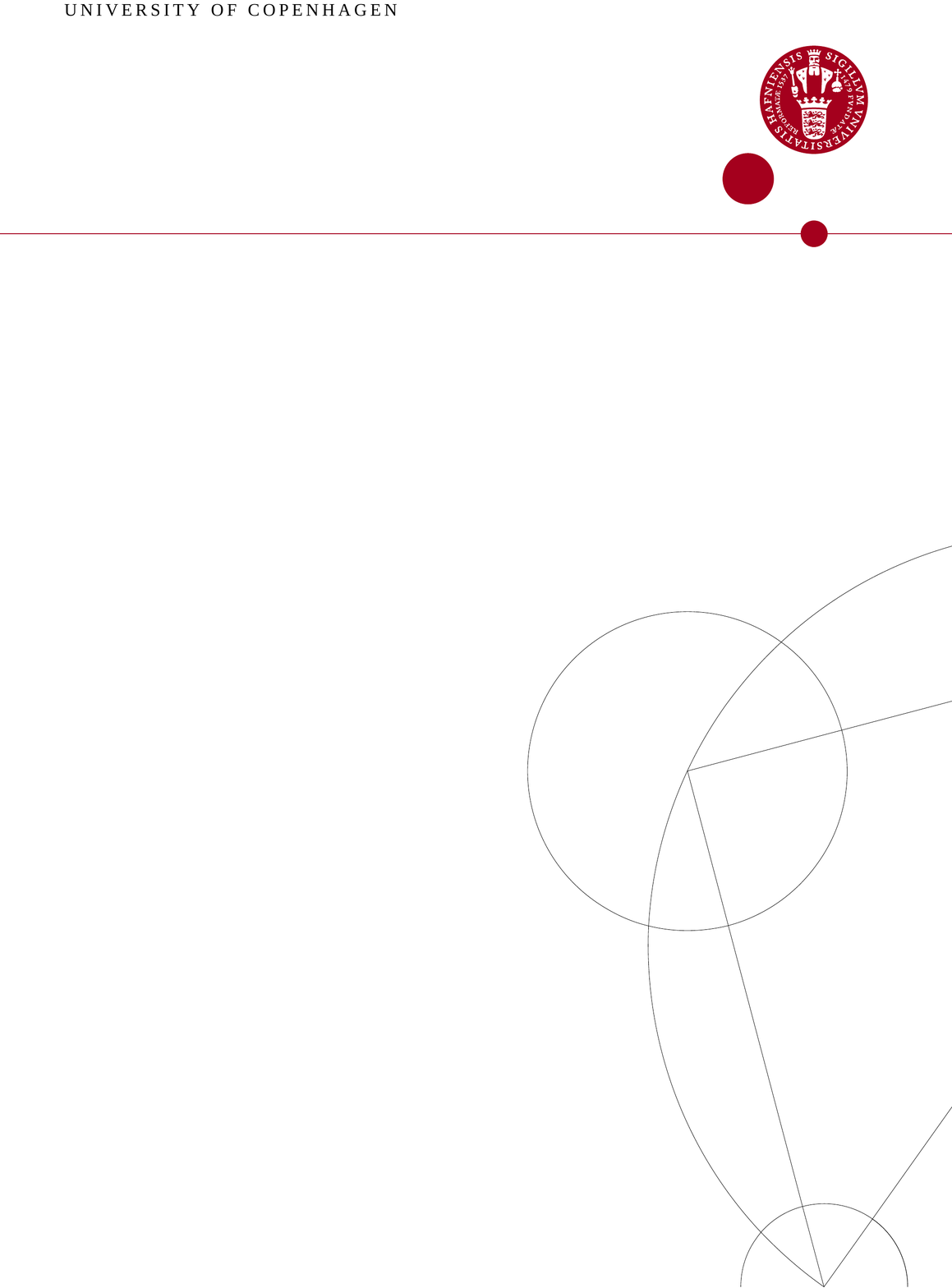}
}}}

\newcommand{\al}[1]{\begin{align}#1\end{align}}

\newcommand{\pdd}[3]{\frac{\partial^2 #1}{\partial #2 \partial #3}}
\newcommand{\pd}[2]{\frac{\partial #1}{\partial #2}}
\newcommand{\PD}[2]{\frac{\partial^2 #1}{\partial #2 ^2}}
\newcommand{\hhat}[1]{\Hat{\Hat{#1}}}

\newenvironment{slant}[1]{
\begin{quote}
   #1 \quad
  \itshape
}{
\end{quote}
}

\title{Supernovae as cosmological probes}
\author{Jeppe~Tr\o st~Nielsen}

\begin{document}

\newgeometry{left=3cm, right=3cm,bottom=0cm}
\AddToShipoutPicture*{\BackgroundPic}

\thispagestyle{empty}
\begin{flushleft}
\vspace*{5cm}
\begin{center}
\Huge{Supernovae as cosmological probes} \\
\end{center}
\vspace*{1.5cm}
Jeppe Tr\o st Nielsen \\ \vspace{6mm}

Niels Bohr International Academy, Niels Bohr Institute, University of Copenhagen\\ \vspace{6mm}
July 24, 2015\\ \vspace{80mm}

 Thesis submitted for the degree of MSc in Physics \\ \vspace{6mm}
Academic supervisors: \\
Subir Sarkar and Alberto Guffanti

\end{flushleft}

\setcounter{tocdepth}{1}

\pagenumbering{gobble}

\clearpage
\restoregeometry

\begingroup
\let\cleardoublepage\relax
\chapter*{Foreword}
\endgroup
\noindent
This work started as a wild goose chase for evidence beyond any doubt that supernova data show cosmic acceleration. Through a study involving artificial neural networks\footnote{This subject is interesting in its own right, but I will not have the space to go into any detail about it.}, trying to find parametrisation free constraints on the expansion history of the universe, we ran into trouble that led us all the way to reconsider the standard method. We encountered the same problem that has been lurking in many previous studies, only in an uncommon disguise. Solving this problem for the neural networks degenerated into solving the original problem. Having done that, it turned out that this result in itself is interesting. This thesis is more or less laying out the article \cite{Nielsen:2015pga} in all gory details. 

The level of rigour throughout is kept, I think, sufficient but not over the top --- particularly the chapter on statistics suffers at the hands of a physicist. I have tried to keep unnecessary details out of the way in favor of results and physical insight. I leave the details to be filled in by smarter people --- well, more interested people.

A bunch of humble thanks to everyone who helped me at the institute, including --- but presumably not limited to --- Laure, Andy, Chris, Anne Mette, Sebastian, Assaf, Morten, Jenny, Christian, Tristan, and the rest of the Academy and high energy groups, and of course Helle and Anette without whom our building would crumble. 

Thanks to Alberto and Subir for company and supervision on this trip through cosmology and data analysis. Obviously I extend my gratitude to Subir for teaching me the most valuable lesson leading to this work: don't believe any analysis you can't understand and, if time permits, carry out the analysis yourself. The following is my attempt at understanding the analysis of supernova data. 

\clearpage
\begingroup
\let\cleardoublepage\relax
\chapter*{Abstract}
\endgroup
\noindent
The cosmological standard model at present is widely accepted as containing mainly things we do not understand. In particular the appearance of a Cosmological Constant, or dark energy, is puzzling. This was first inferred from the Hubble diagram of a low number of Type Ia supernovae, and later corroborated by complementary cosmological probes. 

Today, a much larger collection of supernovae is available, and here I perform a rigorous statistical analysis of this dataset. Taking into account how the supernovae are calibrated to be standard candles, we run into some subtleties in the analysis. To our surprise, this new dataset --- about an order of bigger than the size of the original dataset --- shows, under standard assumptions, only mild evidence of an accelerated universe.

\tableofcontents
\clearpage

\thispagestyle{empty}

\chapter{Introduction}
\pagenumbering{arabic}
The present standard model of cosmology explains quite well a host of observations. The inclusion of a cosmological constant in Einstein's equations combined with the assumed homogeneous and isotropic Friedmann-Robertson-Walker metric description of spacetime gives us the hailed $\Lambda$CDM model. $\Lambda$ for the inferred cosmological constant, more popularly known as \emph{dark energy}, and CDM is the  \emph{cold dark matter}. \emph{Dark} because we can't see it, and \emph{cold} because apparently it behaves like non-relativistic particles --- compared to (almost) massless particles, like neutrinos, which are \emph{hot}. The \emph{baryonic} matter\footnote{This includes all particles of the standard model of particle physics, not just baryons.} is a minor component of the content of the universe. 

The usual starting point of the history of modern cosmology is the two groups studying supernovae at the end of the nineties, \citep{Perlmutter:1998np,Riess:1998cb}. With observations of very far-away supernovae, the two teams independently claimed that the Hubble expansion rate is accelerating and inferred from that a best-fit universe with a cosmological constant density parameter around $0.7$. These results followed a massive experimental effort to find, classify, and calibrate the supernovae. 

The big bang picture of the universe had emerged long before then. From extrapolating the expansion of the universe back in time, it was realised that in the past, the universe will have been much denser and much hotter. Two consequences of this is the cosmic microwave background (CMB) and a particular abundance of light elements, in particular $^4$He, in the early universe --- which is of course altered during the history of the universe. Both these phenomenas are observable today,\footnote{Don't mention the lithium problem! \citep{Cyburt:2008kw}} and confirm to a high degree this picture of a hot plasma filling the universe. Since Penzias and Wilson first saw a glimpse of the cosmic radiation, many experiments have come to the same conclusion. The three latest spaceborne missions, COBE, WMAP, and Planck, have, one after the other, measured to unprecedented precision the spectrum, and lately there has been a spur of interest in detecting gravitational waves in the hopes of information about the \emph{inflationary} stage --- even before the hot plasma!

Since mid-2000, another probe has also come into light. Baryon accoustic oscillations (BAO) are the remnant effects of soundwaves in the primeval plasma, which are supposed to enhance the matter correlation function at a particular scale --- even in the late universe. Other constraints on the model come from more sides than I can hope to do justice here. Large scale structure surveys, gravitational lensing surveys etc., all help to constrain parameters of the model. Supernova observations have since the late nineties been one of the major players in cosmology. They, along with BAO and CMB observations are now the three major pillars of any analysis --- an analysis of one will usually include the constraints of the others when quoting final results. Amazingly, these three observables apparently agree that the universe is indeed mostly cosmological constant and cold dark matter.

In the following I focus on the analysis of supernovae, in particular by performing a maximum likelihood analysis to put constraints on the cosmological model parameters. On the way, we will look at some of the problems of the standard model of cosmology and the standard treatment of the supernova data. I hope to have made the whole thing reasonably self contained. 

I first present all the needed statistical tools in Chap.~\ref{cha:sta}. This is followed by a description of the cosmology we will look at in Chap.~\ref{cha:cos} and the observations of supernovae in Chap.~\ref{cha:sup}. Finally a presentation of the main analysis and result is in Chap.~\ref{cha:put} and some concluding remarks in Chap.~\ref{cha:con}.

\chapter{Statistics}\label{cha:sta}
Statistics is an old, well studied subject, from which physicists take that everything is distributed as gaussians and counting experiments have Poisson statistics. In the present section I hope to clarify why this is the case, and to which extent it is true. The main approach will be what is now known as frequentist, but Bayesian statistics will also be described briefly. For a vivid discussion of the differences between the two, see eg. \citep{fisher1930inverse}.

\section{Probabilities}
I will start with the basics. We write the probability for some event, call it $A$, to happen $P(A)$. One immediate statement is that the universe is \emph{unitary}, which is to say that \emph{something must happen}, so the sum of all probabilities must be one: $\sum_A P(A)=1$. If the outcome $A$ is dependent on some other observation $B$, we write the probability of $A$ to happen, given $B$ as $P(A|B)$. This quantity is in general different from $P(A)$. We can connect the two through summing over the possible outcomes of the event $B$,
\al{\label{eq:stat:sumrule}
P(A) = \sum_B P(A|B)P(B) 
}
We may also consider the joint probability of both events $A$ and $B$ to happen, $P(AB)$. We may now expand this as the probability of just one of the events happening times the probability of the other happening --- given the other. In equations,
\al{\label{eq:stat:bayesdis}
P(AB) = P(A|B)P(B) = P(B|A)P(A)
}
The second step follows from the symmetry of $A$ and $B$. The second equality is known as \emph{Bayes' theorem}. This is what underlies Bayesian statistics --- but it is certainly true whether one is Bayesian or not. 

If we wish to describe outcomes which are not discrete (like heads or tails) but rather continuous, we want to consider instead of just probabilities, a \emph{probability density function} (pdf). To motivate this, consider an infinite number of possible outcomes of an experiment. Then the probability for any individual outcome in general vanishes. This is what the pdf sorts out for us. Say $A$ is a real number we are trying to predict. Then the pdf $f(A)$ is defined to fulfil
\al{
P( A \in [A_{min}, A_{max}] ) =  \int_{A_{min}}^{A_{max}} f(A) dA
} 
This definition is trivially extended to multiple dimensions by simply extending $A$ and generalising the interval. We may write, generally
\al{
P(A\in \Omega) = \int_\Omega f(A)dA,
}
where $\Omega$ is some volume in the space of possible $A$s. As before, the integral over all possible outcomes must be $1$ by unitarity. We note that by putting in delta functions in the above pdfs, we can go back to the discrete picture. Say there are only discrete outcomes $A_i$ of $A$ with probabilities $p_i$, respectively. I can then write the pdf as 
\al{
f(A) = \sum_i \delta(A-A_i) p_i
}
What shall interest us most here are continuous distributions, ie. pdfs. The Eqs.~\eqref{eq:stat:sumrule}-\eqref{eq:stat:bayesdis} extend to
\al{
f(A) =& \int f(A|B) f(B) \ dB \label{eq:stat:intout}\\
f(A|B)f(B) =& f(B|A)f(A) 
}
Note the abuse of notation that $f$ may vary according to the argument. If nothing else is explicit, it is simply to be understood as \emph{the pdf of the argument}.

\section{Expectations}
To any pdf $f(A)$, where $A$ may generally describe a set of multiple parameters, $A=\{ a_1,a_2\dots a_n\}$, we define the \emph{expectation value}\footnote{Note that the \emph{expectation} value is not necessarily what we \emph{expect}. Indeed we may have the situation that $f(\langle A \rangle) = 0$, ie. we have no chance of obtaining the expected value! For this reason, one commonly uses \emph{average} and \emph{mean} to mean the same thing. The most expected value, ie. the value with the highest probability density is called the \emph{mode}.} of a quantity $B(A)$ as
\al{
\langle B \rangle = \int f(A) B \ dA
}
Special cases of this are the average $\mu = \langle A \rangle$ and variance $\sigma^2 = \langle A^2 - \langle A \rangle^2 \rangle$ of a distribution. For some distributions these integrals may not converge, in which case extra care has to be taken. A particular, not immediately interesting, average is the following function of $k$,
\al{
\tilde f(k) = \langle e^{ikA} \rangle = \int f(A) e^{ikA} \ dA,
}
called the \emph{characteristic function}. Obviously, this is just the fourier transform of the pdf\footnote{Up to a constant in front of the integral, depending on your convention.}. The significance of this particular function becomes evident when considering sums of random variables. Take the sum of the independent random variables $\{X_i\}$. The characteristic function of this is the expectation value of $\exp ik\sum_i X_i\equiv\exp ikY$. Writing the exponential in two different ways, we see that the characteristic function of the sum is just the product of the characteristic functions of the summands,
\al{\label{eq:stat:charY1}
\tilde f_Y(k) = \langle \exp ikY \rangle = \prod_i\langle\exp ikX_i \rangle = \prod_i \tilde f_{X_i}(k)
}
Let's see how this works in practice by some examples.

\begin{slant}{The $\chi^2$ distribution}
Consider $\nu$ independent random variables, all drawn from normal distributions. We denote this as\footnote{Seeing $X$ as a vector, I will write $X\sim\mathcal N(\mu,\Sigma) \Rightarrow f_X(X) = |2\pi\Sigma|^{-1/2}\exp(-X^T\Sigma^{-1}X/2)$ to denote a multivariate normal distribution.}
\al{
X_i &\sim \mathcal N(0,1) \nonumber\\
f_X(X_i) &= (2\pi)^{-1/2} \exp (- X_i^2/2),
}
We are now interested in the pdf $f_{\chi^2}$ of $Y = \sum_i^\nu X_i^2$, called the $\chi^2$ distribution with $\nu$ degrees of freedom. We will use that we know how to go back again from the characteristic function, simply by an inverse fourier transform. First writing down the characteristic function, I denote $Z_i=X_i^2$,
\al{
\tilde f_{\chi^2}(k) = \int \prod_i e^{ikZ_i} f_Z(Z_i) \ dZ_i = \prod_i \tilde f_Z(k)
}
since $Y$ is the sum of the $Z_i$s, the characteristic function is just the product of the characteristic functions of the summands. Now we need first the characteristic function for the square of a single normally distributed variable\footnote{Which is the $\chi^2$ distribution with $1$ degree of freedom.}. We find for the pdf of $Z$,
\al{
f_{X^2}(Z) =& \int f_X(X) \delta(Z-X^2) dX = \int f_X(X) \frac{\delta( \sqrt{Z} - X )+\delta( \sqrt{Z} + X )}{2|X|}\ dX \nonumber \\
 =& (2Z\pi)^{-1/2} \exp(-Z/2), \hspace{1cm} Z>0
}
Where the second equality follows from the identity,
\al{\label{stat:eq:diracdelta}
\delta( g(x) ) = \sum_{x_i} \frac{\delta(x-x_i)}{ | g'(x_i) | }
}
where the $x_i$ are the roots of $g$. The proof of Eq.~\eqref{stat:eq:diracdelta} follows by a change of variables in the integral.\footnote{Remember the $\delta$ function only formally makes sense inside an integral.} The characteristic function is then
\al{
\tilde f_{X^2}(k) =& \int e^{ikZ} f_{X^2}(Z)\ dZ =(2\pi)^{-1/2} \int Z^{-1/2} e^{Z(ik-1/2) } \ dZ \nonumber \\
=& (2\pi)^{-1/2} \int e^{(2ik-1) X^2/2} dX = \frac{1}{\sqrt{1-2ik}}
}
From Eq.~\eqref{eq:stat:charY1} we now see by multiplication and taking the inverse fourier transform that
\al{\label{eq:stat:chi2dist1}
\tilde f_{\chi^2}(k) = \frac{1}{(1-2ik)^{\nu/2} } \Rightarrow f_{\chi^2}(Y) = \frac{1}{2\pi} \int dk \frac{\exp(-ikY)}{(1-2ik)^{\nu/2} }
}
This last one is a tricky integral. Anticipating the correct answer, I rewrite it as
\al{
\frac{1}{2} \exp(-Y/2) \left( \frac{Y}{2} \right) ^{\frac{\nu}{2} -1} \frac{1}{2\pi i} \int_{\infty}^{-\infty} e^{Y/2-ikY}\frac{-i Y\ dk}{(Y/2-ikY)^{\nu/2}} 
}
Here I have simply pulled some functions of $Y$ outside the integral and the inverse inside the integral. Changing variables to $s = -ikY+Y/2$, we get
\al{
\frac{1}{2} \exp(-Y/2) \left( \frac{Y}{2} \right) ^{\frac{\nu}{2} -1} \frac{1}{2\pi i } \int_{-i \infty +Y/2}^{i \infty + Y/2} e^s s^{-\nu/2} ds
}
To solve this last integral, we are inspired by how it looks like an inverse Laplace transform, \citep{hankel1864}. Consider first the integral representation of the $\Gamma$ function, which can be moulded to look like a Laplace transform by a change of variables,
\al{\label{eq:stat:gammafun}
\Gamma(z) =& \int_0^\infty t^{z-1} e^{-t} dt = \int_0^\infty (su)^{z-1} e^{-su} s\ du \\
\Rightarrow& \frac{\Gamma(z)}{s^z} =  \int_0^\infty u^{z-1} e^{-su} du = \mathcal L(u^{z-1})
}
We now invert this and find $u^{z-1}$ as the inverse Laplace transform of the left hand side, 
\al{
u^{z-1} =& \frac{1}{2\pi i}\int_{-i \infty +\lambda}^{i \infty + \lambda} e^{su} \frac{\Gamma(z)}{s^z}\ ds \nonumber \\
\Rightarrow \frac{1}{\Gamma(z)} =&\frac{1}{2\pi i} \int_{-i \infty +\lambda}^{i \infty + \lambda} e^{su} (su)^{-z} u\ ds = 
\frac{1}{2\pi i}\int_{-i \infty +\tilde\lambda}^{i \infty + \tilde\lambda} e^{\tilde s} \tilde s^{-z} d\tilde s
}
It is now evident from inserting $z=\nu/2$ and $\tilde\lambda= Y/2$, that we get for Eq.~\eqref{eq:stat:chi2dist1}

\al{
f_{\chi^2}(Y) = \frac{1}{2\Gamma(\nu/2)}  \left( \frac{Y}{2} \right) ^{\frac{\nu}{2} -1}\exp(-Y/2)
}
\end{slant}
The $\chi^2$ distribution is widely used in statistical analysis, and we shall see why later on. 

Another application of characteristic functions is a derivation of the \emph{central limit theorem}, which goes as follows.

\begin{slant}{The central limit theorem}
This theorem states that asymptotically, the sum of many random variables will converge to a normal distribution --- \emph{almost} irrespective of the original distributions! We will again use the fact that the characteristic function of a sum is the product of characteristic functions. Define $Y=\sum_i X_i/\sqrt{N}$, where the $X_i$ are independently, identically distributed \emph{(iid.)} variables,
\al{
f(X_1) = \dots = f(X_N) 
}
We are now interested in $f_Y$ in the limit $N\rightarrow\infty$. Assume first that $f$ has a well defined variance $\sigma^2$ and zero mean $\mu=0$.\footnote{This can always be arranged by simple subtraction.} Now expand the characteristic function to second order in $k$ and write
\al{\label{eq:stat:sumclt}
\tilde f_Y(k)  =& \tilde f(k/\sqrt N)^N =\prod_i \langle e^{ikX_i/\sqrt{N}} \rangle = \left( 1 -\frac{k^2 \sigma^2}{2N} + \mathcal O\left(\frac{k^3}{N^{3/2}}\right) \right)^N \nonumber \\
\approx& \exp\left(- \frac{k^2 \sigma^2}{2} +\mathcal O\left(\frac{k^3}{N^{1/2}}\right) \right)
}
Now we calculate the characteristic function of a general normal distribution,
\al{\label{eq:stat:normchar}
f(x) &=  \mathcal N(a,b) = \frac{1}{b\sqrt{2\pi}}\exp\left( \frac{(x-a)^2}{2b^2} \right) \nonumber \\
\Rightarrow \tilde f(k) &= \int dx e^{ikx}  \frac{1}{b\sqrt{2\pi}}\exp\left( \frac{(x-a)^2}{2b^2} \right)
= \exp\left( iak - \frac{k^2b^2}{2} \right)
}
Comparing Eqs.~\eqref{eq:stat:sumclt} and \eqref{eq:stat:normchar}, we see that the two match if we identify 
\al{
\mu_Y = 0\\ 
\sigma^2_Y = \sigma^2
}
Thus the distribution of a sum of many iid. random variables converges to a normal distribution. This underlies many assumptions made in statistical treatments of errors and uncertainties. 
\end{slant}

A closely related concept to the characteristic function is the \emph{moment generating function}. This is constructed by simply taking $k$ imaginary in the characteristic function,
\al{
M(k) = \int f(x) e^{xk} \ dx = \langle e^{xk} \rangle = \tilde f ( -ik )
}
The nice property of this function is that we can, as the name suggests, generate the moments, $\langle x^n \rangle$ of a distribution. Having all the moments of a distribution defines it uniquely\footnote{This is easily realized with the connection to the fourier transform, which is one-to-one with the original distribution}. To generate the moments, we do the following,
\al{
\langle x^n \rangle = \int x^n f(x) \ dx = \left. \left(\pd{}{k}\right)^n M(k) \right|_{k=0}
}
We can eg. calculate the first two moments of the $\chi^2$ distribution. First, the moment generating function is
\al{
M_{\chi^2} (k)= \tilde f_{\chi^2} ( - ik ) = (1-2k)^{-\nu/2}
}
We then find easily by direct differentiation
\al{
\langle x \rangle_{\chi^2} &= \left. \pd{}{k}(1-2k)^{-\nu/2} \right|_{k=0} = \left. \nu (1-2k)^{-(\nu/2 +1)} \right|_{k=0} = \nu \\
\langle x^2 \rangle_{\chi^2} &= \nu \left. \pd{}{k}(1-2k)^{-(\nu/2+1)} \right|_{k=0}
	= \nu (\nu + 2) \left. (1-2k)^{-(\nu/2+2)} \right|_{k=0} = \nu(\nu+2)
}
Recognising a pattern immediately, we boldly write down the general formula for the $n^{th}$ moment, which can be proven by simple induction,
\al{
\langle x^n \rangle_{\chi^2} = \nu(\nu + 2)\cdots (\nu + 2(n-1) ) = \prod_{i=0}^{n-1} (\nu + 2i)
}

\section{Common distributions}
Some distributions are used more than others, and the normal distribution more than any. In this section, I want to introduce a few common examples of probability distributions. A curious property of the normal distribution is that many other distributions asymptotically converge to it. We will see here exactly how this comes about. This combined with the central limit theorem are the reasons why almost all statistics is carried out with normal distributions.

\subsection{The Poisson distribution}
The Poisson distribution describes the probability of obtaining $N$ successes, eg. a number count of cosmic rays or photons from some cosmic event, in a fixed time interval, if the average rate is fixed and the different successes are uncorrelated. That is, any success is independent from another. Call the rate $\lambda$, then the probability is 
\al{
P(N;\lambda) = \frac{\lambda^N  }{N!} e^{-\lambda}
}
This simply reflects the relative probability of obtaining $N$ successes in the fraction, taking into account combinatorics, along with a normalisation $e^{-\lambda}$, such that $\sum_N P(N;\lambda) = 1$.

\begin{figure}[htb]
\begin{center}
\includegraphics[width=.7\textwidth]{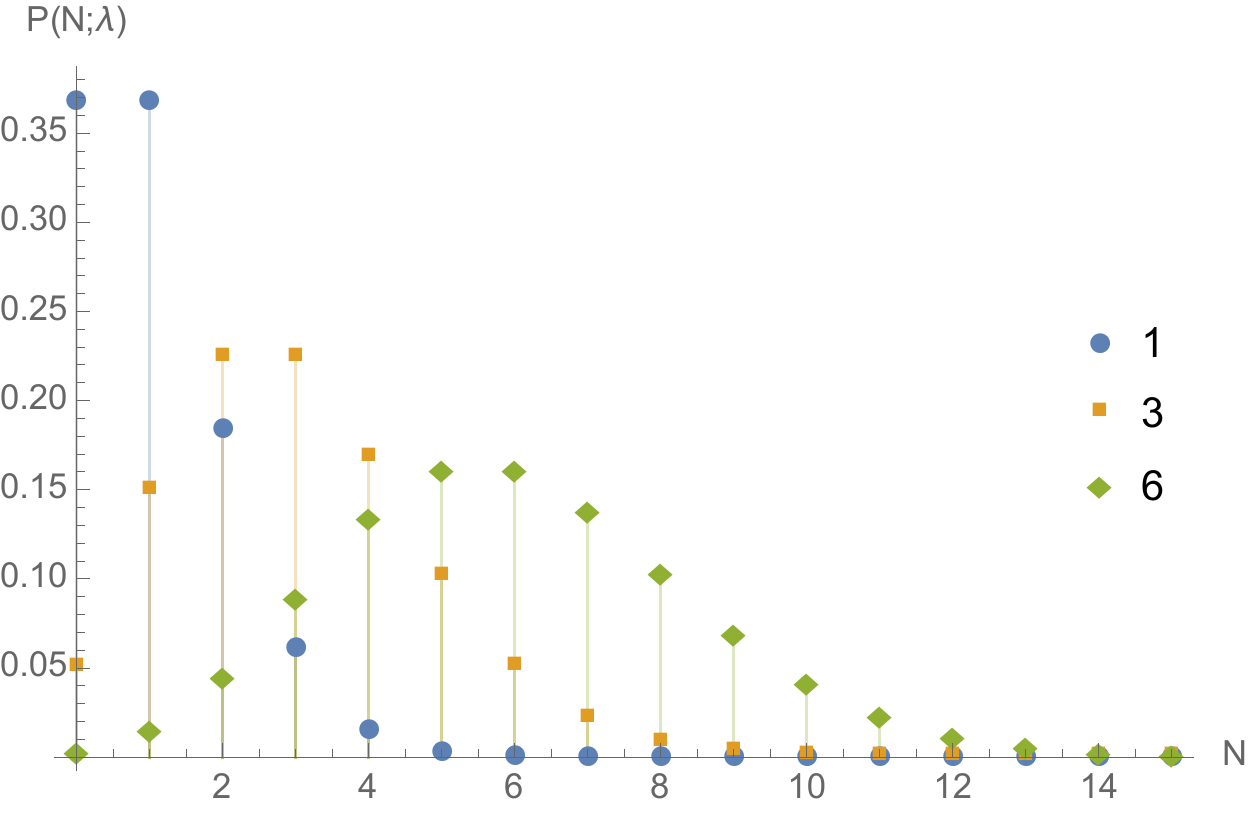}
\caption[Poisson distribution]{Examples of the Poisson distribution for various values of $\lambda$ as described in the legend.}
\label{fig:stat:poisson}
\end{center}
\end{figure}

We can find the mean and standard deviation by direct summation, 
\al{
\langle N\rangle =& \sum_N^\infty N P(N;\lambda) = \lambda e^{-\lambda} \sum_N^\infty \frac{\lambda^N  }{N!} = \lambda \\
\langle N^2\rangle =& \sum_N^\infty N^2 P(N;\lambda) = \lambda e^{-\lambda} \sum_N^\infty (N+1)\frac{\lambda^N  }{N!} \nonumber \\
 =& \lambda e^{-\lambda} (\lambda+1) \sum_N^\infty \frac{\lambda^N  }{N!} = \lambda(\lambda+1) \\
 \Rightarrow \sigma^2 =& \ \lambda
}

Now let's take the limit $\lambda\gg 1$. This means the mean, as we just calculated, is also very large, and we allow ourselves to expand around it, parametrising the distribution with the continuous $N(\delta) = \lambda(1+\delta)$, where the region of interest is $|\delta| \ll 1$. Before things get interesting, we need an intermediate result, known as \emph{Stirling's approximation}. This is basically an expansion of the $\Gamma$ function defined above in Eq.~\eqref{eq:stat:gammafun}. Since $n! = \Gamma(n+1)$, we have
\al{
n! = \int x^n e^{-x} \ dx = \int e^{n\log x-x} \ dx
}
Now I expand the content of the exponential around the maximum at $x_0 = n$. This becomes
\al{
n\log x-x \approx n \log n - n + \frac{1}{2n}(x-n)^2
}
Inserting this into the integral, we have
\al{
n! \approx n^n e^{-n} \int_0^\infty e^{(x-n)^2/2n} \ dx \approx n^n e^{-n} \sqrt{2\pi n}
}
where the last integral is done taking the lower limit to minus infinity, as we take $n\gg 1$. Now put all this back into the distribution function,
\al{
f(\delta;\lambda) &\approx \frac{\lambda^N }{N^N e^{-N} \sqrt{2\pi N}} e^{-\lambda} =
	 \exp\left\{ \lambda\delta - (\lambda[1+\delta]+1/2)\log(1+\delta) \right\}\frac{1}{ \sqrt{2\pi \lambda} } \nonumber \\
	 &\approx \frac{1}{\sqrt{2\pi \lambda}}\exp\left\{ - \frac{\lambda \delta^2}{2}  \right\}
	 = \frac{1}{\sqrt{2\pi \lambda}}\exp\left\{ - \frac{(N-\lambda)^2}{2\lambda}  \right\}
}
where the last approximation expands the content of the exponential to second order in $\delta$ and uses $\lambda\gg 1\gg\delta$. We finally see here the result we might have anticipated, we simply insert the mean and variance of the Poisson distribution in the normal distribution to get the asymptotic expression for the former.

\subsection{The binomial distribution}
This distribution comes about when looking at binary outcomes of a repeated experiment, like a series of coin flips. If the probability of the coin landing heads is $p$, then after $N$ experiments, the probability of obtaining exactly $n$ heads is
\al{
P(n;N,p) = \begin{pmatrix} N \\ n \end{pmatrix} p^n (1 - p)^{N-n}
}
The first factor on the right hand side is the binomial coefficient
\al{
\begin{pmatrix} N \\ n \end{pmatrix} = \frac{N!}{n!(N-n)!},
}
which takes care of the combinatorics of the different orders of obtaining the $n$ heads. Note that here we have a fixed number of repetitions, where in finding the Poisson distribution, we had a fixed time interval.

\begin{figure}[htb]
\begin{center}
\includegraphics[width=.7\textwidth]{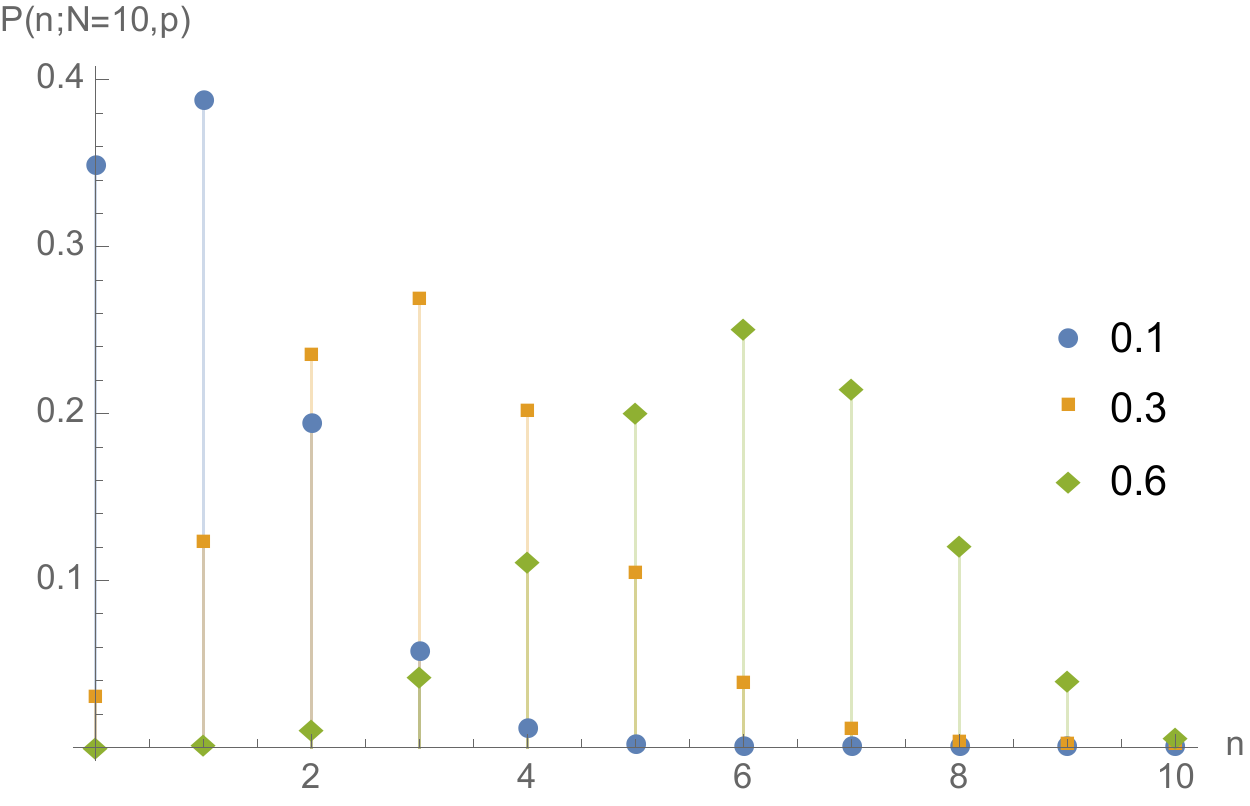}
\caption[Binomial distribution]{Examples of the binomial distribution for various values of $p$, but fixed $N=10$.}
\label{fig:stat:bino}
\end{center}
\end{figure}

We find again the mean and variance
\al{
\langle n \rangle &= \sum_{n=0}^N n P(n;N,p) = Np \sum_{n=1}^N \frac{(N-1)!}{(n-1)!(N-n)!} p^{n-1}(1-p)^{N-n} \nonumber \\
	&= Np \sum_{n=0}^{N-1} P(n;N-1,p) = Np \label{eq:stat:binomialmean} \\
\langle n^2 \rangle &=  Np \sum_{n=0}^{N-1} (n+1) P(n;N-1,p) = Np( [N-1]p +1 ) = (Np)^2 + Np(1-p) \nonumber \\
\Rightarrow \sigma^2 &= Np(1-p) \label{eq:stat:binomialstd}
}

Now consider the double limit $N\rightarrow\infty, p\rightarrow 0$ with the product $Np = \lambda$ fixed. Rewriting the probability distribution using $n\ll N$, we get
\al{
P(n;\lambda) &= \lim_{N\rightarrow\infty} \frac{N!}{n!(N-n)!} \left(\frac{\lambda}{N}\right)^n \left(1-\frac{\lambda}{N}\right)^{N-n} \nonumber \\
&= \frac{\lambda^n}{n!} \lim_{N\rightarrow\infty} \frac{N\cdots(N-n+1)}{N^n} \left(1-\frac{\lambda}{N}\right)^{N-n} \nonumber \\
&\approx \frac{\lambda^n}{n!} e^{-\lambda}
}
which is just the Poisson distribution. That means that for a large amount of trials with vanishing probability per trial, the binomial distribution looks just like the Poisson distribution. This makes sense, since we can exactly interpret the infinite trials as being done in continuous time with vanishing probability, such that $Np$ is the rate of success. Taking $\lambda\gg 1$ of course brings us to the gaussian limit again.

\subsection{The $\chi^2$ distribution}
We have already seen what this distribution is, along with its moments. Here I quickly show how also this distribution asymptotically looks like a gaussian. I again use Stirling's approximation to write, in the limit $\nu\rightarrow\infty$, and writing temporarily $x=\nu(1+\delta)$,
\al{
f(x) &= \frac{\nu/2+1}{\sqrt{4\pi \nu }}\left(\frac{e}{\nu/2}\right)^{\nu/2} \left(\frac{\nu}{2} \right)^{\nu/2 - 1} (1+\delta)^{\nu/2-1} e^{-\nu(1+\delta)/2} \nonumber \\
&\approx \frac{1}{\sqrt{2\pi (2\nu)} } e^{-\nu\delta/2 + (\nu/2-1)\log(1+\delta)} \approx  \frac{1}{\sqrt{2\pi (2\nu)} } \exp\left\{ -\frac{\nu\delta^2}{4} \right\} \nonumber \\
&=  \frac{1}{\sqrt{2\pi (2\nu)} } \exp\left\{-\frac{(x-\nu)^2}{2(2\nu)} \right\}
}
which again is simply a normal distribution with the expected mean and variance.

\begin{figure}[htb]
\begin{center}
\includegraphics[width=.7\textwidth]{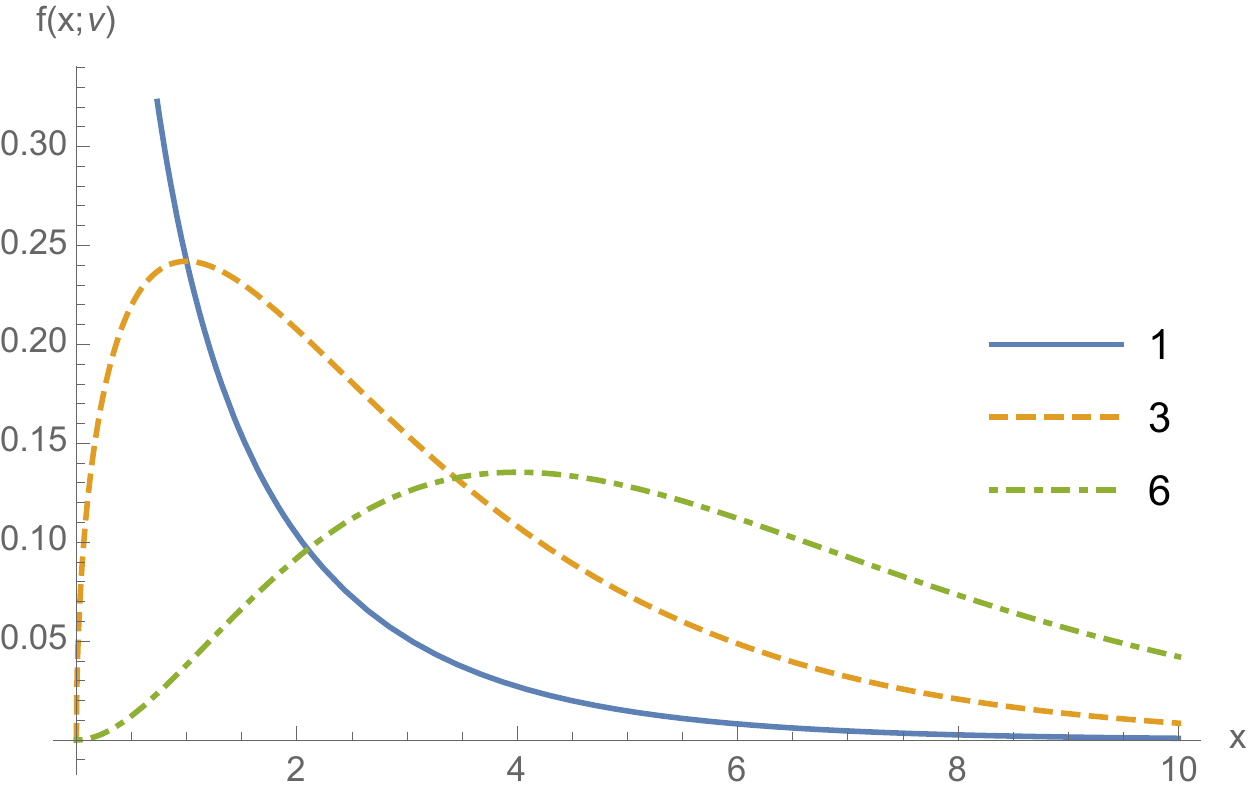}
\caption[$\chi^2$ distribution]{Examples of the $\chi^2$ distribution for various values of $\nu$.}
\label{fig:stat:chi2}
\end{center}
\end{figure}

\section{Parameter estimation}
An ideal theory will naturally explain all constants involved in it. That means we would very simply be able to compare predictions of this theory with an experiment. However, this is usually not the case. What happens most often is that a theory will contain some unexplained parameter(s), which must be \emph{fitted}. Supposing the model is true, we can then \emph{constrain} the parameters of the theory with a particular experiment. This notion of fitting is what the current section explores.

We generally have some experiment, which produces random numbers --- due to noise in the experiment or intrinsic variability in the source. How do we compare our model of the experiment to the data produced and in the process fit the parameters of the model? In general these are two different problems, but by the method we are going to use, they can in general be solved simultaneously. The majority of the current section will be about the \emph{likelihood} and in particular maximising the likelihood, along with finding \emph{estimators} of the model parameters.

The likelihood is defined as the pdf of the data, $\hat X$,\footnote{Hatted variables will generally be either observed data or estimators --- both of which are random variables. Unhatted will usually be the corresponding true variable.} given a specific model, which I generically denote $\theta$,
\al{
\mathcal L(\theta) = f(\hat X | \theta)
}
Note the funny semantics --- it is indeed not a probability density of the model, but we still want to link it to some notion of model selection by probability. This has the potential to confuse. One easily avoids this by simply stating what the likelihood is, and never using it as a probability of the model \cite{fisher1930inverse}. Note right away that the likelihood is itself in general a random variable, as are the estimators we are going to derive from it. 

We now define the \emph{maximum likelihood estimators (MLE)}, $\hat\theta$, to be the model parameters, which maximise the likelihood given the obtained data, ie.
\al{
\pd{\mathcal L(\hat\theta)}{\theta}  = 0
}
These estimators generally have nice properties. The most interesting properties can be found exactly in the context of linear models, which is what I discuss next. In the limit of infinite datasets, these properties extend to non-linear models. I will not discuss this in detail, only illustrate it with an example. For a complete description of the problem and its solution, I refer to textbooks on the subject, eg. \citep{rao2009linear}.

\subsection{Linear models}\label{stat:sub:linmod}
Consider a model describing a dataset $\{ \hat x_i, \hat y_i \}, \ i=1\dots N$ as
\al{
y_i(x_i) = \sum_{j=1}^M a_j A_j(x_i)
}
where $M<N$ and the functions $A_j$ are fixed and linearly independent, ie. $\sum_j a_j A_j(x_i) =0 \Rightarrow a_j=0$. These $A_j$ could be monomials, sines and cosines etc. Now assume we measure $x$ with negligible uncertainty and $y$ with some known uncertainty, which we take to be gaussian, ie. $\hat y_i = y_i + \epsilon_i$, where $\epsilon_i \sim \mathcal N(0,1)$\footnote{It is always possible to absorb the variance of $\epsilon$ into the $A$s and thus have unit variance}. We can now write the likelihood,
\al{\label{eq:stat:likeexp}
\mathcal L \propto \exp\left\{ -\frac{1}{2}\sum_i^N  \left(\hat y_i -\sum_{j=1}^M a_j A_j(\hat x_i) \right)^2 \right\}
}
The constant of proportionality just normalises the likelihood. Now we want to maximise this likelihood as a function of the $a_j$s --- the unknown model parameters. Because the exponential is a bit unwieldy, we take the $\log$ and a factor $-2$ out, and instead of maximising $\mathcal L$, we minimise $-2\log\mathcal L$. The reason for this will hopefully become clear. To find the minimum, we simply solve for the differential to be zero.\footnote{And show that it is indeed a minimum, not a maximum or saddlepoint.} Doing this, we get a set of $M$ equations for the $M$ $a_j$s,
\al{
\pd{\left(-2\log\mathcal L(\hat a_j)\right)}{a_j} = 0 = - 2\sum_i A_j(\hat x_i ) \left( \hat y_i -\sum_{j'=1}^M \hat a_{j'} A_{j'}(\hat x_i) \right)
}
Since we know linear algebra, and this looks an awful lot like it, we drop the indices and see everything as vector-/matrix products. I explicitly define the elements of the matrix $A$ as $A_{ji} = A_j(\hat x_i)$, and the sum now looks like
\al{\label{eq:stat:mleexp}
0 = A (\hat y - A^T \hat a) \Rightarrow \hat a = (AA^T)^{-1} A \hat y
}
The matrix $A$ was defined to have linearly independent rows, which in turn means the inverse of $AA^T$ exists. The proof of this is as follows. Define $S = AA^T$. Any positive-definite matrix is invertible, so I want to show $S$ is positive definite. We have straight forwardly that for any $X$,
\al{
X^T S X = X^T AA^T X = |X^T A |^2 \geq 0
}
which shows it is positive semi-definite. Now we need to show that if the product is exactly $0$, then so is $X$. Remember the functions $A_j$ were assumed to be linearly independent, which means 
\al{
a^T A =0 \Rightarrow a = 0
}
This is exactly what we need, since if we write 
\al{
0 = X^TSX = X^T AA^T X = | X^T A |^2 \Rightarrow X^T A= 0 \Rightarrow X=0
}
This means that $S$ is indeed positive definite and the inverse $(AA^T)^{-1}$ exists.

Now we are interested in two things: the distribution of $-2\log\mathcal L(\hat a_j)$ and of the estimators $\hat a_j$, under repeated (thought-)experiments\footnote{Of course there is only the one actual experiment, but we might imagine performing it again and again. It is under these repetitions that the estimators are random variables, whose pdfs we want to find.}. We first look at the likelihood.\footnote{Note that I have already thrown away a constant normalisation term. This only shifts the distribution, or rather, the distribution we find is that of $-2\log(\mathcal L\sqrt{2\pi}^N)$.}
\al{
-2\log \mathcal L(\hat a) &= 
| \hat y - A^T \hat a | ^2 = | \hat y - A^T(AA^T)^{-1}A \hat y|^2 \nonumber \\
&= \left| \left( \mathbb{1}_N - A^T(AA^T)^{-1}A \right) \hat y \right|^2
}
Here $P = A^T(AA^T)^{-1}A$ is a projection in the sense $P^2X =PX$ for any $X\in\mathbb R^N$ to an $M$ dimensional subspace. By an orthogonal transformation, we can rotate the $\hat y$ to $\tilde y = \mathcal O \hat y$ such that the projection $P\tilde y$ has its elements only in the first $M$ entries, ie. $P(\tilde y_1,\dots,\tilde y_N)^T = (\tilde y_1,\dots,\tilde y_M,0,\dots,0)^T$. Note that since the transformation is orthogonal, we also have $\tilde y_i \sim \mathcal N(0,1)$. Taking now $\bar y = (\mathbb 1_N - P )\tilde y = (0,\dots,0,\tilde y_{M+1},\dots,\tilde y_N)^T$, the likelihood takes the following form
\al{
-2\log \mathcal L (\hat a)= \bar y^T \bar y = \sum_{i=M+1}^N \tilde y_i^2 \sim \chi^2_{\nu=N-M}
}
This result is the origin of two notions, which are often abused in practice. The first is, we simply call $-2\log \mathcal L$ \emph{the chi squared}, $\chi^2$. This may result in a bit of confusion since now one has a random variable called $\chi^2$, which is $\chi^2$-distributed, ie. its pdf is the $\chi^2$ distribution. The other is the idea of a \emph{reduced number of degrees of freedom}, $\nu = N-M$, ie. the number of data points minus the number of fit parameters. These ideas are widely used even when the model is not linear.

Now we turn to the distribution of the estimators $\hat a$. We have already seen the result, which is
\al{\label{eq:stat:mledist}
\hat a &= (AA^T)^{-1}A\hat y \Rightarrow \hat a \sim \mathcal N(a, (AA^T)^{-1}) =\mathcal N(a,\mathcal I^{-1}),
}
where the normal distribution is to be understood in the multivariate sense. We see here a specific example of a more general result. The MLE is \emph{normally distributed} around the true value --- it is \emph{unbiased} --- with covariance matrix described by\footnote{For two matrices $A,B$, we write $A\geq B$ if $A-B$ is positive semi-definite. A proof of this inequality comes later.}
\al{\label{eq:stat:cramerrao}
\Sigma_{\hat a} &\geq \mathcal I(\hat a)^{-1}, \hspace{1cm}\text{where} \\
 \mathcal I(\hat a)_{ij} &= \left\langle\pdd{(-\log \mathcal L(\hat a))}{a_i}{a_j}  \right\rangle
}
where the average is taken over repeated experiments. $\mathcal I=AA^T$ is called the \emph{Fisher Information}. In this case, the double derivative is a constant, so the average is trivial. This bound on the covariance matrix is called the \emph{Cram\' er-Rao bound}, and is the minimal covariance for unbiased estimators. An unbiased estimator with this minimal variance is called \emph{efficient}. We see that the MLE for linear models are all exactly unbiased, normally distributed, efficient estimators for all $N$. 

The linear models are nice because, as we have just seen, practically everything can be done analytically. This gives us a nice starting point for the next discussion. For a general, non-linear model, the results in the example are no longer valid. Let us explore finite sample sizes with a very simple example.

\subsection{A non-linear model}\label{stat:sec:nonlin}
Consider the data set $\{ \hat x_i \},\ i=1\dots N$, drawn from a normal distribution with unknown mean \emph{and} variance, but with no measurement uncertainty, $x_i\sim\mathcal N(\mu,\sigma)$. The likelihood for this experiment is
\al{
\mathcal L = (2\pi\sigma^2)^{-N/2} \exp\left\{ -\frac{1}{2}\sum_i^N \left( \frac{\hat x_i-\mu}{\sigma}\right)^2 \right\}
}
and we are trying to determine $\mu$ and $\sigma^2$. Note how we cannot neglect the normalisation this time, since we are now fitting $\sigma$. The maximum point $(\hat\mu,\hat\sigma^2)$ is
\al{
\hat\mu &= N^{-1}\sum_i \hat x_i \\
\hat\sigma^2 &= \sum_i (\hat x_i-\hat\mu)^2
}
Now consider the distribution of these estimators. The fact that we don't know $\sigma$ complicates things, since this is what set the scale for us before --- we could measure deviations in terms of a fixed number. Now this scale is a random variable. For instance, we immediately see that $\hat\mu\sim\mathcal N(\mu,\sigma/\sqrt{N})$, but here we've used the unknown $\sigma$ to define the variance. 

We turn therefore first to the distribution of the variance $\sigma^2$. I first write out the $\hat\mu$ and rewrite the sum, giving
\al{
\hat\sigma^2 = N^{-2} \sum_{ij} (\hat x_i-\hat x_j)^2
}
We now need a small trick to evaluate this sum. What we really want --- anticipating the answer --- is something like a sum of squares $\sum x_i x_j$, not of squares of differences, as we have. So we recast it to
\al{
\hat\sigma^2 = N^{-1} \sum_{ij} x_i C_{ij} x_j
}
and find the matrix $C$ we need here is
\al{
C = 
\begin{pmatrix}
1-N^{-1} & -N^{-1} & \cdots\\
-N^{-1} & 1-N^{-1} &  \\
 \vdots& & \ddots
\end{pmatrix}, \ |C| = 0
}
We now pseudo\footnote{\emph{Pseudo} since strictly $C$ is only positive semi-definite.} Cholesky factorise $C$, ie. we find an upper triangular matrix $U$, which satisfies $U^TU = C$. This matrix is
\al{\label{aexplicit}
U_{ij} = \left\{
	\begin{matrix}
	\sqrt{ (N-i)/(N+1-i) } & i=j \\
	-1/\sqrt{(N-i)(N+1-i)} & i<j \\
	0 & i>j
	\end{matrix}
	\right.
}
We now use $U$ to find the rank of $C$, which determines the pdf of the sum. Taking the reverse product, we see that
\al{
UU^\text T = 
\begin{pmatrix}
1 & 0 &\cdots & & \\
0 & 1 &  & \\ 
\vdots&  &\ddots &  \\
& && 1 & 0 \\
& && 0 & 0
\end{pmatrix},
}
which immediately tells us the rank of $C$ is $N-1$. This means $U$ is \emph{almost} an orthogonal transformation --- we just lose one degree of freedom. Thus we will define new variables $y_j = \sum_i U_{ji}x_i,\ j=1\dots N-1$, which are also drawn from independent normal distributions. The variance is now given as
\al{
N \hat\sigma^2 &=  \sum_{ij} x_i C_{ij} x_j = \sum_{ijk} x_i U_{ki}U_{kj} x_j \nonumber\\
 &= \sum_i (Ux)_i^2 = \sum_i^{N-1} y_i^2 \sim \sigma^2 \chi^2_{\nu=N-1}
}
This shows that for finite $N$, the estimator is a bit off, as
\al{
\langle \hat\sigma^2 \rangle = \sigma^2 \frac{N-1}{N}
}
This comes about because we fit the mean while calculating it. The \emph{missing} degree of freedom is of course the mean $\hat\mu$ which we now consider. Had we known $\sigma$, we would immediately write $\sqrt N (\hat\mu-\mu) / \sigma \sim \mathcal N(0,1)$. Exchanging $\sigma$ for $\hat\sigma$, the distribution changes a bit. We may write
\al{
\sqrt N (\hat\mu-\mu)/\hat\sigma = \frac{n}{c} 
}
where $n$ is normally distributed $n\sim\mathcal N(0,1)$ and $c$ follows a $\chi$ distribution, $c\sim\chi_{\nu=N-1}$.\footnote{The $\chi$ distribution is simply the distribution of the square root of a $\chi^2$ random variable.} Note how this combination exactly cancels the dependence of $\sigma$. This particular combination of random variables follows a distribution known as \emph{Student's t-distribution} with $\nu=N-1$ degrees of freedom. Its pdf is
\al{
f(x;\nu) = \frac{\Gamma(\frac{\nu+1}{2})}{\sqrt{\nu\pi}\Gamma(\frac{\nu}{2})} \left( 1+\frac{x^2}{\nu}\right)^{-\frac{\nu+1}{2}}
}

We are now in a position to understand the $N\rightarrow\infty$ limit of the MLE. We see that for finite $N$, neither of the two estimators follow a normal distribution, and $\hat\sigma^2$ is even biased. In the asymptotic limit though, both distributions are normal, and we have
\al{
\sqrt N(\hat\mu-\mu)/\hat\sigma \sim\mathcal N(0,1) &\Rightarrow \hat\mu\sim\mathcal N(\mu,\hat\sigma/\sqrt N)	\\
N\hat\sigma^2/\sigma^2 \sim\mathcal N(N,\sqrt{2N}) &\Rightarrow \hat\sigma^2\sim \mathcal N(\sigma^2,\sqrt\frac{2}{N} \sigma^2)
}
It is only in the asymptotic limit the estimators follow an unbiased normal distribution, with variance given by Eq.~\eqref{eq:stat:cramerrao}. As I showed earlier, many distributions tend to a normal distribution for large $N$. This is what is happening here too. In this limit, the likelihood tends to a normal distribution, for which the results from the previous section hold.

\subsection{Cram\' er-Rao lower bound}
Now let us see how the Cram\' er-Rao bound appears. I will follow the proof from \citep{rao2009linear}. Assume we have a set of unbiased estimators $\{\hat g_i\},i=1\dots r$, of the quantities $\{ g_i \}$, ie. $\langle \hat g_i \rangle = g_i$. The likelihood function generally depends on some parameters, say $\theta_j, j =1\dots k$. We now construct another set of variables, $\{\pd{\log\mathcal L}{\theta_j} \}$, and build the $r+k$-vector $\{\hat g_1\dots \hat g_r,\pd{\log\mathcal L}{\theta_1}\dots\pd{\log\mathcal L}{\theta_k} \}$. The covariance matrix of this vector is
\al{
\begin{pmatrix}
\Sigma_{\hat g} & \Delta \\
\Delta^T & \mathcal I
\end{pmatrix}
}
Where $\Sigma_{\hat g}$ is the covariance of the estimators $\hat g$, $\mathcal I$ is the Fisher Information and
\al{
\Delta_{ij} = \int \hat g_i \pd{\log\mathcal L}{\theta_j} \mathcal L \ dx = \int \hat g_i \pd{\mathcal L}{\theta_j}\ dx = \pd{g_i}{\theta_j}
}
By construction, this covariance matrix is positive definite. Furthermore, we have that
\al{
\left| \begin{matrix}
\mathbb{1} & -\Delta \mathcal I^{-1} \\
0 & \mathcal I^{-1}
\end{matrix}\right| = |\mathcal I|^{-1} \geq 0
}
since the Fisher Information is positive definite. This is seen easily since we can rewrite it as
\al{
\mathcal I_{ij} &= \left\langle \pdd{(-\log\mathcal L)}{\theta_i}{\theta_j} \right\rangle = \left\langle	\pd{\log\mathcal L}{\theta_i}\pd{\log\mathcal L}{\theta_j}	\right\rangle \nonumber \\
\Rightarrow q^T \mathcal I q &= \left\langle\left( \sum_i\pd{\log\mathcal L}{\theta_i} q_i \right)^2\right\rangle \geq 0
}
By multiplying the two matrices, we see that
\al{
\left| \begin{matrix}
\mathbb{1} & -\Delta \mathcal I^{-1} \\
0 & \mathcal I^{-1}
\end{matrix}\right|
\times
\left|\begin{matrix}
\Sigma_{\hat g} & \Delta \\
\Delta^T & \mathcal I
\end{matrix}\right|
=
\left| \begin{matrix}
\Sigma_{\hat g} - \Delta \mathcal I \Delta^T & 0 \\
\mathcal I^{-1}\Delta^T & \mathbb{1}
\end{matrix}\right| = \left| \Sigma_{\hat g} - \Delta \mathcal I^{-1} \Delta^T \right| \geq 0,
}
which holds for any subset of the estimators $\hat g$. From this it follows that all eigenvalues of $ \Sigma_{\hat g} - \Delta \mathcal I^{-1} \Delta^T$ are positive or zero, or equivalently that the matrix is positive semi-definite. Looking at unbiased estimators of the $\theta$s, we see that $\Delta$ reduces to an identity matrix and the bound dictates the matrix $\Sigma_{\hat\theta} - \mathcal I^{-1}$ is positive semi-definite. This is exactly what is meant in Eq.~\eqref{eq:stat:cramerrao}.

Note however, that in deriving this bound, we rely on the estimator being unbiased. It is easy to think of estimators with lower variance, say $\hat g = 1$. This has obviously zero variance, but is not a particularly good estimator of anything. It is also worth noting that this bound does not require that the estimator follows a normal distribution. It sets a bound on the variance of \emph{any unbiased estimator}. However, it is only a lower bound, and by no means a guarantee --- only in special cases, like the MLE of a linear model, does an estimator saturate the bound exactly.

\subsection{Confidence regions}
Having found the distributions of the estimators of the parameters of a theory, I now want to define the notion of \emph{confidence regions}. Loosely speaking, these are regions in which we are confident the true value of the parameter lies. This confidence is usually defined in terms of a \emph{coverage probability}, $p_c$. That is, if we define our confidence regions in the same way in repeat experiments, then for every repetition we have the probability $p_c$ that $\theta$ is inside our confidence region. The usual objection here is that once the experiment is done, we can no longer speak of a probability that the true $\theta$ is inside or outside the confidence region --- it either is or is not! The probability as such is defined prior to the experiment. This distinction shall not worry us too much. 

To begin the discussion on confidence regions, we have to understand the concept of a \emph{p-value}, which is closely related to the coverage probability. This is very simple. The p-value of some event is the probability of seeing something more extreme or as extreme as what is observed. In different scenarios this may be computed in a variety of ways, depending on the difficulty of the problem at hand. In some cases, p-values can be computed analytically, while for others one resorts to Monte Carlo (MC), ie. random simulations. As such, the p-value is entirely dependent on the model being tested, and is only telling us \emph{how unlikely something is, given a specific model}. Let us see how this works in an example.

\begin{slant}{A fair coin?}
Consider tossing the same coin $N$ times. We now ask ourselves the question "is the coin fair?", and we can address the answer with a p-value. Say the coin lands heads up $M$ times, where without loss of generality, $M\geq N/2$. To calculate the p-value, we now simply add up the probabilities of getting $M$ or more heads \emph{when tossing a fair coin},
\al{
p &= \sum_{m=M}^N \begin{pmatrix}N\\m\end{pmatrix}0.5^m 0.5^{N-m} =
\sum_{m=M}^N \begin{pmatrix}N\\m\end{pmatrix}0.5^N \nonumber \\
&= 0.5^N \begin{pmatrix}N\\M\end{pmatrix} \phantom{}_2 F_1(1,M-N,1+M;-1), \label{stat:eq:binpval}
}
where $\phantom{}_2 F_1$ is the hypergeometric function, whose form is not particularly enlightening. To make things more clear, let's take a specific example. In Fig.~\ref{stat:fig:binpval} I take various values for $N$ and plot the p-value one would obtain as a function of $M$. The line across denotes the custom $95\%$ confidence level, ie. everything under the line is excluded at more than $95\%$ confidence. It is evident that the as $N$ goes up, we need a smaller and smaller \emph{relative} deviation from $M=N/2$ before we can exclude that the coin is fair. 
\begin{figure}[htb]
\begin{center}
\includegraphics[width=.7\textwidth]{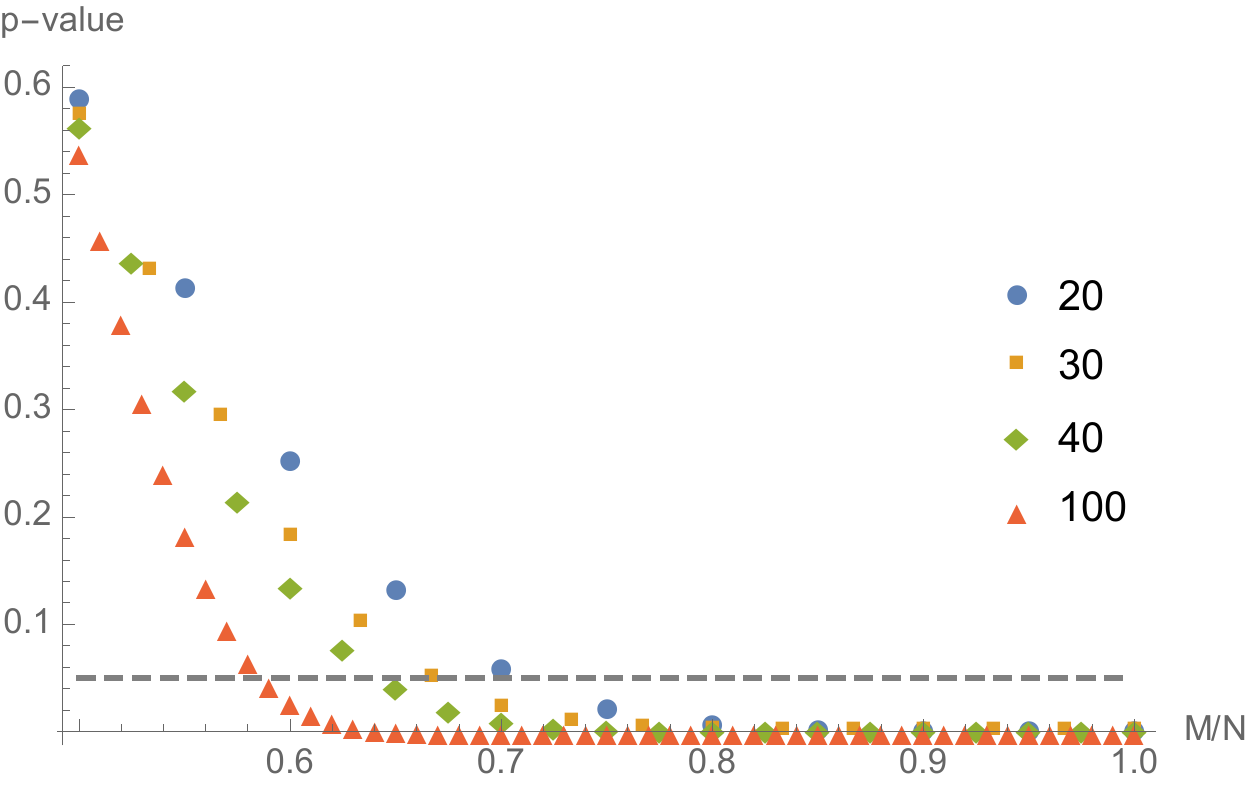}
\caption[Biased coin exclusion by p-value]{p-value, given by Eq.~\eqref{stat:eq:binpval}, of different outcomes $M$ from tossing a coin $N$ times for different values of $N$ as labeled in the legend. This tests the hypothesis that the coin is fair.}
\label{stat:fig:binpval}
\end{center}
\end{figure}
\end{slant}

Originally we wanted to constrain our parameters. With the p-value at hand, we just need \emph{Wilks' theorem}, which tells us the distribution of a likelihood ratio in terms of a $\chi^2$ distribution. This was first shown in \citep{wilks1938large}. First I go through the proof of the theorem, and following that, we will see how this constrains our parameters through confidence regions. I will here just look at a linear model, and I simply argue that the results we find extend to non-linear models in the asymptotic limit --- and that we abuse this fact and use Wilks' theorem always. 

Consider the type of model from Sec.~\ref{stat:sub:linmod}. Take a space for the possible coefficients, $\Omega$, and a subset $\bot\in\Omega$ of dimensions $N$ and $M$ respectively, so $0\leq M<N$. Now call the true parameters $a_\Omega\equiv\{a_{\bot},a_\omega\}$, where $a_\omega\in\omega,a_\bot\in\bot = \omega^\bot$. We can see $\bot$ as the remaining part of $\Omega$, when we fix $a_\omega$. Now we have both the MLE $\hat a_\Omega=\{\hat a_\bot, \hat a_\omega\}\in\Omega$ and a restricted MLE $\hhat a_\bot\in\bot$, which satisfy
\al{
\pd{\log\mathcal L(\hat a_\Omega)}{a_i} &= 0, \hspace{1cm} i=1\dots N \\
\pd{\log\mathcal L(\hhat a_\bot,a_\omega)}{a_i} &= 0, \hspace{1cm} i=1\dots M
}
The quantity $\mathcal L_p(a_\omega) = \mathcal L(\hhat a_\bot,a_\omega)$ is called the profile likelihood. $\hhat a_\bot$ is given by
\al{
\hhat a_\bot = \hat a_\bot - \mathcal I_\bot^{-1} \tilde{\mathcal I} (a_\omega - \hat a_\omega)
}
where I have partitioned the Fisher Information as
\al{
\mathcal I_\Omega =
\begin{pmatrix}
\mathcal I_\bot & \tilde{\mathcal I} \\
\tilde{\mathcal I}^T & \mathcal I_\omega
\end{pmatrix}
}
Now I define the likelihood ratio
\al{
\lambda = \frac{\mathcal L(\hhat a_\bot,a_\omega)}{\mathcal L(\hat a_\Omega)}
}
and seek the distribution of this under the hypothesis that $a_\omega$ are indeed the true parameters. Take $-2\log$ of this and insert factors of the true likelihood $\mathcal L(a_\Omega)$,
\al{\label{eq:stat:likerat}
-2\log\lambda = -2\log \frac{\mathcal L(\hhat a_\bot,a_\omega)}{\mathcal L(a_\Omega)} + 2\log \frac{\mathcal L(\hat a_\Omega)}{\mathcal L(a_\Omega)}
}
Each of the terms on the right hand side can be reduced to the forms
\al{
-2\log \frac{\mathcal L(\hhat a_\bot,a_\omega)}{\mathcal L(a_\Omega)} &=- (\hhat a_\bot-a_\bot)^T \mathcal I_\bot (\hhat a_\bot-a_\bot) \\
2\log \frac{\mathcal L(\hat a_\Omega)}{\mathcal L(a_\Omega)} &=(\hat a_\Omega-a_\Omega)^T \mathcal I_\Omega (\hat a_\Omega-a_\Omega)
}
This is seen by simply inserting the MLE, Eq.~\eqref{eq:stat:mleexp} into Eq.~\eqref{eq:stat:likeexp} and collecting terms. Now write the derivative of the log-likelihood at the true parameters $a_\Omega$, split into the $\bot$ and $\omega$ parts as
\al{
\begin{pmatrix} \eta \\ \xi \end{pmatrix}_i = \pd{\log\mathcal  L(a_\Omega)}{a_i} 
}
This gives two expressions for $\eta$ and one for $\xi$, 
\al{
\begin{pmatrix} \eta \\ \xi \end{pmatrix} &= \mathcal I_\Omega (\hat a_\Omega - a_\Omega ) \\
\eta &= \mathcal I_\bot (\hhat a_\bot - a_\bot)
}
Remember, since the estimators follow the distribution in Eq.~\eqref{eq:stat:mledist}, these variables follow a normal distribution $(\eta,\xi)_i\sim \mathcal N(0,\mathcal I_\Omega)$. Inserting this into Eq.~\eqref{eq:stat:likerat}, we have
\al{\label{eq:stat:likerat1}
-2\log\lambda = 
\begin{pmatrix} \eta \\ \xi \end{pmatrix}^T \mathcal I_\Omega^{-1} \begin{pmatrix} \eta \\ \xi \end{pmatrix} 
- \eta\mathcal I_\bot^{-1}\eta
}
Using the following block inversion identity
\al{
\mathcal I_\Omega^{-1} = \begin{pmatrix}
\mathcal I_\bot^{-1}+\mathcal I_\bot^{-1}\tilde{\mathcal I}(\mathcal I_\omega - \tilde{\mathcal I}^T\mathcal I_\bot^{-1} \tilde{\mathcal I})^{-1}\tilde{\mathcal I}^T\mathcal I_\bot^{-1} &
-\mathcal I_\bot^{-1}\tilde{\mathcal I}(\mathcal I_\omega - \tilde{\mathcal I}^T\mathcal I_\bot^{-1} \tilde{\mathcal I})^{-1}\\
-(\mathcal I_\omega - \tilde{\mathcal I}^T\mathcal I_\bot^{-1} \tilde{\mathcal I})^{-1}\tilde{\mathcal I}^T\mathcal I_\bot^{-1} &
(\mathcal I_\omega - \tilde{\mathcal I}^T\mathcal I_\bot^{-1} \tilde{\mathcal I})^{-1}
\end{pmatrix}
}
we can write the first product in Eq.~\eqref{eq:stat:likerat1} as 
\al{
\eta^T\mathcal I_\bot^{-1}\eta + 
(\tilde{\mathcal I}^T \mathcal I_\bot^{-1}\eta - \xi)^T 
(\mathcal I_\omega - \tilde{\mathcal I}^T\mathcal I_\bot^{-1} \tilde{\mathcal I})^{-1}
( \tilde{\mathcal I}^T\mathcal I_\bot^{-1}\eta - \xi)
}
The first term here is subtracted in the likelihood ratio, and we have 
\al{\label{eq:stat:likeratlong}
-2\log\lambda = (\tilde{\mathcal I}^T \mathcal I_\bot^{-1}\eta - \xi)^T 
(\mathcal I_\omega - \tilde{\mathcal I}^T\mathcal I_\bot^{-1} \tilde{\mathcal I})^{-1}
( \tilde{\mathcal I}^T\mathcal I_\bot^{-1}\eta - \xi)
}
This combination of variables, $\tilde{\mathcal I}^T\mathcal I_\bot^{-1}\eta - \xi$ again follows a normal distribution, for which the covariance is easily seen to be 
\al{
\left\langle(\tilde{\mathcal I}^T\mathcal I_\bot^{-1}\eta - \xi)
(\tilde{\mathcal I}^T\mathcal I_\bot^{-1}\eta - \xi)^T \right\rangle 
= &
\left\langle \xi\xi^T 
- 2 \tilde{\mathcal I}^T\mathcal I_\bot^{-1}\eta\xi^T 
+ \tilde{\mathcal I}^T\mathcal I_\bot^{-1}\eta\eta^T\mathcal I_\bot^{-1} \tilde{\mathcal I} 
\right\rangle \nonumber
\\ = &
(\mathcal I_\omega - \tilde{\mathcal I}^T\mathcal I_\bot^{-1} \tilde{\mathcal I})
}
Meaning the likelihood ratio is simply the sum the squares of $N-M$ --- the number of fixed dimensions --- independent gaussian random variables
\al{
-2\log\lambda\sim\chi^2_{\nu=N-M}
}
To test the hypothesis that $a_\omega$ are the true parameters, we now simply find the p-value of getting the particular $-2\log\lambda$ value for that $a_\omega$. This p-value is given by
\al{\label{eq:stat:pvalwilks}
\text{p-value} = \int_{-2\log\lambda}^\infty \chi^2_{\nu=N-M}(x) \ dx
}
To illustrate this, let's look at an example.

\begin{slant}{Constraining a one-parameter linear model}
Consider drawing from a gaussian distribution with known variance, say $\sigma=1$, but unknown mean $\mu$. The likelihood is of the form Eq.~\eqref{eq:stat:likeexp}, specifically
\al{
\mathcal L(\mu) \propto \exp\left\{ -\frac{1}{2}\sum_i ( \hat y_i - \mu)^2 \right\}
}
and we want to say something about $\mu$ given some experimental result. For a particular outcome of the experiment, say $N$ datapoints, we use Wilks' theorem in the following way. We take as $\Omega$ the full range of the $\mu$, for which we find the MLE as
\al{
\hat\mu = N^{-1}\sum_i \hat y_i
}
and for every possible value of $\mu$, we take $\omega$ as just that $\mu$. Since there are no parameters left, the restricted MLE in $\bot$ is trivial. The p-value is calculated according to Eq.~\eqref{eq:stat:pvalwilks},
\al{
\text{p-value}(\mu) &= \int_{-2\log\lambda(\mu)}^\infty \chi^2_{\nu=1}(x) \ dx & \text{where} \\
-2\log\lambda(\mu) &= -2\log[ \mathcal L(\mu)/\mathcal L(\hat\mu) ] = N(\mu-\hat\mu)^2
}
I now choose to look at the values $\mu_n=\hat\mu(1\pm n/\sqrt N)$ for various $n$. This gives us the integral, for $n=\{1,2,3\}$,
\al{
\text{p-value}(\mu_n) = \int_{n^2}^\infty\chi^2_{\nu=1}(x) \ dx = \{0.32, 0.046, 0.0027\} 
}
Or in words, we can exclude these values with confidence $\{0.68, 0.954, 0.9973\}$. Say we want to be at least $68\%$ confident, then our confidence region is $\hat\mu\pm\frac{\hat\mu}{\sqrt N}\equiv [\hat\mu(1-\frac{1}{\sqrt N}),\hat\mu(1+\frac{1}{\sqrt N})]$, ie. \emph{no values inside this interval can be excluded with confidence greater than 68\%}.

Because of the gaussian nature of the likelihood ratio, this limit is usually called the $1$-$\sigma$ confidence interval, as it is exactly one standard deviation away from the mean, and the standard deviation is usually denoted $\sigma$. We can in the same fashion construct the $n$-$\sigma$ interval for the other $n$s.
\end{slant}
The previous example simply shows the general use of Wilks' theorem. Another subtle thing we can do is to eliminate parameters, which are not of immediate interest. Such parameters are usually called \emph{nuisance parameters}. To see how this works, we just have to have one more parameter. The following example is trivially extended to $N$ parameters of which $M<N$ are nuisance parameters. Unfortunately the 2 dimensional nature of paper only allows for easy visualisation of 2 dimensions.
\begin{slant}{Eliminating nuisance parameters}
Consider a two-parameter linear model with the general likelihood, in vector notation,
\al{
\mathcal L(a) \propto \exp\left\{ -\frac{1}{2} ( \hat y - A^T a)^2 \right\}
}
with $a=\{a_\bot,a_\omega \}$ and $y\in\mathbb R^N$. As stated before, the MLE is given by Eq.~\eqref{eq:stat:mleexp}, $\hat a = (AA^T)^{-1}A \hat y$. First, let's do the same thing we did before, and let $\Omega$ be the entire space of $a$, while $\omega$ fixes both parameters, ie. $\Omega = \omega$. That makes the likelihood ratio
\al{
-2\log\lambda(a) = (a-\hat a)^T\mathcal I (a-\hat a)
}
a random $\chi^2_{\nu=2}$ variable for which we again calculate p-values according to Eq.~\eqref{eq:stat:pvalwilks}.

Now one of the parameters $a_\bot$ is a nuisance parameter. This means that we only fix $a_\omega$, and find the constrained maximum over $a_\bot$. So we look at the quantity
\al{\label{eq:stat:consmax}
-2\log\tilde\lambda(a_\omega) = -2\log\frac{\mathcal L(\hhat a_\bot, a_\omega)}{\mathcal L(\hat a)},
}
Now, by Wilks' theorem, this quantity is a random $\chi^2_{\nu=1}$ variable. The last two points are illustrated in Fig.~\ref{fig:stat:proflike}. For the sake of illustration, the parameters are taken to be very correlated.
\begin{figure}[tbh]
\begin{center}
\includegraphics[width=.5\textwidth]{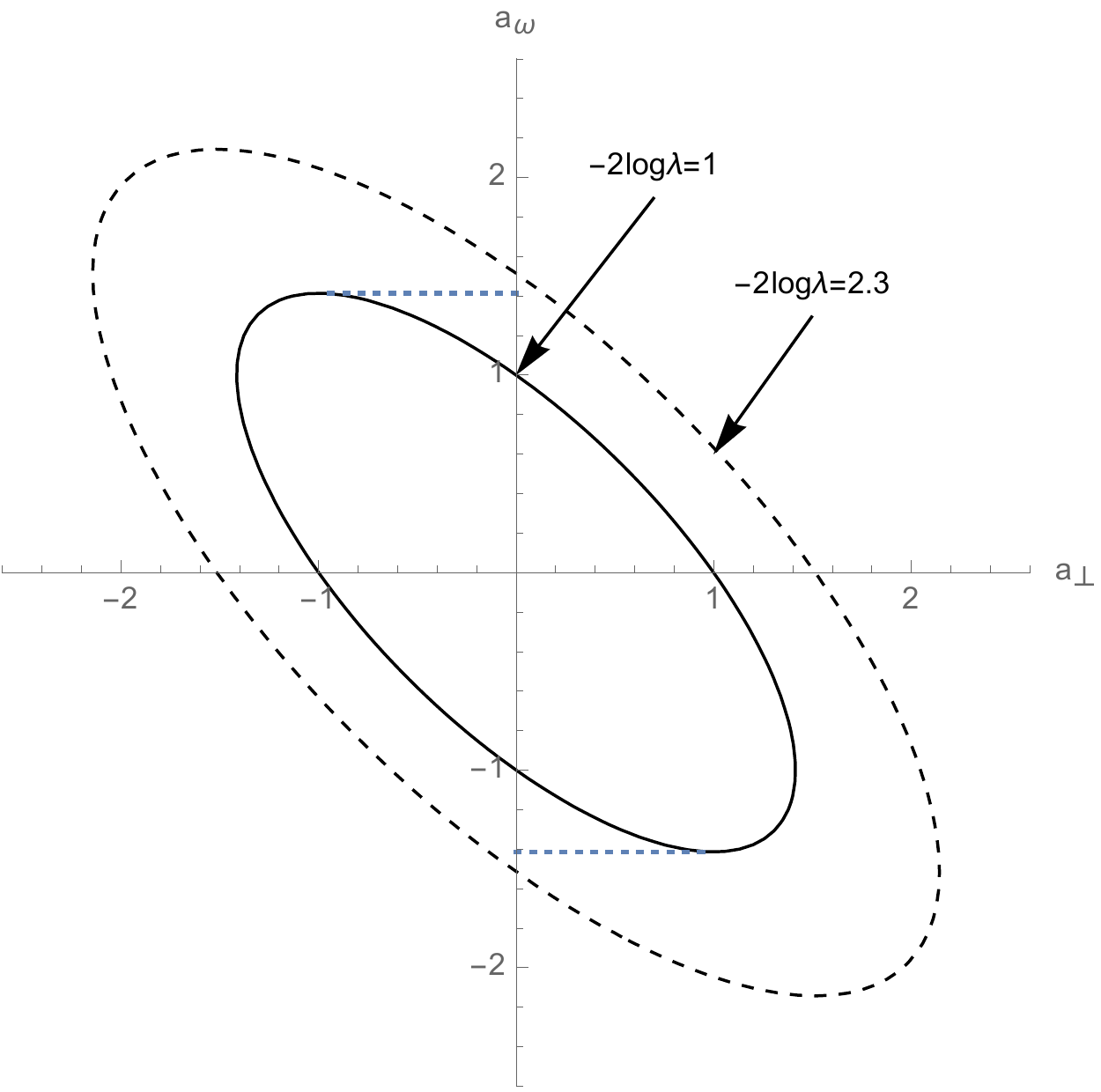}
\caption[Illustration of distinction between confidence region of parameters in subspaces and the full space]{Illustration of confidence regions for two parameters with $\mathcal I_\bot=\mathcal I_\omega =1, \tilde{\mathcal I} = 1/\sqrt 2$. The dashed contour shows the $68\%$ confidence region of both parameters, while the dotted lines are the boundaries of the $68\%$ $a_\omega$ confidence interval, taking $a_\bot$ to be a nuisance parameter. As shown, these dotted lines mark the extreme $a_\omega$ for which any $a_\bot$ gives $-2\log\lambda\leq 1$. The number $2.3$ is the solution $y$ to the equation $\int_y^\infty \chi^2_{\nu=2}(x)\ dx = 1-0.68$. For higher dimensions, one could also give the boundaries of the joint contour in lower dimensions --- here the boundary would be at $\pm\sqrt {2.3}$ instead of $1$. It is important though, to remember the difference in meaning. The bigger one also contains information on the other parameter, while the small one take all but $a_\omega$ as nuisance parameters.}
\label{fig:stat:proflike}
\end{center}
\end{figure}
\end{slant}
We see that the question which Wilks' theorem helps us answer is if we can confidently exclude some parameters $a_\omega$ \emph{for all values of the remaining parameters $a_\bot$}. Even if there is just a single set of parameters $\{a_\bot,a_\omega\}$ such that the p-value is big enough, ie. $-2\log\lambda$ is small enough, then $a_\omega$ cannot be excluded. From Fig.~\ref{fig:stat:proflike} we see exactly how for $a_\omega=\sqrt 2$, we only have $-\log\tilde{\lambda}\leq 1$ when $a_\bot=1$. This \emph{still} means $a_\omega= \sqrt 2$ \emph{cannot be excluded at $1\sigma$}. Said differently, for every $a_\omega$ we test the hypothesis that this is the true value, regardless of what the $a_\bot$ parameter is.

\subsection{Marginalisation}\label{sec:stat:marg}
In the previous derivation, I strictly refer to maximisation of likelihoods. Even so, one will often encounter the term \emph{marginalised} likelihood. The use of this should be kept to a minimum outside Bayesian reasoning, which is described briefly in Sec.~\ref{sec:stat:bayes}. Marginalising the likelihood in simply integration instead of maximisation. That is, instead of using $\mathcal L_p$, we define the marginal likelihood
\al{
\mathcal L_m( \theta ) = \int \mathcal L(\theta, \phi )\ d\phi
}
A trivial exercise is to show that the confidence regions determined from this quantity is \emph{in general not the same} as one would get with the profile likelihood. The objection is now that, obviously, the marginal likelihood is \emph{not} reparametrisation invariant, ie. for some other parametrisation of the nuisance parameters $\Phi = f(\phi)$,
\al{
\int \mathcal L(\theta, \phi )\ d\phi \neq \int \mathcal L(\theta, \Phi )\ d\Phi
}
The two integrands differ by a jacobian $J = d\phi/d\Phi$. This means that when you pick your parametrisation for the likelihood, you assume in some sense that this is a \emph{good parametrisation}. This again reflects the issue that the likelihood is \emph{not} a pdf of the model --- that is why the meaning of this integral is not immediate.

Now it is an equally easy exercise to convince oneself that the maximisation procedure is completely free of this caveat. The maximum likelihood for some $\theta$ cannot depend on the chosen parametrisation of $\phi$, so obviously $\max_\phi\mathcal L(\theta,\phi) = \max_\Phi\mathcal L(\theta,\Phi)$.

\section{Monte Carlo methods} \label{sec:stat:MC}
The previous sections have mostly described linear models, and in one case a very simple non-linear model, whose answer can be found analytically. This, unfortunately, is not always the case. For some random variables, it can be impossible to find explicit expressions for their distributions. When this happens, as is often the case, one way around it is to simply simulate the distribution. This approach is broadly called Monte Carlo (MC) methods, and underlies many results of modern physics. The approach can also be applied to numerical evaluation of integrals. To see this, let's go through the classic example, where we find $\pi\approx 3.14$ by MC integration.
\begin{slant}{Estimating $\pi$}
We know the ratio of areas of a unit circle to a square with side length $2$ to be $\pi/4$. Now as an exercise we want to find the value of this numerically. We look at a single quadrant, $x\in[0,1],\ y\in[0,1]$, where the ratio of areas is the same. We now draw $N$ points inside this region and for every point check if it is inside or outside the circle. So for every point, check if $\sqrt{x^2+y^2}\leq 1$. Finally, we count the number inside the circle, call it $M$, and divide by $N$. The ratio $M/N$ estimates $\pi/4$ (since the region from which we draw has unit area). 

Now, since we are doing this as MC, the estimate we get has an associated error, which we must also estimate. Namely, for every point we draw, it has the probability $\pi/4$ to be inside the circle. That means $M$ will be binomial distributed with $p=\pi/4$ with $N$ draws. From our previous calculations \eqref{eq:stat:binomialmean}-\eqref{eq:stat:binomialstd}, we get immediately
\al{
\langle M \rangle = N \cdot \pi/4 \\
\sigma_M^2 = N \cdot \pi/4 ( 1 - \pi/4 )
}
or, if we look at the quantity $M/N$, and approximate the binomial with $N$ very large as a gaussian,
\al{
M/N \sim \mathcal N(\pi/4 , \sqrt{ \pi/4(1-\pi/4) / N} ).
}
We see here a very general (approximate) result: the error on the estimate falls off as $\sqrt N^{-1}$. So, not surprisingly, the larger we take $N$, the better the approximation we get. This is illustrated in Figs.~\ref{fig:stat:piillu} and \ref{fig:stat:piintegral}. This technique is in its most naive form extended trivially to any integral in any number of dimensions. Of course, as the parameter space becomes larger, computing time increases, but the basic picture remains.
\begin{figure}[tb]
\begin{center}
\includegraphics[width=0.50\textwidth]{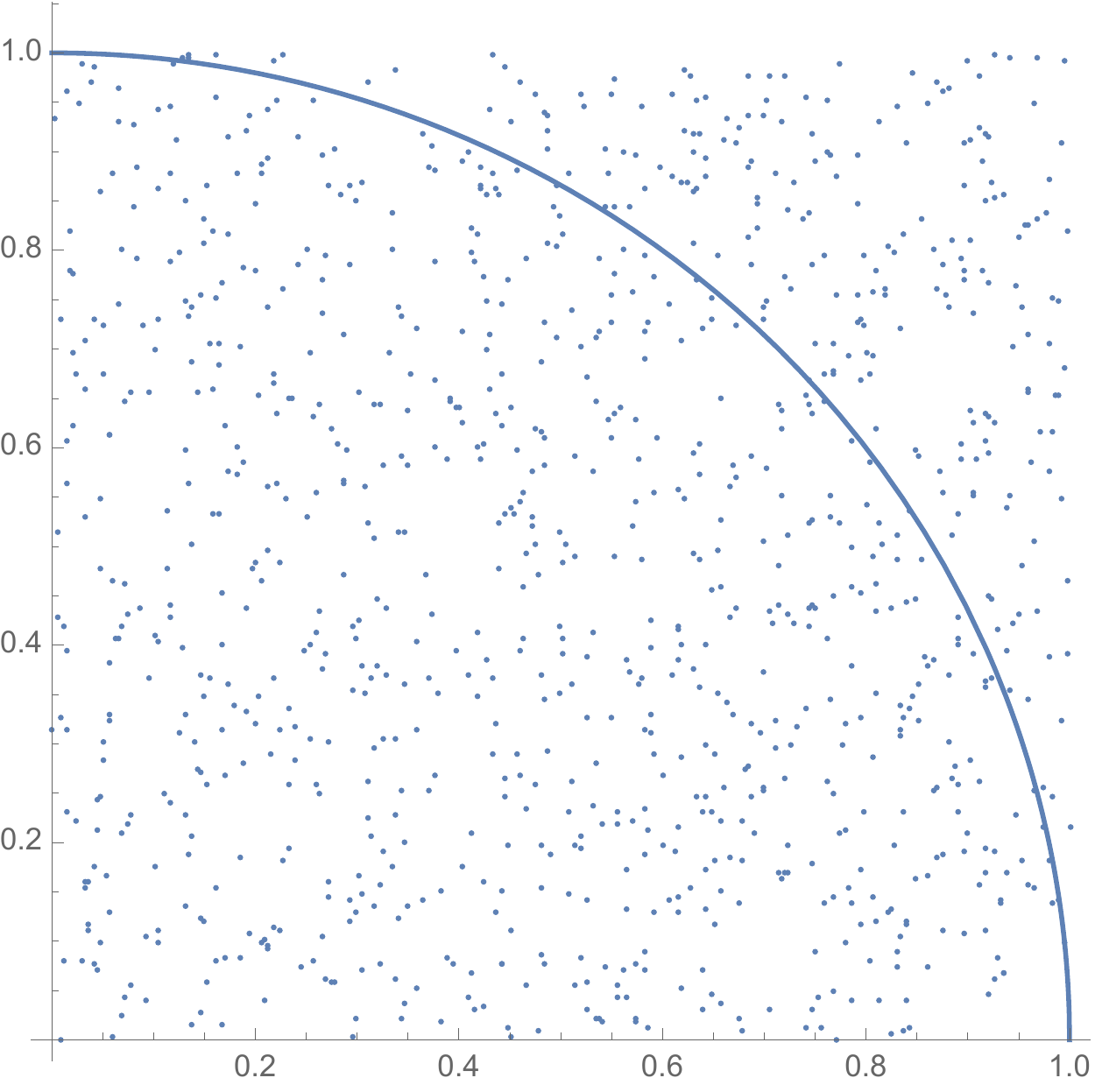}
\caption[$\pi$ by MC]{Example of MC integration. Each point is drawn at random. In this case, $N=10^3,\ M=781$. This means the $1\sigma$ confidence interval for the integral is approximately $(\pi/4)_{\text{MC}} = 0.781\pm0.013$, compared to the true value, $\pi/4=0.7853$, we see that this is indeed a reasonable estimate.}
\label{fig:stat:piillu}
\end{center}
\end{figure}
\begin{figure}[htb]
\begin{center}
\includegraphics[width=0.70\textwidth]{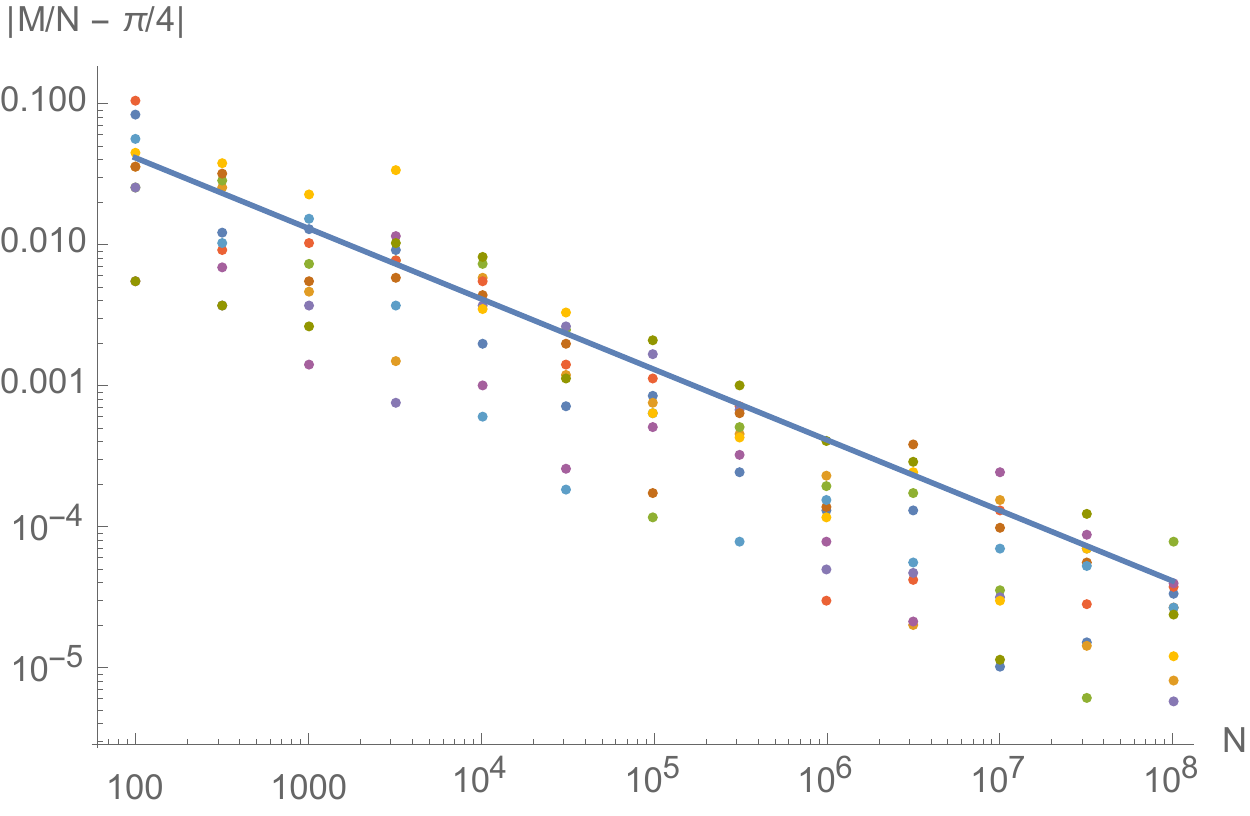}
\caption[Error on $\pi$ by MC]{Errors from the computation of $\pi$ by MC integration. We see that all the errors are of expected magnitude (notice that it is plotted on log-log axes). For every $N$, I perform $10$ MC simulations, simply to show the intrinsic variability in the estimate.}
\label{fig:stat:piintegral}
\end{center}
\end{figure}
\end{slant}

So we can do integrals numerically. This is comforting! As mentioned earlier, we also might want to find distributions for which we cannot find an analytic expression. This is heavily used when finding p-values for some non-trivial quantity. What one does is to simulate an experiment a number of times, say $N$, and for every simulation find the desired quantity. The distribution of these simulated quantities then answers the same question as would the analytic expression, \emph{given this model, how (un)likely is the observed outcome}, simply by numerical comparison between the MC results and the real experiment. I will now extend the previous non-linear model of Sec.~\ref{stat:sec:nonlin} very slightly, and we shall see that we immediately lose the analytic expression for the estimators. We will then use MC to regain control.
\begin{slant}{Unequal errors on measurements}
Take again the estimation of a normal distribution with $(\mu,\sigma^2)=(0,1)$, but this time add distinct measurement errors, $\sigma_i$, on all $\hat x_i$s. This means the likelihood is
\al{
\mathcal L = \prod_i^N (2\pi[\sigma^2+\sigma_i^2])^{-1/2} \exp\left\{ -\frac{1}{2} \frac{(\hat x_i - \mu)^2}{\sigma^2+\sigma_i^2} \right\}
}
Looking for the MLE $(\hat\mu,\hat\sigma^2)$ of this model, we get
\al{
\hat\mu =& \frac{\sum \hat x_i/(\sigma^2+\sigma_i^2)}{\sum 1/(\hat\sigma^2+\sigma_i^2)} \\
\sum 1/(\hat\sigma^2+\sigma_i^2) =& \sum \frac{(\hat x_i - \hat\mu)^2}{(\hat\sigma^2+\sigma_i^2)^2}
}
The appearance of $\sigma_i$ in these sums prohibits the nice manipulations we could do before, and at this point we're stuck on the analytic side. What we do is to simply solve these two equations numerically, for a number of simulated experiments and find an empirical distribution. It is immediate that the distribution of $\sigma_i$ has a lot to say about the distribution of the MLE.

Now let's do the concrete MC for two different experiments. The only difference between the two is the distribution of the individual, \emph{known} errors $\sigma_i$. We will take $N=100$ datapoints in every experiment, and $10^4$ simulations. The first experiment is just like the old one, we take all $\sigma_i=1$ equal. The other has uniformly distributed errors $\sigma_i\sim U(0.1,1.9)$, and $\langle \sigma_i \rangle = 1.02$, ie. almost $1$ like the other. The exact distribution is not of huge importance. Now let's see what difference this makes. Simulating the experiment $10^4$ times, we get the distributions shown in Fig.~\ref{stat:fig:mcdists}. We see that while both are hitting the right answer on average, the tails are different in the distribution of $\hat\sigma^2$. 

\begin{figure}[htb]
\begin{center}
\includegraphics[width=0.45\textwidth]{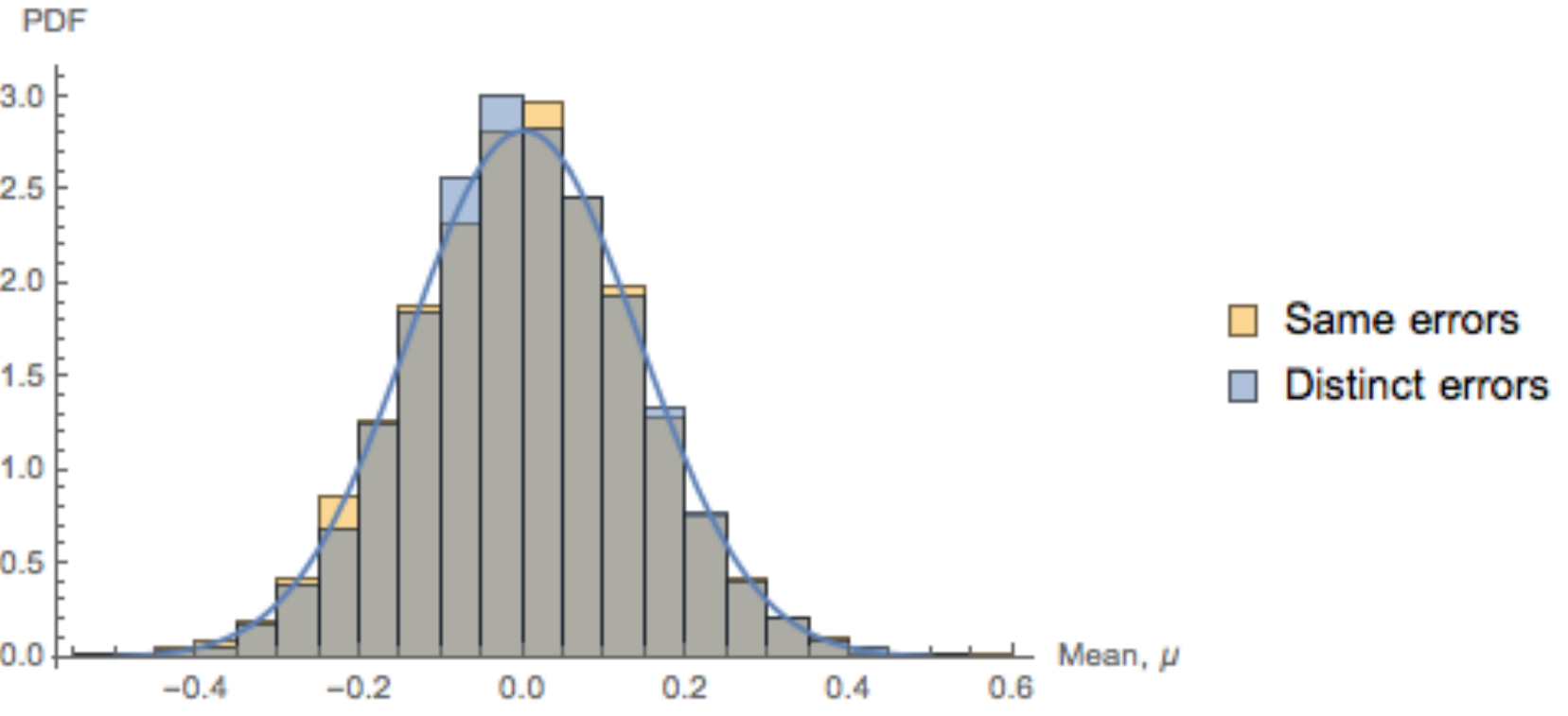}
\includegraphics[width=0.45\textwidth]{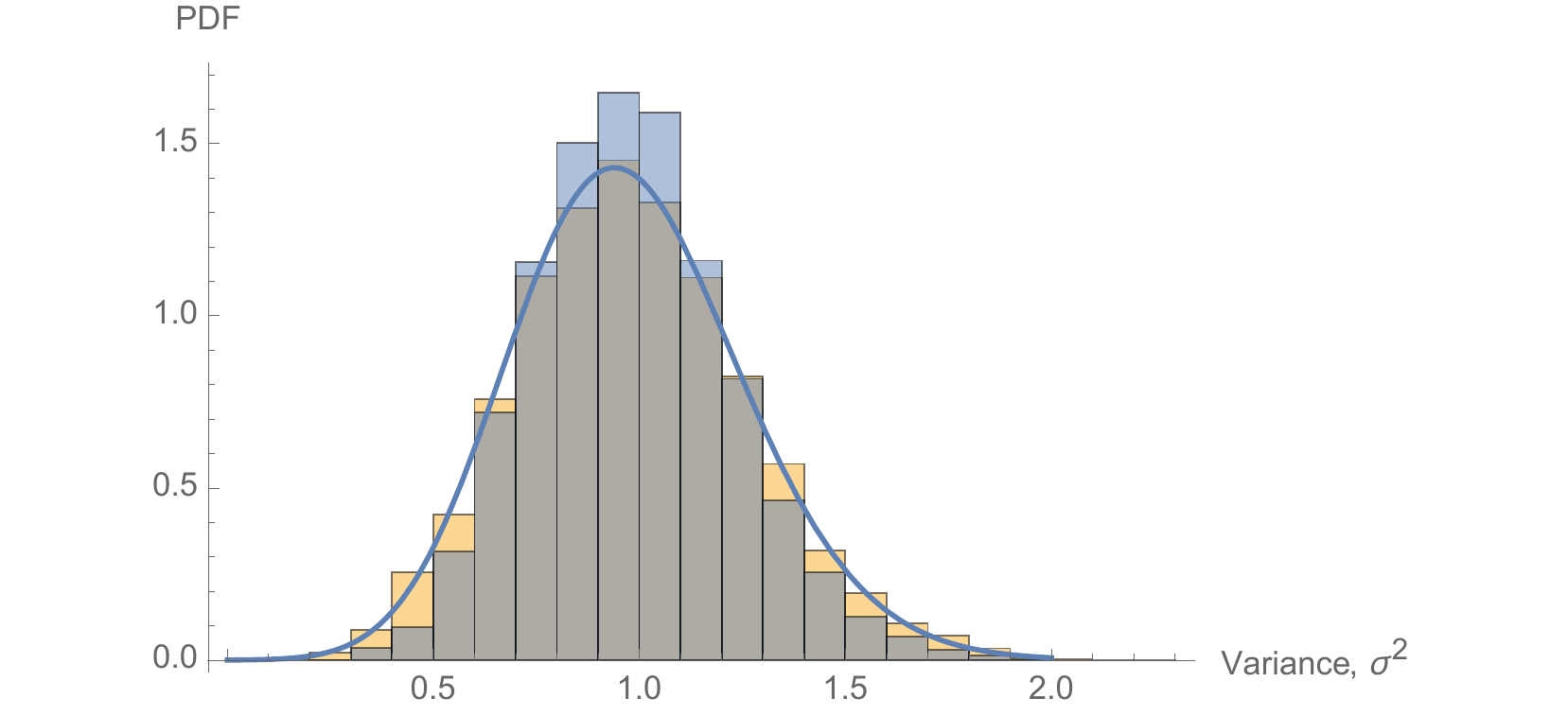}
\caption[Distinction between known and unknown errors]{Distribution of $\hat\mu$ and $\hat\sigma^2$ from $10^4$ MC simulations. Orange shows the original experiment with only the same errors, while blue shows the distribution with errors $\sigma_i$ distributed uniformly between $0.1$ to $1.9$. We see clearly that while the distribution of the mean is more or less unchanged, the distribution of $\hat\sigma^2$ is altered, and no longer follows the $\chi^2$ distribution derived earlier. The two histograms have the expected distribution for the original experiment superimposed.}
\label{stat:fig:mcdists}
\end{center}
\end{figure}

Another interesting distribution to see from this experiment is the distribution of the $\chi^2=\sum(\hat x_i - \hat\mu)^2/(\sigma^2+\sigma_i^2)$. This is shown in Fig.~\ref{stat:fig:notchi2}. It is immediate that the $\chi^2$ distribution does not describe this distribution very well. We can interpret this as exchanging variability is the $\chi^2$ for variability in the $\hat\sigma^2$. Had we set all $\sigma_i^2\ll\sigma^2$, then we end up with the situation from Sec.~\ref{stat:sec:nonlin}, and the $\chi^2$ is always \emph{perfect}, and all variability is in the $\sigma^2$. If we instead have $\sigma_i^2\gg\sigma^2$, then all the errors are practically fixed and we end up with an almost linear model, ie. the $\sigma^2$ does nothing to the fit, and we just fit $\mu$. This gives us a fixed $\sigma\approx 0$ and a $\chi^2$ which is distributed, well, as a $\chi^2$. The situation here is a kind of middle ground, where both are of the same order, and so the $\chi^2$ holds some of the variation, while also the $\hat\sigma^2$ varies. 

Most importantly, this shows that when the errors on the datapoints are not equal, the MLE is not always a \emph{perfect fit}, ie. $\chi^2\neq N$. Even when fitting the error, some variation remains.

\begin{figure}[bht]
\begin{center}
\includegraphics[width=0.5\textwidth]{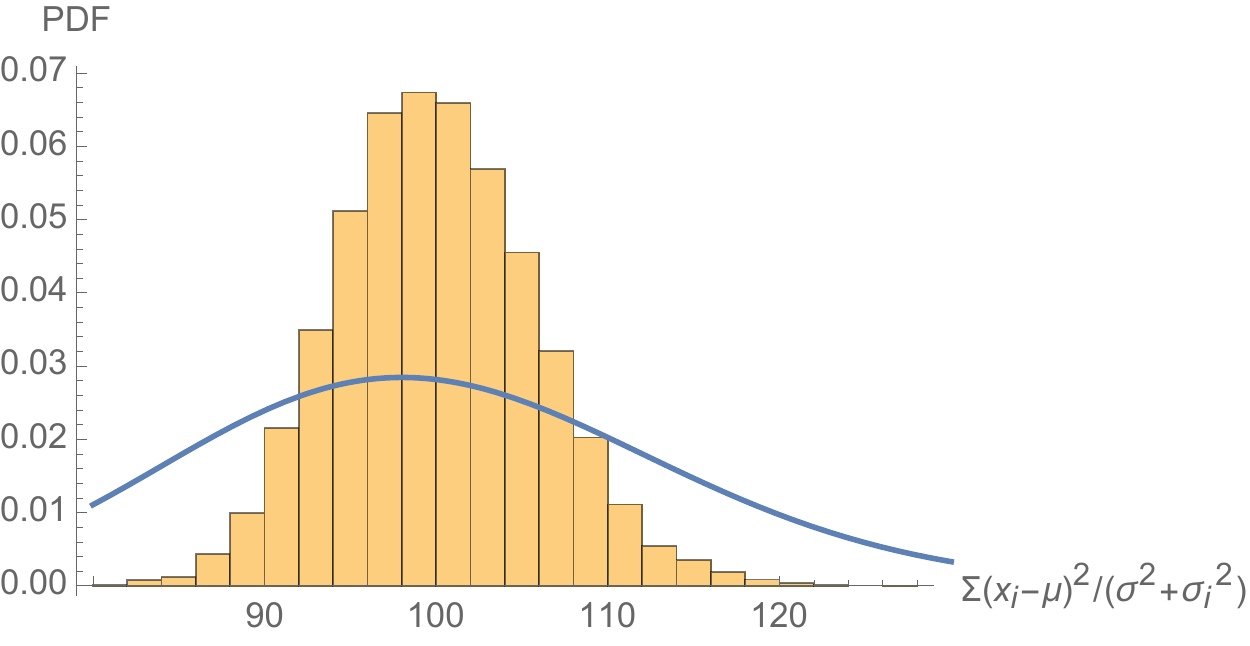}
\caption[Distribution of obtained $\chi^2$s]{Distribution of $\chi^2=\sum(\hat x_i - \hat\mu)^2/(\sigma^2+\sigma_i^2)$ from MC simulations with distinct errors. Superimposed is a $\chi^2$ distribution with $100$ degrees of freedom.}
\label{stat:fig:notchi2}
\end{center}
\end{figure}
\end{slant}

These two examples show the very basics of MC simulations, and the types of problems they solve. This section is by no means exhaustive. It is mostly meant as a very soft introduction to the subject of \emph{stuff we can't calculate exactly}, which unfortunately is a very big one.

\section{Bayesian statistics}\label{sec:stat:bayes}
All statistical analysis in our work is \emph{frequentist}. An objection to what I have shown so far, as I already mentioned, is that the p-values we get out are not probabilities in the sense we would like them to be --- they do not represent model probabilities. If one is unsatisfied by this, then we may use Bayes' theorem to go from the likelihood, which is the pdf of data given a model, to a \emph{posterior pdf}, say $f(\theta|\hat X)$, which is the probability of a certain model given the obtained data. By Eq.~\eqref{eq:stat:bayesdis}, this is done as
\al{
f(\theta | \hat X) = \frac{\mathcal L(\theta) f(\theta)}{f(\hat X)}
}
where, given $f(\theta)$, $f(\hat X) = \int \mathcal L(\theta) f(\theta) \ d\theta$. $f(\theta)$ is called the \emph{prior}, and $f(\hat X)$ is called the \emph{evidence}. Note however, that using Bayes' theorem requires a prior, for which we in most cases of interest in fundamental physics have no idea what should be. In particular, the pdf changes under change of variables, so if we were to pick something boring, in the sense of being uninformative, then the very same function in another variable might be very restrictive --- recall the discussion in Sec.~\ref{sec:stat:marg}.

With Bayesian statistics, we get exactly what we like --- a direct measure of the pdf of a model given the data we see. No hypothesis testing and no ambiguous p-values. The price one has to pay is the choice of a prior, which in some cases is less trivial than other. In a sense, the Bayesian method is trying to answer the unanswerable --- doing fundamental physics, there is \emph{no way} we can pick the \emph{true} prior, since all our knowledge on any subject is derived from experience, which again would have to have been interpreted with some prior.

\chapter{Cosmology}\label{cha:cos}
Today's cosmological studies are by and large interpreted within the bounds of the so-called Concordance or Standard cosmological model. In this section, I will give a summary of the theory with some examples of links to observables and experiments constraining it. I cannot hope to give a textbook introduction to cosmology, but instead refer to one of the many excellent books written on the subject, \citep{peebles1993principles,carroll2004spacetime,kolb1990early,Weinberg:1972aa}.

\section{General relativity}
The foundation of modern cosmology is Einstein's general theory of relativity. Here I aim to introduce main motivations and concepts necessary for the framework of cosmology\footnote{The following derivation follows \citep{Weinberg:1972aa}, including his conventions.}. This describes not only how matter moves in space and time, but also how matter influences, or perhaps more famously \emph{bends}, spacetime. 

The geometry of spacetime is described by the \emph{metric}, which tells the distance between neighbouring points. We define the proper time interval as
\al{\label{eq:cos:line}
d\tau^2 \equiv - g_{\mu\nu} dx^\mu dx^\nu,
}
which defines for us the metric. The equations of motion for a test-particle in spacetime is, in a freely falling, locally inertial coordinate system, a straight line, or more specifically a curve of extremal proper time. In this coordinate system, call it $\xi$, this means we differentiate the coordinates of the particle two times with respect to the proper time and require it be zero,
\al{\label{eq:cos:free}
\PD{\xi^\mu}{\tau} = 0.
}
By reparametrisation invariance --- loosely the statement that Nature doesn't care what coordinates we use --- we can translate the coordinates $\xi$ to any coordinate system $x$ we find convenient, leaving all physics invariant. In particular, the line-element Eq.~\eqref{eq:cos:line} doesn't change,
\al{\label{eq:cos:reparinv}
- \eta_{\mu\nu} d\xi^\mu d\xi^\nu  = - g_{\mu\nu} dx^\mu dx^\nu
}
In the $\xi$ coordinates, the metric takes the very special form $\eta = diag(-1,1,1,1)$.\footnote{Note that I omit any factors of the speed of light $c$. This factor can be restored by dimensional analysis.} In the $x$ coordinates Eq.~\eqref{eq:cos:free} takes the form
\al{\label{eq:cos:geodes}
\pd{}{\tau}\left( \pd{\xi^\mu}{x^\nu}\pd{x^\nu}{\tau} \right) &= 
\pd{\xi^\mu}{x^\nu}\PD{x^\nu}{\tau} + \pdd{\xi^\mu}{x^\nu}{x^\rho}\pd{x^\nu}{\tau}\pd{x^\rho}{\tau} = 0 \nonumber \\
&\Rightarrow \PD{x^\mu}{\tau} + \Gamma^\mu_{\rho\sigma}\pd{x^\rho}{\tau}\pd{x^\sigma}{\tau} = 0
}
where the second line follows from multiplying with $\pd{x^\lambda}{\xi^\mu}$ and renaming indices. I also introduce the affine connection 
\al{\label{eq:cos:condef}
\Gamma^\mu_{\rho\sigma} \equiv \pdd{\xi^\nu}{x^\rho}{x^\sigma}\pd{x^\mu}{\xi^\nu} \\
\Rightarrow 
\pdd{\xi^\lambda}{x^\rho}{x^\sigma} = \pd{\xi^\lambda}{x^\mu} \Gamma^\mu_{\rho\sigma} \label{eq:cos:conre}
}
Eq.~\eqref{eq:cos:geodes} is known as the \emph{geodesic equation}.

There is a subtlety here, which I brushed over. For massless particles --- \emph{radiation} --- we cannot use the proper time as independent variable to label the path, since this vanishes identically. Instead use the zero-component of the coordinate vector, $\xi^0$. The following derivation is like before and we end up with
\al{
0= \PD{x^\mu}{(\xi^0)} + \Gamma^\mu_{\rho\sigma}\pd{x^\rho}{\xi^0}\pd{x^\sigma}{\xi^0}
 }
 We will need these equations to describe the propagation and properties of particles in the universe. Before doing that, we must know how spacetime reacts to matter. First, let's rewrite the connection. Rewrite Eq.~\eqref{eq:cos:reparinv}
 \al{\label{eq:cos:difmet}
 g_{\mu\nu} = \pd{\xi^\alpha}{x^\mu}\pd{\xi^\beta}{x^\nu}\eta_{\alpha\beta}
 }
 and differentiate with respect to the $x$ coordinates
 \al{
 \pd{g_{\mu\nu}}{x^\lambda} &= 
 \left\{ 
 \pdd{\xi^\alpha}{x^\mu}{x^\lambda}\pd{\xi^\beta}{x^\nu}  + 
 \pd{\xi^\alpha}{x^\mu}\pdd{\xi^\beta}{x^\nu}{x^\lambda}
 \right\}\eta_{\alpha\beta} \nonumber \\
 &= \left\{
 \Gamma^\sigma_{\mu\lambda} \pd{\xi^\alpha}{x^\sigma}\pd{\xi^\beta}{x^\nu} + 
 \pd{\xi^\alpha}{x^\mu}	\Gamma^\sigma_{\nu\lambda}	\pd{\xi^\beta}{x^\sigma}
 \right\} \eta_{\alpha\beta} \nonumber \\
 &= \Gamma^\sigma_{\mu\lambda}g_{\sigma\nu} + \Gamma^\sigma_{\nu\lambda}g_{\sigma\mu},
 }
 where line 2 and 3 follow from Eq.~\eqref{eq:cos:conre} and Eq.~\eqref{eq:cos:difmet} respectively. Next, add three of these with mixed indices, 
 \al{
 \pd{g_{\mu\alpha}}{x^\nu} + \pd{g_{\nu\alpha}}{x^\mu} - \pd{g_{\mu\nu}}{x^\alpha} 
 &= 
 \Gamma^\sigma_{\mu\nu}g_{\sigma\alpha} +  \Gamma^\sigma_{\alpha\nu}g_{\sigma\mu} \nonumber \\
 &+  \Gamma^\sigma_{\nu\mu}g_{\sigma\alpha} + \Gamma^\sigma_{\alpha\mu}g_{\sigma\nu} \nonumber \\
 &- \Gamma^\sigma_{\mu\alpha}g_{\sigma\nu} - \Gamma^\sigma_{\nu\alpha}g_{\mu\sigma} \nonumber \\
 &=2 \Gamma^\sigma_{\mu\nu}g_{\sigma\alpha} \label{eq:cos:halfmetcon},
 }
where I use that the connection is symmetric in the two lower indices, as is clear from the definition Eq.~\eqref{eq:cos:condef}. Defining the inverse of the metric, $g^{\mu\nu}$, 
\al{
g^{\mu\nu}g_{\nu\lambda} = \delta^\mu_\lambda
}
we multiply Eq.~\eqref{eq:cos:halfmetcon} by $g^{\lambda\alpha}$ and get 
\al{
\Gamma^\lambda_{\mu\nu} = \frac{ 1}{2}g^{\lambda\alpha}  \left\{ \pd{g_{\mu\alpha}}{x^\nu} + \pd{g_{\nu\alpha}}{x^\mu} - \pd{g_{\mu\nu}}{x^\alpha}  \right\}
}
 This expression is entirely free from the coordinates $\xi$, and can be readily calculated given the metric $g^{\mu\nu}$ in any coordinate system. 
 
 Now we want to write tensors describing the spacetime. Using just the metric and its first and second derivatives, one can show that the unique tensor which is linear in second derivatives of the metric, is the \emph{Riemann(-Christoffel curvature-)tensor},
\al{
R^\lambda_{\mu\nu\rho} = \pd{\Gamma^\lambda_{\mu\rho}}{x^\nu} + \pd{\Gamma^\lambda_{\mu\nu}}{x^\rho} +
\Gamma^\lambda_{\rho\eta} \Gamma^\eta_{\mu\nu} + \Gamma^\lambda_{\nu\eta}\Gamma^\eta_{\mu\rho}
}
Of course we can also take contractions of this tensor, of which the two we will need are the \emph{Ricci tensor},
\al{
R_{\mu\nu} = R^\lambda_{\mu\lambda\nu}
}
and the curvature scalar
\al{
R = R^\mu_\mu
}
 In general, a non-vanishing Riemann tensor signifies the presence of a gravitational field. If the Riemann tensor is strictly zero, then some transformation takes one back to Minkowski space, which has the metric $\eta_{\mu\nu}$. Any non-zero component of the Riemann tensor prohibits such a transformation. With these tensors, Einstein's field equations (EFE) take the form\footnote{Note that sign different conventions for $g_{\mu\nu}$ and $R^\lambda_{\mu\nu\rho}$ may lead to different signs here!}
 \al{
 &R_{\mu\nu} - \frac{1}{2} g_{\mu\nu}R - \Lambda g_{\mu\nu} = - 8\pi G T_{\mu\nu}, \\
 \Leftrightarrow &R_{\mu\nu} = -8\pi G\left( T_{\mu\nu} - \frac{1}{2} T^\lambda_\lambda g_{\mu\nu} \right) - \Lambda g_{\mu\nu}
 }
 where $T_{\mu\nu}$ is the energy stress tensor, $G=6.67\cdot 10^{-11} \text{Nm}^2\text{/kg}^2$ is Newton's constant and $\Lambda$ is the infamous \emph{Cosmological Constant}. I return to this in Sec.~\ref{sec:cosmo:lambda}. The second equation above follows from tracing the first.
 
\begin{slant}{Newtonian mechanics}
As everyone learned in school, Newton predicted the trajectories of planets, combining his $F\propto r^{-2}$ law of gravity with $F=ma$. Let's see how this is the limiting case of the geodesic equation and a specific geometry --- as of course it should be.

The limit we will take is a stationary weak field, and a slowly moving test particle. This translates to the following expressions
\al{
g_{\mu\nu} = \eta_{\mu\nu} + h_{\mu\nu} \\
| h_{\mu\nu} | \ll 1 \\
\pd{h_{\mu\nu}}{t} = 0 \\
\left| \pd{t}{\tau}\right| \gg \left| \pd{x^i}{\tau}\right| \label{eq:cos:slow}
}
Using Eq.~\eqref{eq:cos:slow}, we write the geodesic equation \eqref{eq:cos:geodes} as
\al{\label{eq:cos:geonew}
\PD{x^\mu}{\tau} = \Gamma^\mu_{00} \left(\pd{t}{\tau}\right)^2
}
Calculating the connection, we use that all time  derivatives of the metric vanish, and derivatives only act on the small, $h$-part. To first order in $h$ we have
\al{
\Gamma^\mu_{00} = - \frac{1}{2} g^{\mu\nu}\pd{g_{00}}{x^\nu} = - \frac{1}{2}\eta^{\mu\nu}\pd{h_{00}}{x^\nu}
}
Putting this into Eq.~\eqref{eq:cos:geonew} we get, 
\al{
\PD{t}{\tau} &= 0 \\
\PD{x}{\tau} &= \frac{1}{2} \left(\pd{t}{\tau}\right)^2 \nabla h_{00} \Rightarrow \PD{x}{t} = \frac{1}{2} \nabla h_{00}
}
This looks an awful lot like the Newtonian result, 
\al{
ma = - m\nabla \phi
}
where $\phi$ is some Newtonian potential. For eg. a spherical mass distribution of mass $M$, this takes the familiar form $\phi = -GM/r$. We see that setting $h_{00} = -2\phi$ gives us the Newtonian solution. To check that our approximation holds for typical potentials, put in values for the Sun- and Earth-radius and mass,
\al{
| \phi_\text{Sun} | = \frac{GM_\text{Sun}}{R_\text{Sun}} = 2.12 \cdot 10^{-6} \\
| \phi_\text{Earth}| = \frac{GM_\text{Earth}}{R_\text{Earth}} = 6.95 \cdot10^{-10}
}
Evidently the approximation is very good even at astrophysical scales!
\end{slant}

 \section{The cosmological principle}
 The EFE are in general very hard to solve. Given $T_{\mu\nu}$, they describe 10 coupled partial differential equations for the metric $g_{\mu\nu}$. As such, any exact solution typically has a lot of simplifying symmetry. The cosmological principle is one such set of symmetries. In short, it states that our or anyone else's place and orientation in the universe shouldn't be special\footnote{Or stated otherwise, the universe is homogeneous and isotropic.}. Any translation or rotation must therefore leave the metric invariant. Obviously, the universe isn't exactly homogeneous or isotropic. These properties are meant to be approximately true only on cosmological scales\footnote{The canonical length scale is $100$ Mpc $\approx 3\cdot 10^{24}$ m.}, meaning when we average matter and geometry over large enough scales, this description is suitable.
 
This high degree of symmetry forces the line element \eqref{eq:cos:line} to take the form
 \al{\label{eq:cosmo:FLRWmet}
 d\tau^2 = dt^2 - a(t)^2 \left( \frac{dr^2}{1-kr^2} + r^2 d\Omega^2 \right)
 }
 where $d\Omega^2 = d\theta^2+ d\phi^2\cos^2\theta$ and $k\in \{-1,0,+1\}$\footnote{Another convention takes $a(t_0)=1$ and lets $k$ describe the curvature. One can go back and forth by rescaling $k,r$ and $a$, leaving invariant the combination $ka^{-2}$, the curvature of the space, which is a physical quantity --- conventions don't affect observables. I find it instructive to keep both explicit.}. The different signs of $k$ correspond to an open, flat and closed universe, respectively. The metric is known as the Friedmann-Lema\^ itre-Robertson-Walker (FLRW) metric. The function $a(t)$ is some so far unspecified function of cosmic time $t$, called the scale factor. To find this function, we must solve the EFE. The source must also be maximally symmetric in space, and so takes the form of a perfect fluid\footnote{Fluid in the sense of \emph{fluid dynamics}.},
 \al{
 T_{\mu\nu} = p g_{\mu\nu} + (p + \rho )U_\mu U_\nu,
 }
where $p$ and $\rho$ are the pressure and energy density of the fluid, and $U$ is the fluid velocity, which in the cosmic rest-frame is given by
\al{
U^0 = 1 \nonumber \\
U^i = 0, \nonumber
}
that is to say, the contents of the universe are, on cosmological scales, relatively quiet. Because of the high degree of symmetry in the problem, only two independent equations remain of the EFE. The first is the \emph{Friedman equation},
\al{\label{eq:cos:friedsimple}
\dot a^2 + k = \left( \frac{8\pi G}{3}\rho + \frac{\Lambda}{3} \right)a^2
}
and the second I take as conservation of energy, and write as
\al{\label{eq:cos:consen}
\frac{d}{da}(\rho a^3) + 3pa^2 = 0
}
To close the set of equations, we need an \emph{equation of state}, describing the pressure as a function of the energy density
\al{
p = p(\rho)
 }
 Two equations of state are of particular importance. These are of non-relativistic matter, or \emph{dust}, and ultra-relativistic matter, or equivalently, radiation. The two are
 \al{
 p_\text{matter} \ll \rho \\
 p_\text{radiation} = \rho / 3
 }
 For the two we find, according to Eq.~\eqref{eq:cos:consen} the dilution of the energy density is
 \al{
 \rho_\text{matter} \propto a^{-3} \\
 \rho_\text{radiation} \propto a^{-4}
 }
These factors should not come as a surprise. Thinking in terms of an expanding universe, matter is simply spread over greater volumes and dilutes as $1/V$, whereas radiation is not only diluted, but also stretched by the expansion. One can in general think of some perfect fluid with equation of state
\al{
p = w\rho.
}
I will in the following keep the radiation and matter factors explicit, but all calculations can be made with arbitrary $w$.\footnote{There are some subtleties in what values of $w$ are physical. I will not address these issues here.}

With these expression for the energy density, we can in principle solve the Friedmann equation. It is customary to rewrite the equation a bit. First introduce the Hubble parameter and critical density,
\al{
H &= \dot a / a, \hspace{1cm} H_0=H(\text{today})=100h \frac{\text{km}}{\text{s}\cdot\text{Mpc}}\\
\rho_c &= \frac{3 H_0^2 }{8\pi G}.
 }
Dividing Eq.~\eqref{eq:cos:friedsimple} through by $a^2$ we get
\al{\label{eq:cosmo:friedcomp}
H^2 = H_0^2 \left( \frac{\rho}{\rho_c} + \frac{\Lambda}{3H_0^2} - \frac{k}{a^2H_0^2}  \right) 
}
The density $\rho$ can now be matter, radiation or both. Taking into account how the two densities scale, write
\al{
\rho = \rho_m + \rho_R = \frac{a_0^{3}}{a^3}\rho_{m0}  +\frac{a_0^4}{a^{4}} \rho_{R0} 
}
where $a_0 = a(t=t_0)$ is the scale factor today. Now define the density parameters $\Omega_i$ as
\al{\label{eq:cos:denspar}
\Omega_m &= \frac{\rho_{M0} }{\rho_c}, \hspace{1.0cm} \Omega_{R}  = \frac{\rho_{R0}}{\rho_c} \nonumber \\
\Omega_\Lambda &= \frac{\Lambda}{3H_0^2}, \hspace{1cm} \Omega_k = -\frac{ k}{a_0^2 H_0^2}
}
and finally, write the Friedmann equation as
\al{\label{eq:cos:freid1}
H^2 = H_0^2 \left\{ \Omega_m (a/a_0)^{-3} + \Omega_R (a/a_0)^{-4} + \Omega_\Lambda + \Omega_k (a/a_0)^{-2} \right\}
}
Inserting $t=t_0$ we easily see that the density parameters obey the sum rule
\al{\label{eq:cosmo:sumrule}
\Omega_m  + \Omega_R + \Omega_\Lambda + \Omega_k = 1
 } 
Widely accepted, \emph{concordance}, values for the values of these parameters in the present universe are (\citep{Ade:2013zuv}) 
\al{\label{eq:cosmo:params}
\Omega_m &\approx 0.3 \hspace{1cm} \Omega_R\approx 0 \nonumber \\
\Omega_\Lambda &\approx 0.7 \hspace{1cm} \Omega_k \approx 0 \nonumber \\
H_0 &\approx 70 \text{Mpc}^{-1}\text{km/s} \approx 1.3\cdot 10^{-41} \text{GeV}
}
which is why the current setting is called $\Lambda$CDM. $\Lambda$ for a Cosmological Constant, CDM for cold dark matter. The actual baryonic matter we are all made of is in this picture a mere $5\%$, which is included in the $\Omega_m$ here.
 
\begin{slant}{Single component universes}
For the sake of intuition, let's work through some examples of single component universes. In particular, consider the four immediate possibilities --- matter, radiation, curvature, and Cosmological Constant-dominated universes, with each of the four density parameters $\Omega_i=1$ and all others $0$. This corresponds to solving the equation 
\al{
\frac{\dot a}{a} = \frac{\dot a_0}{a_0} \left( \frac{a}{a_0}  \right)^{-n/2} \Rightarrow
	 \frac{\dot a}{\dot a_0} = \left( \frac{a}{a_0} \right) ^{1-n/2}
}
for $n\in \{ 3,4,2,0 \}$, respectively. Assume now a power-law form, $a\propto t^m$. Putting this in our equation, we get the condition
\al{
m = \frac{2}{n}, \ \ \ n\neq 0
}
For the Cosmological Constant, this solution fails, but we see immediately for $n=0$ the answer must be an exponential function. For the four different single component universes we have the following solutions
\al{
a(t) = a_0 \times
\left\{ \begin{matrix}
\left( \frac{t}{t_0}\right) ^{2/3} & \text{matter dominated (Einstein-de Sitter) }\\
 \left( \frac{t}{t_0}\right) ^{1/2} & \text{radiation dominated }\\
 t/t_0 &  \text{curvature dominated (Milne) }\\
 \exp(H_0 t) &  \text{Cosmological Constant dominated (de Sitter)}
\end{matrix}\right.
}
Finally, extrapolating $a \rightarrow 0$, we get the following expressions for the age of the universe in terms of the present Hubble constant,
\al{
t_{a=0} = \frac{1}{H_0}\times 
\left\{ \begin{matrix}
2/3 & \text{matter dominated (Einstein-de Sitter) }\\
1/2 & \text{radiation dominated }\\
1 &  \text{curvature dominated (Milne) }\\
\infty &  \text{Cosmological Constant dominated (de Sitter)}
\end{matrix}\right.
}
\end{slant}
 
Since we observe neither cosmic time, nor the absolute scale factor, it would be nice to have a proxy for the two. To this end, we introduce the \emph{cosmological redshift},\footnote{Not to be confused with the Doppler redshift.} denoted $z$. This is the  fractional amount the wavelength of radiation has been stretched by the universe expanding. To see how this comes about, place an observer at $r=0$ and let a wave crest be emitted at $t_1$ propagating radially inwards from some radius $r_1$. For a lightlike test particle, the proper time is zero, and we have
\al{\label{eq:cos:nullgeo}
0 = dt^2 - a(t)^2 \frac{dr^2}{1-k r^2}
}
Call the time it is observed $t_0$, we then have the equation
\al{\label{eq:cos:propdis}
\int_{t_1}^{t_0} \frac{dt}{a(t)} = \int_{0}^{r_1} \frac{dr}{\sqrt{1-kr^2}} = \frac{1}{\sqrt{k}} \sin^{-1}(\sqrt{k}r_1)
}
Notice the sign in taking the square root is fixed by the direction of propagation. The next wave crest is emitted shortly after, follows the same path and obeys the same equation but with slightly shifted time coordinates,
\al{
\int_{t_1+1/\nu_1}^{t_0+1/\nu_0} \frac{dt}{a(t)} = \frac{1}{\sqrt{k}} \sin^{-1}(\sqrt{k}r_1),
}
where $\nu_i$ is the frequency at $r_i$. For  frequencies much larger than $H \approx 3.2\cdot 10^{-18}h\text{s}^{-1}$ we get
\al{
0 =& \int_{t_1}^{t_0} \frac{dt}{a(t)} - \int_{t_1+1/\nu_1}^{t_0+1/\nu_0} \frac{dt}{a(t)} \approx \frac{1}{a(t_1)\nu_1} - \frac{1}{a(t_0)\nu_0} \nonumber \\
\Rightarrow& \hspace{1cm} \frac{\nu_1}{\nu_0} = \frac{a(t_0)}{a(t_1)}
}
We now define the redshift as the fractional increase in wavelength,
\al{
z = \frac{\lambda_0 - \lambda_1}{\lambda_1} = \frac{\nu_1}{\nu_0}-1 = \frac{a(t_0)}{a(t_1)} -1
}
This is a nice quantity to work with because it is readily observable through analyses of spectra. We can rewrite Eq.~\eqref{eq:cos:freid1} trading $t$ and $a$ for $z$, giving
\al{
H(z)^2 = H_0^2 \left\{  \Omega_m(1+z)^3 + \Omega_R (1+z)^4 + \Omega_\Lambda + \Omega_k (1+z)^2 \right\}
}

 \section{Cosmography}\label{cos:sec:mography}
 On cosmological scales, the intuitive notion of distances fails. Depending on the question you ask, distances to the same object may differ --- by a lot. In this section, I explore the different measures of distance and try to clarify their meaning. 
 
 First, let us connect the $r$ coordinate to the physical redshift. Take an observer and an emitter, say a galaxy or a supernova, at relative proper distance $r_1$. Emitting a single photon at $t=t_1$, we observe it at $t=t_0$. The photon follows the path described in Eq.~\eqref{eq:cos:nullgeo}, and upon inverting Eq.~\eqref{eq:cos:propdis} we get, with a change of variables,\footnote{The following expression holds, by analytic continuation of $\sin$, for all $k$.}
 \al{\label{eq:cos:r1sin}
 r_1 = \frac{1}{\sqrt{k}} \sin\left( \frac{\sqrt{k}}{a_0} \int_0^z \frac{dz'}{H(z')} \right).
 }
 
 Usually though, a single photon is not enough. What we might hope to measure is a stream of light from a source of known luminosity. Considerations from Euclidian space lead us to define the \emph{luminosity distance}, $d_L$ as
 \al{ \label{eq:cosmo:lumdis0}
 F = \frac{L}{4\pi d_L^2},
 }
where $F$ is the measured flux from an object of luminosity $L$. Now we seek the relation between this definition and the proper distance --- and hence the redshift. Note that $F$ and $L$ are \emph{bolometric} quantities, ie. integrated over all frequencies. First, consider the area over which the emitted light is spread. Integrating the angular part of the metric, we get a total area, at time $t_0$ --- when the light is observed
\al{
A = 4\pi a(t_0) r_1.
}
Travelling across the universe has its price, though. First, the emitted light is redshifted, which reduces the energy per observed  photon by one factor $(1+z)$, and second, the distance between individual photons is increased, also by a factor $(1+z)$. This means the observed flux is reduced by a total factor $(1+z)^2$, giving
\al{\label{eq:cos:lumdis1}
F = \frac{L}{4\pi a(t_0)^2 r_1^2 (1+z)^2} \Rightarrow d_L = (1+z)a(t_0) r_1.
}
 
 Next we look at an object or a feature, which is extended across the sky in some angle $\delta\theta\ll 1$ at proper distance $r_1$. Looking again to Euclidean geometry, we expect the measured angle to be the length of the object, $D$, divided by the distance $d_A$,
 \al{
\delta\theta=\frac{ D }{ d_A }
 }
To find the relation between the angular diameter distance and the proper distance, we arrange our coordinate system appropriately and integrate only $\theta$ in the metric. Doing this we get that the proper distance between the two ends of the object at $t_1$ is
 \al{\label{eq:cos:angdis}
 D = a(t_1)r_1 \delta\theta \Rightarrow d_A = a(t_1) r_1 = (1+z)^{-1} a(t_0) r_1.
 }
 An equivalent definition in terms the solid angle $\delta\Omega$, filled by an object of proper area $\delta A$ is
 \al{ \label{eq:cosmo:fromhui}
  d_A = \sqrt{\frac{\delta A}{\delta\Omega}}
 }

The transverse comoving distance is defined as the ratio of the proper transverse motion of a particle to the angular motion we see
\al{
d_M = \frac{\Delta D / \delta t_1}{\delta\theta / \delta t_0} = d_A(1+z) = a_0 r_1.
}
Note that it is not, as the angular diameter distance, the physical length of an object.

\begin{figure}[t]
\centering
\includegraphics[width=.45\textwidth]{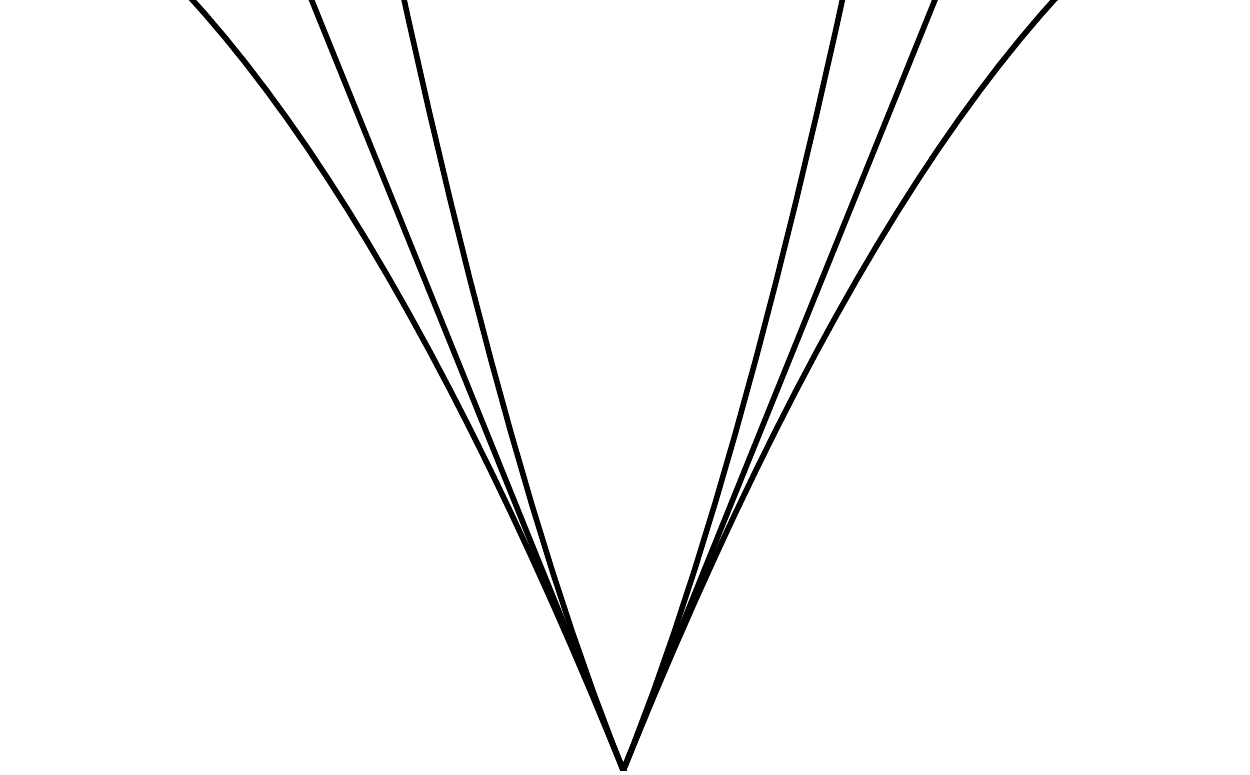}
\caption[Straight lines curve in curved space]{Sketch of lines with equal incident angle at the observer point, propagating in differently curved spaces. From outside in, the \emph{universes} have $k=-1,0,+1$. This shows how the distance measured is affected by the curvature of space, as the length between the lines at the top --- at the source position --- is changed by the warping of the geodesics.}
\label{fig:cos:curve}
\end{figure}

\begin{slant}{Curved space}
To gain a bit of intuition for curved space, consider measuring $d_M$. Without looking to the equations, we ask ourselves "are we going to measure more or less than we think?". Recall that in positively curved space, parallel lines get closer and closer, while in negatively curved space, they grow further apart --- the first point is most easily seen by imagining a 2-sphere, where lines that are parallel at and orthogonal to the equator will intersect at the poles. Now, we observe some angle, which is to say at our position, the two lines going to each of the two sources we observe have some incident angle at our position. As we just argued, the separation between two lines changes in curved space compared to flat space. This means that in positively curved space, the two lines going to the two sources will get closer as they go along, and the distance $d_M$ is smaller than in flat space. Conversely, in negatively curved space, the lines get further apart and $d_M$ is larger, see Fig.~\ref{fig:cos:curve}. This effect is exactly the effect of the $\sin$ function in the expression Eq.~\eqref{eq:cos:r1sin}.
\end{slant}

We see that the angular diameter distance and luminosity distance are not independent from the proper distance --- they satisfy the \emph{Etherington reciprocity relation},
\al{\label{eq:cosmo:ether}
d_A (1+z) = d_M = d_L (1+z)^{-1}
}

We finally want to know what part of the universe can ever have had an effect at our position, given that the current dynamics are what have always been at play. That is, at a given time in the history of the universe, how big was the causally connected part. The proper distance to this \emph{horizon} is just the integral of the square root of the radial part of the metric
\al{\label{eq:cos:horizonint}
d_H = a(t) \int_0^{r_H} \frac{dr}{\sqrt{1-kr^2}} = a(t) \int_0^{t} \frac{dt'}{a(t')} 
}

\begin{slant}{The horizon problem}
Consider a matter-dominated universe, which for the following calculation will simulate the universe we live in. Calculating the distance to the horizon is straight-forward, and we get
\al{
d_{H,\text{matter}} = \frac{2}{H_0}(1+z)^{-3/2}.
}
Watching this horizon on the sky from far away, we expect that any two points further apart than $d_H$ will not be in causal contact --- and will not a priori know anything about one another. Let's calculate the size on the sky of such a horizon patch. The angular diameter distance is in the matter dominated universe given by 
\al{
d_{A,\text{matter}} = \frac{2}{H_0} \frac{1- (1+z)^{-1/2}}{1+z}
}
The angular size of the patch for a given redshift is then
\al{
\delta \theta = d_H / d_A
}

The cosmic microwave background (CMB) radiation is the leftover thermal bath of photons from the early universe. The photons decoupled at redshift $1+z\approx 1100$ and have been free streaming since then. The size of a horizon patch at this decoupling redshift is
\al{
\delta \theta \approx \frac{1100^{-1/2}}{1-1100^{-1/2}} = 0.031 \text{rad} = 1.8^o
}
Note that this is significantly less than $180^o$. This means that patches on the sky separated by more than $1.8^o$ should be completely independent --- the exact number changes slightly for different universes, but the point remains. The great surprise is the fact that the temperature of these photons is to very high precision constant over \emph{the whole sky}. This means that apparently, the entire observed universe has been in causal constant at some point, yet our calculations show, that given the present expansion of the universe, there is no way it could have been. This is known as the \emph{horizon problem}. One solution to this problem --- \emph{inflation} --- is to insert a sudden de Sitter period, which blows up the horizon, while keeping the Hubble parameter constant. This can leave the observed universe inside the horizon distance. The problem is of course formulated assuming the metric description holds to $t=0$, in order that the integral \eqref{eq:cos:horizonint} can be calculated.
\end{slant}

For distances at low redshift, $z\ll1$, we can expand the expressions previous and do the integrals. I will illustrate this with the luminosity distance, and on the way introduce the \emph{deceleration parameter}. Take the expression Eq.~\eqref{eq:cos:lumdis1} and taylor expand the integral of in $r_1$. Since $\sin(x) = x +\mathcal O(x^3)$, we can get to second order in $z$ while just considering the expression
\al{
H_0 d_L =& (1+z) \int_0^z \frac{dz'}{
\sqrt{ \Omega_m (1+z)^{3} + \Omega_\Lambda + (1-\Omega_m-\Omega_\Lambda) (1+z)^{2} }
} \nonumber \\
=& (1+z) \left(  z - \frac{z^2}{2} (1-q_0) +\mathcal O(z^3) \right) , \hspace{1cm} q_0 \equiv \Omega_m/2 - \Omega_\Lambda
} 
where $q_0$ is the deceleration parameter, and I have ignored radiation, as is justified in the late universe. This measures the degree to which the universe is decelerating --- named so since historically it was believed the universe was decelerating and positive numbers are pleasing. Let's see how the acceleration of the FRW universe is related to this parameter. We turn to the expansion of the scale factor around the present time, $t-t_0\ll H_0^{-1}$, 
\al{
a(t) &= a_0 + \dot a_0 (t-t_0) + \frac{\ddot a_0}{2}(t-t_0)^2 +\mathcal O(t^3) \nonumber \\
&= a_0 \left( 1 + [t-t_0]H_0 -\frac{1}{2} \left( -\frac{\ddot a_0 a_0}{\dot a_0^2} \right) ([t-t_0]H_0)^2 +\mathcal O(t^3)  \right)
}
The coefficient in front of the second order term is just what we're looking for. To see this, reorder and differentiate the Friedmann equation, \eqref{eq:cos:freid1} with respect to time,
\al{
\pd{}{t} \dot a &= \pd{}{t}H_0 a \sqrt{\Omega_m (a/a_0)^{-3} + \Omega_\Lambda + \Omega_k (a/a_0)^{-2} } \nonumber \\
\Rightarrow \ddot a &= H_0 \dot a \left( \sqrt{\Omega_m (a/a_0)^{-3} + \Omega_\Lambda + \Omega_k (a/a_0)^{-2} } 
\right. \nonumber \\ 
&\left. + \frac{-3\Omega_m (a/a_0)^{-3} -2 \Omega_k (a/a_0)^{-2} }{2\sqrt{\Omega_m (a/a_0)^{-3} + \Omega_\Lambda + \Omega_k (a/a_0)^{-2} }}\right) \nonumber \\
\Rightarrow \ddot a_0 &= H_0 \dot a_0 \left(1 + \frac{-2 - \Omega_m +  2\Omega_\Lambda}{2} \right) = - H_0 \dot a_0 \left( \Omega_m/2 -\Omega_\Lambda \right) \nonumber \\
\Rightarrow  - q_0 &= \frac{\ddot a_0 a_0}{\dot a_0^2}.
}
We can see $q_0$ as a scale-free measure of the deceleration of the universe --- the scale of expansion is set by $H_0$ and the scale of the universe by $a_0$. Note that $q_0$ only describes the deceleration of the universe \emph{today}. Generally, $q$ changes throughout the course of the universe. Only in very special cases the universe is forever non-accelerating.

\subsection{Moving emitter and observers}\label{sec:cos:moving}
The Doppler effect, being a well established phenomenon, also has to be taken into account when measuring the universe. Typical \emph{peculiar velocities} of galaxies, which is to say the velocity in excess of the Hubble recession, are expected to be of the order a few hundred kilometers per second, ie. $v\approx 10^{-3}$ in units of the speed of light. This includes both us as observers and eg. SNe as emitters. The problem I wish to address in the present subsection is what difference this makes to the light we receive. Now, since these velocities are only mildly relativistic, we shall only look at the first term in an expansion around zero velocity. The following derivation follows the work in \citep{Hui:2005nm}.\footnote{Note that this particular article follows a different notation --- the bars are non-bars here and vice versa --- and only treats flat space.}

First we realise that the redshift we see is not only redshifted by the expanding universe, but also by normal relativistic Doppler shifting. Denoting the expected redshift in a completely still universe by $z$ as before, we write $\bar z$ for the corrected redshift in the universe where everyone is moving around. By normal Doppler shifting, $\bar z$ is given by 
\al{\label{eq:cos:redshiftshift}
1+\bar z = (1+z)(1+ n\cdot [ v_e - v_o ] ) + \mathcal O( v^2)
}
where the $v_i\ll1$ are the velocities of the emitter and observer, respectively, and $n$ is a unit vector point from the observer to the emitter. From here onwards, anything but the first $v_i$ term is neglected. Now, beaming effects also come into play, in particular the solid angle of the emitter is changed by relativistic beaming as
\al{
\delta\Omega\rightarrow \delta\Omega( 1 - 2 n\cdot v_o )
}
Note that this only depends on the observer-velocity, not the emitter. This changes the angular diameter distance, Eq.~\eqref{eq:cosmo:fromhui}, which we can in turn link to the luminosity distance through Eq.~\eqref{eq:cosmo:ether}. We see that the changes are the following
\al{
\bar d_A( \bar z) &= d_A(z) ( 1 + n\cdot v_o) \\
\Rightarrow \bar d_L(\bar z) &= d_L(z) ( 1 + n\cdot v_o) \frac{(1+\bar z)^2}{(1+ z)^2} \approx d_L(z) (1+n\cdot [2v_e-v_o] )
}
We are still not done yet, as this last equation does not relate directly observable quantities. The redshift we observe is naturally $\bar z$, so we will have to also evaluate $d_L$ at this slightly shifted redshift. What I will do is a simple Taylor expansion of the function. This means we take
\al{
d_L( z ) = d_L(\bar z) + \pd{d_L(\bar z)}{z}(z - \bar z),
}
where we can write $z-\bar z = -(1+\bar z)n\cdot[v_e-v_o]$, and so we just miss the derivative. From Eq.~\eqref{eq:cos:lumdis1} we get
\al{
\pd{d_L(\bar z)}{z} = \frac{ d_L( \bar z)}{1+\bar z}+
	\frac{1+\bar z}{H(\bar z)} \cosh \left[ \sqrt \Omega_k d_C / d_H\right]
}
Putting this into the former expression we finally have
\al{
\bar d_L(\bar z,n) =&\ d_L(\bar z) \left[ 1-  n\cdot v_e\right ] - \frac{(1+\bar z)^2}{H(\bar z)} \cosh\left[ \sqrt \Omega_k d_C(\bar z) / d_H\right]  n\cdot(v_e-v_o)  \\
\xrightarrow{\Omega_k = 0}&\ d_L(\bar z) \left[ 1-  n\cdot v_e\right ] - \frac{(1+\bar z)^2}{H(\bar z)}  n\cdot(v_e-v_o)
}

Now, the random movement of emitters will induce an uncertainty of this sort. I return to this point later. This also means that since the Earth is not completely still in the universe, we will have to correct for this effect. This movement of the Earth can be estimated by assuming there is no intrinsic cosmic dipole in the CMB, and then looking at how big the observed dipole is. This dipole must then be the result of a doppler shift, from which one deduces the velocity $v_{\text{Earth through space}} \approx 369 \text{km/s}$, (\citep{Kogut:1993ag,Aghanim:2013suk}). Since this is a constant effect, it is usually subtracted from data sets before publication.

Effects of this kind in relation to SNe have been addressed in eg. \citep{Colin:2010ds,Bonvin:2006en,Feindt:2013pma,Hui:2005nm,Davis:2010jq} regarding both uncertainty estimation and direct searches for bulk flows.

\section{The Cosmological Constant}\label{sec:cosmo:lambda}
I will now devote a single section to an unfairly brief discussion of a problem, whose formulation is maybe more subtle than its answer. The assumed detection of a Cosmological Constant of order $H_0^2$, ie. $\Omega_\Lambda = \mathcal O(1)$ is puzzling for many reasons. I will try to sum up the problem, and refer to some of many reviews on the subject for a deeper analysis, see eg. \citep{Weinberg:1988cp,Carroll:1991mt,Martin:2012bt} and their many references.

The first issue with this particular problem is, that it is not immediate what the problem actually is. We have measured some value of a particular constant in our theory, namely $\Lambda$ in $\Lambda$CDM --- and so what? The first question we might ask ourselves is, \emph{why $\mathcal O(1)$?} How come that the Cosmological Constant $\Lambda$ knows about the hubble scale \emph{today}, and is just about that value. Now, this may just be a coincidence.\footnote{There is a related problem, called the \emph{coincidence problem}.
This is the observation that living in a universe with comparable matter and dark energy densities, $\Omega_m\approx\Omega_\Lambda$, seems somewhat unlikely. Extrapolating back to, say recombination, the matter density has since then been diluted by a factor $\approx 10^9$, while the dark energy density is forever fixed. Yet just now, when we are here, they are almost equal. See \cite{Velten:2014nra} for a more precise definition and discussion of this problem.} Even if it were, things are not this simple. 

The value of the hubble constant, as we saw in Eq.~\eqref{eq:cosmo:params} is \emph{very small} compared to energies of eg. masses of standard model particles, $m_e\approx 10^{-3} GeV$ for the electron to heavier particles like the Higgs, which is $m_H \approx 100 GeV$. Now, the masses of particles come in since the EFE have both a left and right hand side. The \emph{bare} Cosmological Constant is the term $\Lambda$ on the left. But the stress energy tensor, when considering a quantum field theory living in your theory of gravity, gets vacuum contributions, which we might denote $\langle T^{\mu\nu}\rangle$. By Lorenz invariance of the vacuum, this contribution must be of the form $-\rho_\text{vacuum} g^{\mu\nu}$ --- it looks exactly like the $\Lambda$ term. Now the problem is not just that the Cosmological Constant has a peculiar value, but that two distinct physical effects cancel such as to make the sum $\Lambda + 8\pi G\rho \approx H_0^2$. To see why this seems unreasonable, we have to look at the natural sizes of the individual terms.

The classic tale of $\rho$ is vacuum fluctuations of the Standard Model fields. As free fields in a quantum field theory are quantized as an infinite sum of harmonic oscillators, for which the zero-point energy is $\omega/2$, the zero-point energy of a single field is in some sense the sum of these individual terms. As this sum of course diverges, one may be inclined to put in a cut off, with the argument that eg. \emph{we don't know what happens above the Planck scale $E_p$}, and so the sum only goes to energies of order $E_p\approx 10^{18}GeV$. This naive argument gives vacuum contributions of the order $\rho\approx E_p^4\approx 10^{72}GeV^4$ --- one power from the energy of the oscillators, and one power from each of the spatial dimensions we integrate over. This is to be compared with the energy 'density' of the $\Lambda$ term, which is about the critical value, $\rho_\Lambda \approx \rho_c \approx 10^{-46} GeV^4$. The discrepancy between these two numbers is the famous $72-(-46)\approx120$ orders of magnitude between theory and observation.

There is however a flaw in our previous derivation. We introduced an energy cutoff, which explicitly breaks Lorentz invariance --- yet we are trying to calculate a manifestly Lorentz invariant quantity. This is not so good. Doing the calculation more carefully also shows that what we did before would lead to an equation of state $w=+1/3$. It looks like radiation! This is nothing like what we want. It is immediate that we have to abandon the sharp cutoff. What we must do is find a Lorentz invariant way to get rid of the UV --- the high energy modes, which we do not know exactly how behave. Taking a clue from particle physics, we can do dimensional regularization. This is doing the calculation in a general dimension, $d$. Of course the original answer will still diverge, but doing the calculation like this, we can exactly see where and how the infinities occur. That means we can meaningfully subtract an infinity from our result to get something observable. Doing this calculation, we get that it is not the cutoff to the fourth power, but \emph{the mass of the individual fields to the fourth power}, summed, up to some constants. 

But we're still not done. Another term contributing to the vacuum energy density is the zero point of any potential of any particle. There's only one obvious one in the Standard Model, which is the Higgs potential --- the, now famous, Mexican hat. This has a peculiar effect attached to it, since its zero point is different in the past, very hot universe and in the present, cold universe. Namely, when the universe is \emph{very hot}, the potential does not actually look like a mexican hat, but like a normal $\phi^2$ potential because of thermal effects. This in turn means that the difference between the potential energies of the vacuum before and after the \emph{phase transition} is $m_H^4/(4\lambda)$, where $m_H$ is the Higgs mass and $\lambda$ is the Higgs self coupling. If we interpret the potential energy as contributing to the vacuum energy density, this means that either before or after, we are going to have a massive contribution from the Higgs potential. A similar thing happens when chiral symmetry in QCD\footnote{Quantum Chromo Dynamics, the theory of quarks, gluons and their \emph{color} interactions.} is spontaneously broken \cite{Rugh:2000ji}. Inserting standard model values for these quantities, we get
\al{
| \rho_{\text{EW phase transition}}| \approx 10^8 GeV^4 \\
| \rho_{\text{QCD phase transition}}| \approx 10^{-2} GeV^4
}

Collecting all the terms so far lands us at (\cite{Koksma:2011cq}),
\al{\label{eq:cosmo:vacen}
\rho_{\text{vacuum}} \approx& \pm | \rho_{\text{EW phase transition}}| \pm | \rho_{\text{QCD phase transition}}| \nonumber \\
	&+ \rho_\Lambda + \sum_{\text{SM field degrees of freedom}} (-) \frac{m_i^4}{64\pi^2}
}
where the $\pm$ show that there is no a priori preference for what should be the zero point of the phase transition energies, and the minus in the sum is only there for fermion fields. Since the top is so heavy, this sum evaluates to something negative of the order $\sum_{\text{SM fields}}\rho \approx - 10^8 GeV^4$. Thus, the problem has been ameliorated a bit from the initial 120 orders of magnitude fine tuning to a mere $8-(-46) = 54$ orders of magnitude. Fine tuning here means that we have at least the four terms in Eq.~\eqref{eq:cosmo:vacen}, maybe more, all of which are \emph{very big}, and cancel, apparently not exactly, to 54 decimal places, to give us the value $\Omega_\Lambda\approx 1$ today. A very long explanation of all this is found in \cite{Martin:2012bt}.

\emph{This} is the Cosmological Constant problem. The apparent almost-cancellation to an unreasonable number of decimal places of quantities that should know nothing about one-another --- eg. why would the Higgs potential know what the hubble scale is, and why would an arbitrary constant, the Cosmological Constant, know what the top-mass is?

\section{Alternative views}
The story of cosmology in text books is fairly straight forward. Here I want to present some views opposing the very optimistic approach of the perturbed FLRW metric as a valid description for the entire universe. I hope to summarise the idea behind some select points of view in recent literature, but this is by no means meant as even a fair introduction to the subjects, each of which could have been the subject of an entire thesis. As such, I will be skipping technical details, and simply appeal to the idea behind and intuition about the approaches. The nature of the different subjects varies a lot, from changing gravity itself to doing more careful studies of the existing gravity, and the nearby universe.

Because of the large and ever increasing number of cosmological datasets, there is a host of constraints on any model. I will mostly address issues regarding supernovae, while reminding that other non-trivial constraints exists.

\subsection{Changing gravity}
To see the start of this approach, we have to reformulate the derivation of the EFE a bit. As it turns out,\footnote{I will not do the computation, which is messy and not very enlightening. I instead refer to eg. \citep{carroll2004spacetime} for a thorough walkthrough of the results.} the field equations can be found from the principle of least action, given the Lagrange density
\al{\label{eq:cosmo:EHL}
L = \frac{1}{16\pi G} R
}
We then define the action as $S=\int d^4x \sqrt{-g} L$, and the sourceless EFE follow from requiring
\al{
\frac{\delta S}{\delta g_{\mu\nu}} = 0
}
By adding a matter term $L_m$ to Eq.~\eqref{eq:cosmo:EHL} we get the sourced EFE when we identify $T^{\mu\nu} =-2\delta L_m/\delta g_{\mu\nu}+g^{\mu\nu} L_m$. We may also add the constant $\Lambda$ with proper normalisation, which is the Cosmological Constant. This means we get the total Lagrange density
\al{
L = \frac{R-2\Lambda}{16\pi G} + L_m
}
Now inspired by the effective field theory approach of particle physics, we simply consider adding more $R$-like terms to the action. Without specifying further, we just have \emph{some function} of $R$, and we have the lagrange density
\al{
L = \frac{1}{16\pi G}f(R) + L_m
}
from which these kinds of theories derive their name \emph{$f(R)$-gravity}. This is fundamentally changing gravity. Without some great insight, all we are now left with is fitting not just $L_m$ and $\Lambda$, but also the infinite dimensional function $f(R)$, which may or may not be parametrised in some way.  In particular the FLRW metric is still viable, and so this really extends the Cosmological Constant. Note how in the above Lagrange density, $\Lambda$ has been absorbed as the constant part of $f(R)$.

An interesting observation is that Starobinsky inflation (\citep{Starobinsky:1980te}) is an $f(R)$ extension\footnote{Although it is hidden away in the original article --- the $R^2$ term is put into the $T^{\mu\nu}$.}, which --- although having its own problems --- solves problems related to inflation.

Of course, constraints on deviations from general relativity are tight, see eg. \citep{Bertotti:2003rm}, so constraints on reasonable functions $f(R)$ are too. For a comprehensive review of these theories see eg. \citep{Sotiriou:2008rp}.

\subsection{Averaging problem}
The following approach questions what it means that the universe is homogeneous and isotropic \emph{on average} \citep{Buchert:1999er}. The first problem becomes the actual averaging process. It turns out that averaging anything but scalars is a problem, since in general the average of a tensor field does not transform as a tensor. What was started in \citep{Buchert:1999er} was the study of averaged scalar fields, in particular the matter density of the universe. One starts by defining the \emph{spatial average} of a scalar field over a particular region $\mathcal V$ of the universe as 
\al{
\langle \Psi(x,t) \rangle = \frac{1}{\mathcal V} \int_\mathcal V \sqrt{\det h} \ d^3x \Psi
}
where $h$ is the spatial part of the metric and the volume is given by 
\al{
\mathcal V (t)= \int_\mathcal V \sqrt{\det h} \ d^3x
}
This allows us to define an effective scale factor for $\mathcal V$ as 
\al{
a_\mathcal V(t) = \left(\frac{\mathcal V(t)}{\mathcal V_0}\right)^{1/3}
}
One can now average eg. the Friedmann equations with no Cosmological Constant, which gives
\al{
\left(\frac{\dot a_\mathcal V(t)}{a_\mathcal V(t)}\right)^2 = 
	8\pi G \langle \rho\rangle -\frac{1}{2}\langle \mathcal R \rangle - \frac{1}{2} Q_\mathcal V
}
On comparison with Eq.~\eqref{eq:cosmo:friedcomp}, we recognise both the density and $\mathcal R$ term, which is disguised as $k$, but also notice the appearance of a new term $Q_\mathcal V$, which in the simplest case is defined as\footnote{In the interest of intuition, I am skipping a lot of definitions, in particular here is a slight abuse of the original notation. I use here $\Theta^\mu_\mu = 3H$ instead of $\langle\Theta^\mu_\mu\rangle = 3H$.}
\al{
Q_\mathcal V = 6( \langle H^2 \rangle - \langle H\rangle ^2 )
}
Ie. $Q_\mathcal V$ is a measure of the inhomogeneity of the expansion of space, and this term feeds into the Friedmann equation. As it turns out, looking at the equation of state of this new term\footnote{It was found in \citep{Buchert:2006ya} that this can also be interpreted as a scalar field called the \emph{morphon}.} we find that it behaves just like a Cosmological Constant, $w_Q = -1$. Furthermore, it also feeds into the sum rule of Eq.~\eqref{eq:cosmo:sumrule}. These points mean that neglecting this term naturally leads to a biased parameter estimation.

A nice feature of this approach is that the beginning of cosmological acceleration in the FLRW sense seems to coincide with structure formation. This has an immediate interpretation in this formalism, since now the inferred acceleration is linked to the inhomogeneous nature of the universe, \citep{Buchert:2007ik}.

\subsection{Exact inhomogeneous spacetimes}
The Lem\^ aitre-Tolman-Bondi (LTB) metric is an exact general solution to the same questions as the FLRW metric was, except homogeneity, as found very early in \citep{Tolman:1934aa} and later again in \citep{Bondi01121947}. Inspired by the isotropy of the CMB, this was first rediscovered as a physical model of cosmology in \citep{Celerier:1999hp}. Actual fitting of mass profiles to various datasets, including SNe, has been carried out in eg. \citep{Nadathur:2010zm}, which also introduces the various concepts I use below in a simple way. This approach abandons the exact cosmological principle and suggests that our immediate neighbourhood does not have the same density as the rest of the universe, eg. we could be living in an underdensity.

The LTB metric is given by, when molded to a suggestive FLRW-like form,
\al{
ds^2 = -dt^2 + \frac{A'^2(r,t)}{1+K(r)} dr^2 + A^2(r,t) d\Omega^2,
}
where $A'\equiv\pd{A}{r}$. Comparing to Eq.~\eqref{eq:cosmo:FLRWmet}, we notice that putting $A=a(t)r$ and $K=-r^2k$, we obtain again the FLRW metric, which is of course a special case of the LTB metric. 

We can again derive a Friedmann-like equation for this spacetime, by putting in a suitable matter term in the EFE, and we get
\al{
\left(\frac{\dot A}{A}\right)^2 = H_0 \left\{	\Omega_m (A/A_0)^{-3} + \Omega_K (A/A_0)^{-2}	\right\}
}
for some reasonable definitions of $\Omega_i$ --- which of course reduce to the versions we already saw in the homogeneous limit. What we are now left to do is determine the properties of the various functions involved. In particular determining $A$, which can be thought of as a spatially varying scale factor.

The intuitive picture of how an inhomogeneous universe might resemble a universe with a Cosmological Constant can be thought of as follows. What was initially claimed in the SN data was that the far-away SNe were fainter than what was predicted in cosmologies with no $\Lambda$, ie. they were further away. This was interpreted as a recent onset of acceleration of the expansion rate, which in FLRW can only be explained by a $\Lambda$ term. In an inhomogeneous universe, this accelerated expansion is instead explained as the far away universe simply not having the same matter densities as the nearby one. This makes it possible to have different expansion rates at equal times in the universe without invoking a Cosmological Constant. 

There have been arguments over the physical validity of the Earth being the \emph{centre of the universe}, when taking the zero-point of the LTB coordinates to be us, here, see eg. \citep{GarciaBellido:2008gd,Bull:2011wi}. This point though, is taking the LTB too literally, \citep{Celerier:2011zh,Krasinski:2009qq}. In its form here, it should still be thought of as an approximation to what is really going on. This includes the immediate idea that we are, most likely, not the center of the universe. It might be the case in some average sense, that an inhomogeneous metric captures the real world better than a perfectly symmetric one, \citep{Bolejko:2010wc}.

Other exact solutions exist, like the Szekeres model, \citep{Szekeres:1974ct,Bolejko:2010eb} and more contrived examples like patching together FLRW- and LTB-metrics in a kind of \emph{Swiss cheese model}, \citep{Biswas:2007gi}.

\subsection{Dark flow}\label{sec:cos:darkflow}
 Everything we cannot immediately explain the origin of is called \emph{dark}. Dark matter is also dark because we have no evidence that it interacts with light, but dark energy is simply dark because we have no idea what it is. In the same way, there have been claims that there exists an unexplained large scale bulk flow --- a dark flow --- of the nearby universe, see eg. \citep{Kashlinsky:2008ut,Watkins:2008hf}. The first problem is explaining such a large bulk flow in what is supposed to be a very still --- maximally symmetric --- spacetime. Assuming this is done, the presence of the dark flow may mimic cosmic acceleration, \citep{Tsagas11062010,Tsagas:2011wq}. 
 
 The argument is, that the observed acceleration, originally parametrised by $q_0$, is affected by a large bulk flow. First of all, one realises that the apparent hubble constant changes according to the size and magnitude of the bulk flow. This allows one to write the deceleration parameter in the dark flowing frame --- in which we are supposed to reside. Supposing the universe is only non relativistic matter, the global deceleration parameter is $q_0 = \Omega_m/2$, while the local deceleration parameter takes the following form
 \al{
 1+ \tilde q_0 = (1+\Omega_m/2)\left(1+\frac{\theta}{3H}\right)^{-2}\left(1+\frac{\dot\theta}{3H'}\right)
 }
where $\theta$ is a measure of the bulk flow. The difference between dots and primes is a change of frame, but are both time derivatives. The problem now becomes translating from bulk velocities to these quantities, especially $\theta$ and $\dot\theta$. In \citep{Tsagas:2011wq}, using the most optimistic values for these parameters, one gets a change in the deceleration parameter as one goes from the still frame to the bulk flowing one as large as $-0.3$. More conservative estimates diminish this by about an order of magnitude. Even if one is just interested in vanilla $\Lambda$CDM, this is an effect one cannot neglect in a proclaimed era of \emph{precision cosmology}

\chapter{Supernovae}\label{cha:sup}
Studying supernovae (SNe) dates back hundreds, if not thousands of years. If one is lucky, as were the Chinese and Tycho Brahe at different times in history, these exploding stars can be seen as clearly as every other star. Indeed the one observed by Tycho was called \emph{Stella Nova}, the new star. The first systematic attempt of scientific research with these stars was done at the Palomar observatory, \citep{Minko1941}. Already early on, the SNe were split into different types, I and II, depending on their primary element abundances. Later, with more data, the types Ia, Ib and Ic were distiguished, and today many more subclassifications exist, Iax, IIn and IIP, IIL. See eg. \cite{Turatto:2003np} for more on the classification and \cite{Foley:2012tu} for the new Iax class. For a history of SNIa observations, see eg. \cite{Clocchiatti:2011fw}.

What will be the subject of the present section is strictly Type Ia SNe for cosmological purposes. This class was early on seen to be relatively homogeneous, ie. their absolute luminosities are very similar. Having a standard luminosity would mean one could map out the distances in the universe using the relations derived in Sec.~\ref{cos:sec:mography}. This can easily be understood. Take a Euclidean, flat, space and scatter, say $60W$ lightbulbs in it. Measuring the flux, $F$, from a bulb, we easily find the distance to it using $F=L/(4\pi d^2)$. This \emph{luminosity distance} is exactly what we found an expression for in Eq.~\eqref{eq:cosmo:lumdis0} for a general spacetime. Obviously, there are a host of complications in this procedure, and here I want to illuminate some problems and their proposed solutions.

\section{Supernova progenitors}
I will not try to review the history of stars here, but simply state that when stars such as our Sun, a so-called main sequence hydrogen burning star, ends its life, it becomes a \emph{white dwarf} \cite{Koester1990}. The main point we shall consider about white dwarfs is that they are supported against gravity mainly by electron degeneracy pressure. This is a pressure coming from Pauli's exclusion principle --- two or more electrons cannot be in the same state, and so if we squeeze electrons enough, they will fight back. Detailed calculation of a degenerate electron gas shows that the radius of a white dwarf shrinks as we put more mass in it. This leads us to the limit beyond which the degeneracy pressure cannot support the star --- this is called the Chandrasekar limit \cite{Chandra1931}. The numerical value depends on the distribution of mass in the star and the ionisation degree in the gas, and is around
\al{
M_C \approx 2.86 \cdot 10^{30} \text{kg} \approx 1.44 \text{ sun masses}
}
These white dwarfs are what we think is going to be type Ia SNe. The leadup to the SN explosion is still uncertain, but one story, for which \cite{Dilday:2012cy} recently found concrete evidence, goes as follows. Take a system with one white dwarf and another star. The white dwarf may now, over time, suck in matter from the companion star. This only continues as long as the white dwarf is stable --- at most until the Chandrasekar limit, at which point the white dwarf heats up, collapses and initiates a thermonuclear reaction, releasing more than enough energy to blow the white dwarf apart.

The main point of the story is that we have reason to believe that all SNe of this type came from stars of about equal masses. Now if all the SNe have similar boundary conditions and similar evolutions, then we might expect that these can be used as our standard $60W$ bulbs in the universe, \cite{Mazzali:2007et}.\footnote{Of course, this is not the only possible scenario for the progenitor of SNe, and different scenarios might lead to different energy outputs and evolutions. This is an immensely important point, which I will neglect for the main part of the analysis. Different \emph{unidentified} classes of SNe Ia could indeed bias the results of an analysis, which does not identify them as such \cite{Karpenka:2015vva}.}

The result of this violent explosion is a lot of highly radioactive material flung in every direction.\footnote{Although not necessarily isotropically!} The light from the radioactive decay of this debris is what we observe, and is what contests entire galaxies in luminosity.

\section{Observing supernovae}	\label{sec:sn:obs}
\begin{figure}[htb]
\begin{center}
\includegraphics[width=0.8\textwidth]{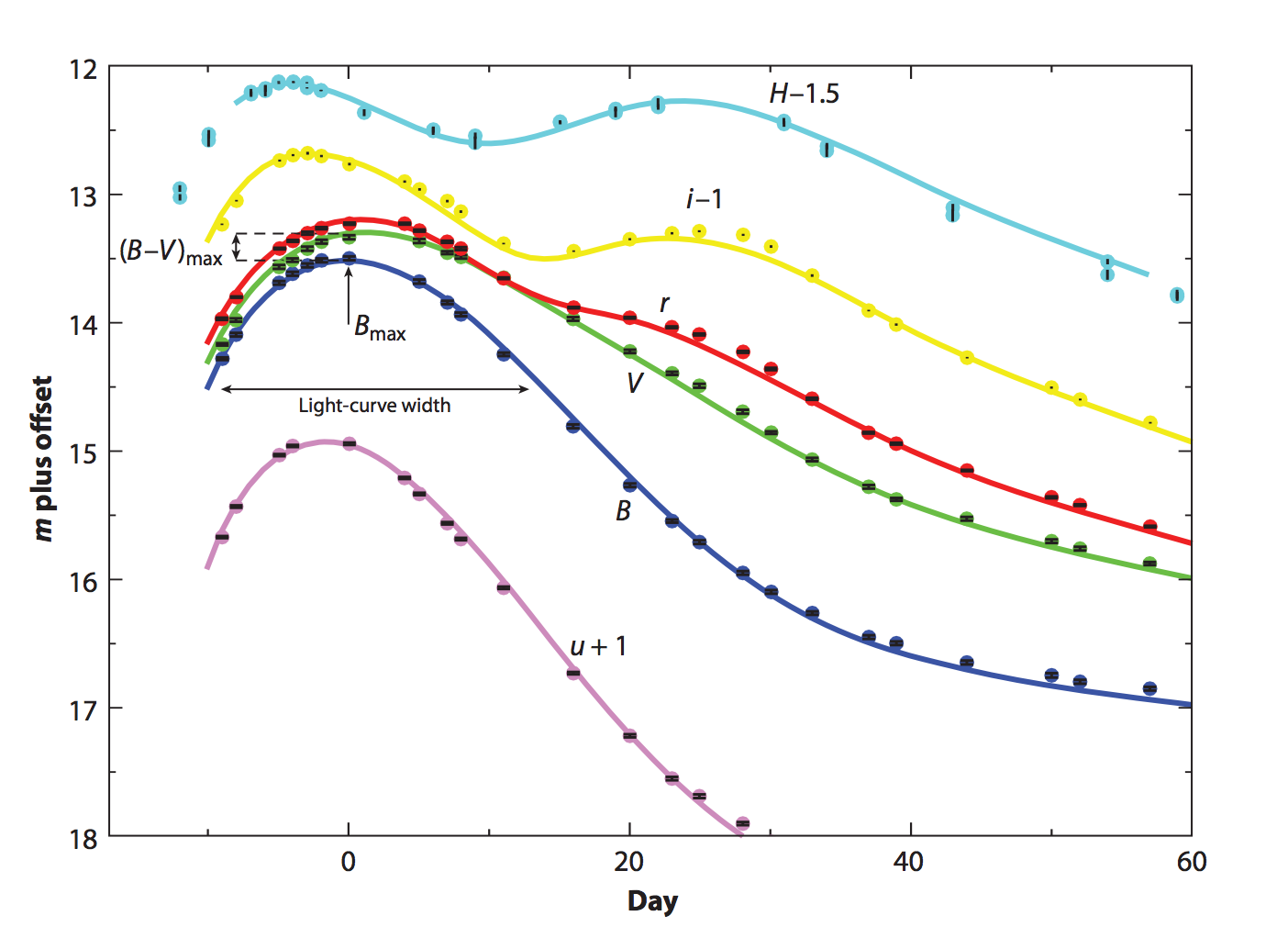}
\caption[Lightcurves of SN 2007af]{Optical and near-infrared lightcurves of SN 2007af from the Carnegie Supernova Project. The mean wavelength of the bandpasses ranges from $350\text{nm}$ (u band) to $1600\text{nm}$ (H band). The y-axis is in apparent magnitudes, and the time axis is shifted so day $0$ is at maximum B band brightness. Figure is taken from \cite{Goobar:2011iv}.}
\label{fig:sn:snexample}
\end{center}
\end{figure}

Once the hurdle of actually finding SNe is overcome,\footnote{This point is naturally very non-trivial, but will not concern us too much.} the observation is a timeseries of photometric measurements, ie. fluxes through various colour filters. These timeseries are called lightcurves. The classic photometric system is the Johnson-Cousins or UBV \emph{(ultraviolet, blue, visual)} system \cite{Johnson:1953zz}. \emph{Many} more systems exist today, see eg. \cite{Bessell:2005zz}. Such a photometric system is a series of window functions on the allowed frequencies/wavelengths of the observed light. An example of such an observation in an extended system is shown in Fig.~\ref{fig:sn:snexample}.

\begin{figure}[htbp]
\begin{center}
\begin{turn}{-0.5}
\includegraphics[width=0.8\textwidth]{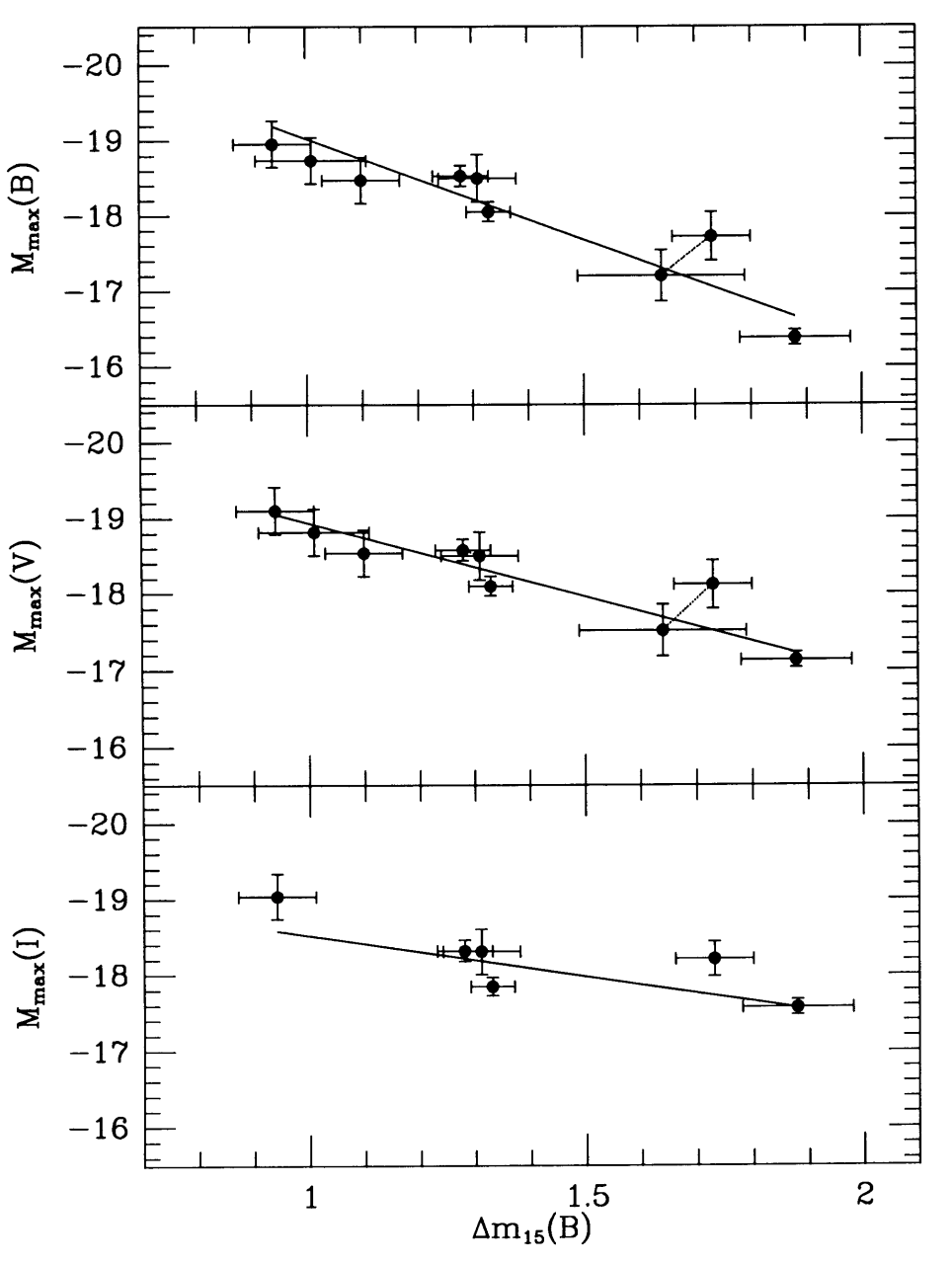}
\end{turn}
\caption[Plot of the original Phillips' relation]{The plot from \cite{Phillips:1993ng}, where Phillips noticed the trend that as the lightcurves get wider, the SNe are brighter. The trend is present in all three bands considered here, but most prominent in the B band. The x axis, $\Delta m_{15}(B)$ is the decline in B band apparent magnitude after 15 days, ie. small numbers correspond to wide lightcurves. The y axis shows the derived maximum absolute magnitudes in different bands.}
\label{fig:sn:phillips}
\end{center}
\end{figure}

Now to do cosmological studies, we need a way to convert from the observed fluxes to distance measurements. To do this, we need three conventional things: a flux, a luminosity and a distance. These three quantities naturally must obey Eq.~\eqref{eq:cosmo:lumdis0}. Measuring another flux, as we do, and assuming this comes from a standard candle, ie. a class of sources of equal luminosity, as we hope SNe are, we have the following relationship, 
\al{
\frac{F/F_\text{ref}}{L/L_\text{ref}} = \frac{1}{(d_L/d_{L,\text{ref}})^2}
}
We now define the apparent magnitude $m=-2.5 \log_{10}{ F/F_\text{ref} }$ and absolute magnitude $M = -2.5\log_{10} L/L_\text{ref}$, which gives us the expression
\al{	\label{eq:sn:dismod1}
\mu = m-M = 5 \log_{10}\frac{d_L}{d_{L,\text{ref}}}
}
For historical reasons, $d_{L,\text{ref}} = 10\text{pc}$. $\mu$ is called the distance modulus. It is now apparent that if $M$ is the same for all SNe, then measurements of the flux are directly linked to the luminosity distance, which reveals to us the expansion history of the universe. Of course we don't expect the intrinsic scatter in the luminosity to be exactly zero, even if the progenitor scenarios are similar. The earliest observations of SNe were too scattered for a good cosmological study. But they did show a remarkable feature --- the width of the lightcurve was tightly linked to the absolute magnitude. This effect is known as the \emph{Phillips relation} \cite{Phillips:1993ng}, and the very first plot used just 9 observations, see Fig.~\ref{fig:sn:phillips}. Later, Tripp \cite{Tripp:1997wt} found another correlation with the colours of the SNe. These two quantities are marked in Fig.~\ref{fig:sn:snexample}. The hope is still today that one will be able to reduce the scatter in the Hubble diagram even further by finding more observables with significant correlation to the Hubble residuals, see eg. \cite{Kelly:2009iy,Hayden:2012aa} for some examples of this.
Taking the corrections of the absolute magnitude to be linear in the new observables --- as is approximately observed in Fig.~\ref{fig:sn:phillips} and corroborated in later studies, \cite{Guy:2005me} --- we have for the two-parameter model, writing in modern notation $x_1$ for the shape of the lightcurve and $c$ for the colour correction,
\al{	\label{eq:sn:measmu}
\mu = m-M \rightarrow m - ( M - \alpha x_1 + \beta c ),
}
where $\alpha,\beta$ are unknown coefficients --- we still have no good theoretical model of what these should be. This parametrisation of the corrections is the one used by the SALT (\emph{Spectral Adaptive Lightcurve Template}) analysis, see \cite{Guy:2005me,Guy:2007dv} for detail about the fitting and the exact meaning of the parameters $x_1,c$. In short, higher $x_1$ are broader lightcurves, and higher $c$ are redder colours.

This means that we now see the SNe as \emph{standardisable candles}, their luminosity can be corrected to be more or less standard (we will see later to what degree they actually are).\footnote{Other parametrisations exist, most prominently the MLCS(2k2) \emph{(Multicolor Light Curve Shapes)} \cite{Riess:1996pa,Jha:2006fm} and the newer SiFTO \cite{Conley:2008xx}. A more complete list of older methods is found in the introduction of \cite{Conley:2008xx}.} Naturally, these measurements come with some uncertainty due to experimental noise. This means that our dataset of the maximum B band apparent magnitude, shape and colour correction, $(m^*_B,x_1,c)$ comes with some covariance matrix. One part of this is the statistical uncertainty --- the noise --- and the other is various systematic uncertainties. Determining these uncertainties is a big part of any analysis. This also shows up in how surveys are done, since early time coverage of the SNe is important to precisely determine the parameters, especially the width of the lightcurve. This means that ideally one wants to take pictures of the sky where there is no SN, since if there is going to be one in ten days, you want to see the very early time light --- notice how in Fig.~\ref{fig:sn:snexample} the observations start about ten days before maximum brightness.

\subsection{K-correction}
When we derived the luminosity distance, we considered the \emph{bolometric luminosity}, ie. the luminosity integrated over all colours of the light. Yet, when calculating the distance estimate we only have light in certain bands. Now, redshifting the spectrum, but keeping the filters fixed --- for obvious reasons --- we introduce a redshift dependent bias in the distance estimate, given by (\cite{Oke1968,Hamuy1993})
\al{
K = 2.5 \log_{10}(1+z)+ \log_{10}\left( \frac{\int F(\lambda) S(\lambda) \ d\lambda}{\int F(\lambda/[1+z]) S(\lambda) \ d\lambda}  \right),
}
where $F(\lambda)$ is the differential flux and $S(\lambda)$ is the filter response. Now, given $F$, we can correct for this by taking instead of the bolometric absolute magnitude,
\al{
M \rightarrow M + K
}
This is usually done beforehand in data releases, and so we do not need to worry about it.

\section{Comparison with the cosmos}
As calculated in the previous section, we find for every SN a distance modulus, which we want to compare to our cosmological model. From Eq.~\eqref{eq:sn:dismod1} and \eqref{eq:cos:lumdis1} we see that the expected distance modulus is
\al{\label{eq:sn:dismod2}
\mu_\mathcal{C} = 5 \log_{10}( d_L / 10\text{pc} )= 5 \log_{10}\left( \frac{(1+z)}{\sqrt{\Omega_k}} \sinh\left[ \sqrt{\Omega_k} \int_0^z \frac{H_0 dz'}{H(z)} \right] \right) \phantom{,}
	\nonumber \\
	+ 25  - 5\log_{10}{h}+ 5\log_{10}\left(\frac{c}{100 \text{km/s} }\right),
}
where the last log evaluates to $\approx 3.477$. The take-away here is that when comparing this with Eq.~\eqref{eq:sn:measmu}, both contain unknown constants. The measurement has the absolute magnitude, and the theoretical expression contains the Hubble constant. Importantly, these combine to a single constant which factorises completely from the rest of the expression when taking the difference --- as we shall do later. That means we cannot with SNe alone constrain either of these! What one does is to set the Hubble constant to something reasonable, eg. $h=0.7$,\footnote{This choice is manifestly arbitrary, and can always be changed after the analysis.} and then fit the absolute magnitude. It is important to remember though, that without a direct measurement of the absolute magnitude or the Hubble constant, we can't break this degeneracy.\footnote{What has been refined for decades is the so-called distance ladder. This is a series of classes of objects, each with its own defining feature which allows determination of the distance to it. Every class is then a rung on the distance ladder, see eg. \cite{Weinberg:1972aa}. Putting the SN observations on this ladder allows one to determine the Hubble constant and in turn the absolute magnitude.} 

Let's now go through some of the cosmological effects adding uncertainty to the measurements.

\subsection{Peculiar velocities}
Deriving the luminosity distance, we assumed that the source was stationary. Using the results from Sec.~\ref{sec:cos:moving}, we may estimate the error we commit when doing this. From independent measurements, we estimate the variance of the isotropic velocity field to be about $\sigma_v \approx 150 \text{km/s}$.\footnote{Isotropy here is a rather rough approximation, since this is extrapolated down to small scales --- another method takes into account the correlations in the velocity field, see eg. \cite{Davis:2010jq}. Here I just aim to illustrate the physics behind the effect.} By Eq.~\eqref{eq:cos:redshiftshift}, this leads to a redshift uncertainty, given in terms of the variance of the peculiar velocity along the line of sight. This is of course just $\sigma_v$, and so
\al{
\sigma_{z,\text{pecvel}} = (1+z) \sigma_v
}
The $(1+z)$ term here is usually neglected with the reason that these errors are important only at low $z$, when the term is small anyway. Now we just need to convert this redshift uncertainty to an uncertainty in the distance modulus. This is computed approximately as
\al{\label{eq:sn:pecunc}
\sigma_{\mu,\text{pecvel}} = \sigma_{z,\text{pecvel}} \pd{\mu}{z} = \sigma_{z,\text{pecvel}} \frac{5}{\log(10)} \pd{d_L}{z} \frac{1}{d_L}
}
The usual procedure now is to take \emph{some} cosmology and calculate $\pd{d_L}{z}$ explicitly. Which cosmology one choses doesn't matter much, since they are all similar at low $z$ where the error is important, \cite{Davis:2010jq}. So let us choose the empty universe, $\Omega_m =\Omega_\Lambda= 0$, which has luminosity distance
\al{
d_{L,\text{empty}} = \frac{1+z}{H_0} \sinh\left( \log(1+z) \right) \\
\Rightarrow \pd{d_L}{z} \frac{1}{d_L} = \left( \frac{1}{1+z} + \frac{1}{(1+z)\tanh\log(1+z)} \right)
}
Evaluating $\tanh\log(1+z) = z(z+2)/[2+z(z+2)]$ reduces Eq.~\eqref{eq:sn:pecunc} to
\al{
\sigma_{\mu,\text{pecvel}} \approx \sigma_{z,\text{pecvel}} \frac{5}{\log(10)}\left( \frac{1+z}{z(1+z/2)} \right)
}
This is then usually added in quadrature to other errors, since we assume this effect is entirely uncorrelated to all other errors. This effect is why cosmological datasets usually have a lower redshift limit. We only want to look at SNe, which are safely in the \emph{Hubble flow}, ie. where peculiar velocity effects are not dominant. The exact lower limit varies from analysis to analysis, but is usually of order $10^{-2}$.

\subsection{Weak gravitational lensing}
Gravitational lensing, a subject in its own right \cite{Schneider2006}, also affects SN measurements \cite{Kantowski:1995bd,Frieman:1996xk,Gunnarsson:2005qu,Jonsson:2010wx}. Light running through the universe is bent by the inhomogeneous large scale structure. This means that the flux we infer is also contaminated by the distorted image --- eg. a demagnified SN will appear fainter, ie. further away. This effect is greater for far-away SNe, as is intuitively clear --- the further away, the bigger the \emph{optical depth}, ie. the more lenses to distort the image.

For a precise determination of the lensing effect, one needs not only properties of the universe, but also of dark matter haloes --- the profile of dark matter surrounding galaxy clusters. These are hard to determine, and so the exact numbers for the lensing uncertainties vary from work to work. An early study \cite{Holz:2004xx} found a linear relation between the noise and redshift, quoting an error of $\sigma_\text{lens} \approx 0.088z$, while a newer study \cite{Jonsson:2010wx}, dedicated to the data of \cite{Betoule:2014frx}, finds $\sigma_\text{lens} \approx 0.055z$. This is indeed the value used in \cite{Betoule:2014frx}, which we also take. This error is also added in quadrature, even though the actual lensing bias is not expected to be gaussian.

Future surveys hope to be able to correlate large scale structure with the lensing bias --- ie. to look at the line of sight through which the SN was found and try to determine separately the expected lensing, and as such make the once uncertainty a new signal. These possibilities will be explored by eg. the \emph{Dark Energy Survey}.\footnote{Thanks to Tamara Davis for insight on this point.}

\chapter{Putting the pieces together}\label{cha:put}
Now we combine the last three sections into an analysis of SN data. The ultimate goal of this analysis is of course to lay out the expansion history of the universe, potentially unravelling the mysteries of the cosmos. In less grandiose terms, we wish to constrain the parameters $\Omega_m,\Omega_\Lambda$ of our favourite, maximally spatially symmetric space-time, the FLRW model.

The data I use is the \emph{Joint Lightcurve Analysis (JLA)} catalogue \cite{Betoule:2014frx} --- a combination of data from Sloan Digital Sky Survey (SDSS-II), the SuperNova Legacy Survey (SNLS), SNe from the Hubble Space Telescope (HST) and some low redshift SNe from a selection of other surveys. See also \cite{Conley:2011ku} for description of selection criteria and outlier rejections. In this work, a lot of issues of combining SN surveys have been adressed, in particular calibration issues between telescopes and the empirical \emph{training} of the SALT procedure. 

This section follows closely our recent work \cite{Nielsen:2015pga}, only in more detail.
\section{The dataset}
\begin{figure}[htbp]
\begin{center}
\includegraphics[width=\textwidth]{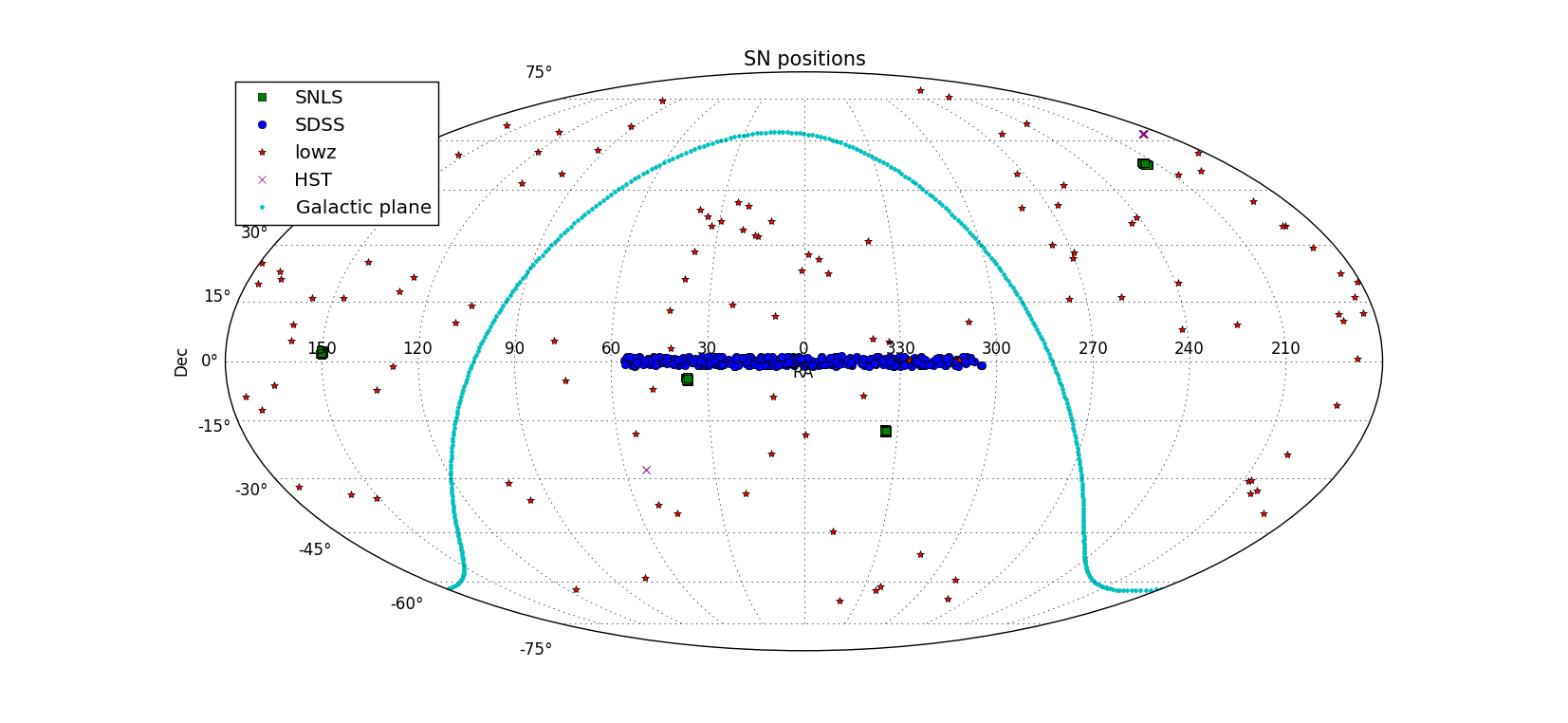}
\caption[Distribution of SNe on the sky]{Mollweide projection of the distribution on the sky of the four sets of SNe. The cyan line going across the sky is the galactic plane. It is immediately obvious that not many SNe are seen through the Galactic plane. Notice how the SDSS and SNLS surveys are constrained to very small regions of the sky, where the observers look over and over again. This helps them get nice early time coverage of the lightcurves.}
\label{fig:ev:skycov}
\end{center}
\end{figure}

\begin{figure}[htbp]
\begin{center}
\includegraphics[width=0.49\textwidth]{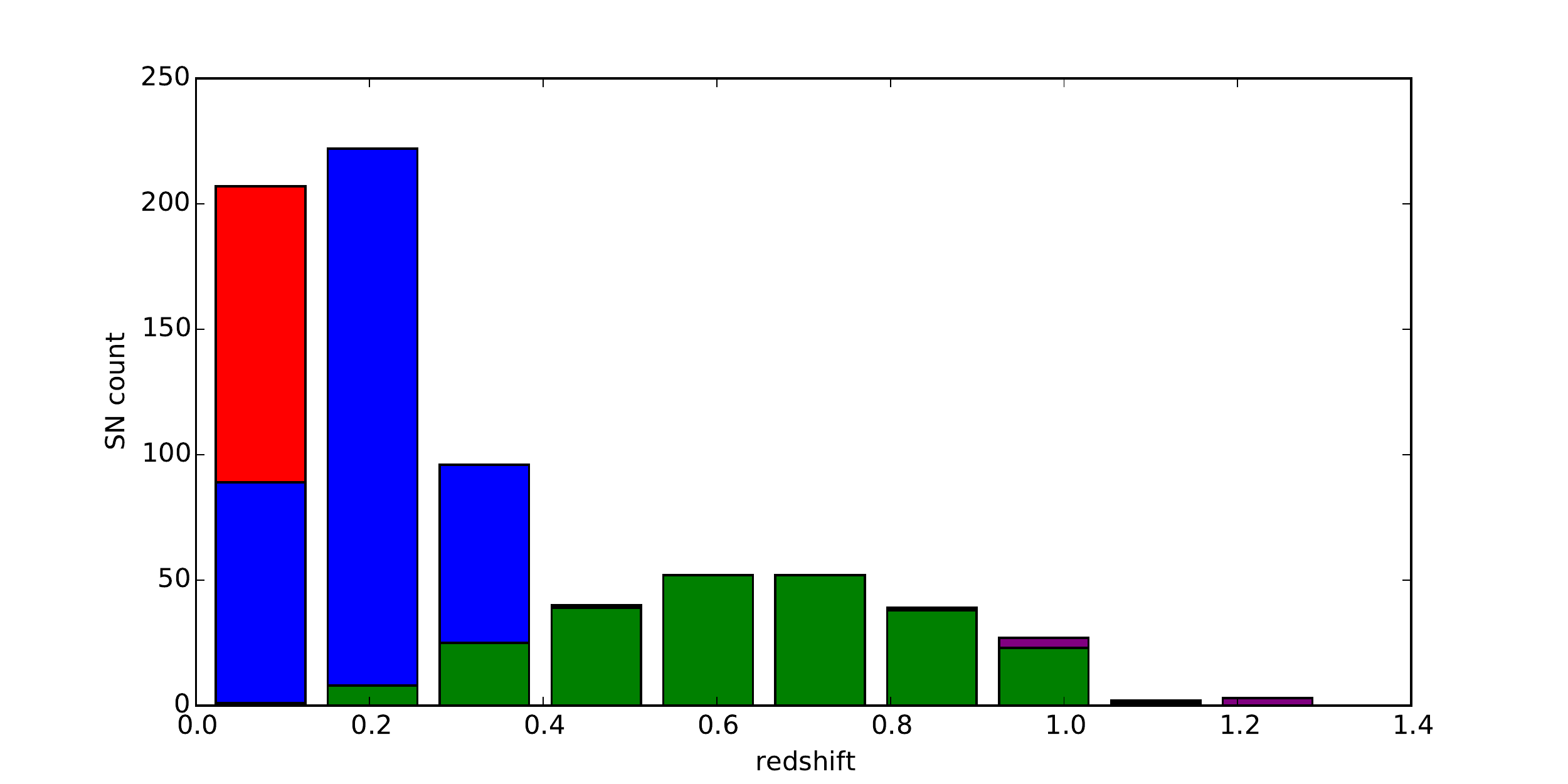}
\includegraphics[width=0.49\textwidth]{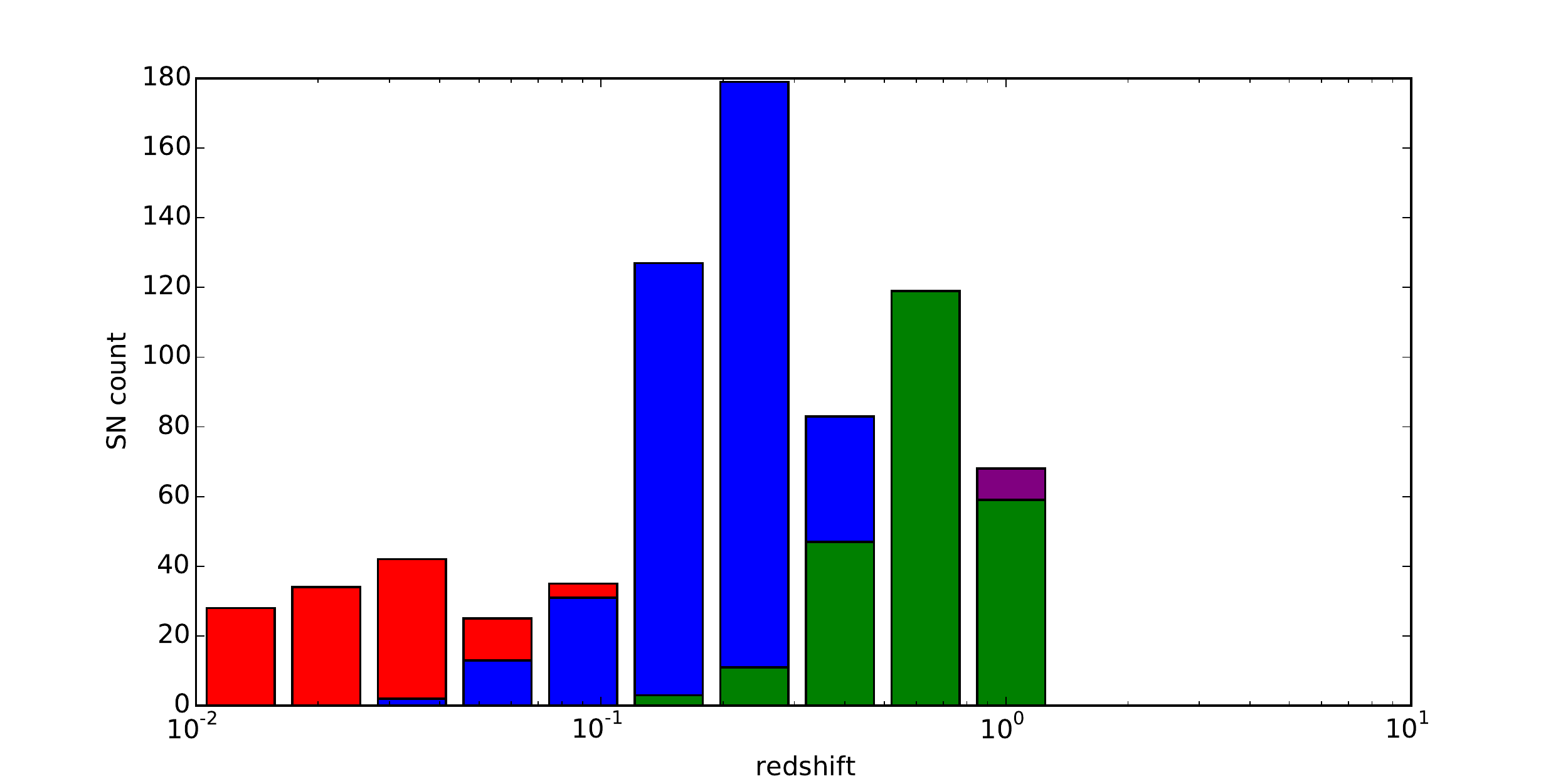}
\caption[Redshift distribution of SNe]{Distribution of SN redshift from different surveys. The right histogram has logarithmic redshift axis. From low to high redshift (red, blue, green, purple, same colour coding as Fig~\ref{fig:ev:skycov}) is low-z, SDSS, SNLS, HST. }
\label{fig:ev:zcov}
\end{center}
\end{figure}

All data I use, including covariances, is available through the website of the JLA collaboration.\footnote{\url{http://supernovae.in2p3.fr/sdss_snls_jla/ReadMe.html}} Here I wish to give a brief overview of how it looks and feels. Fig.~\ref{fig:ev:skycov} shows the distribution of the SNe on the sky in equatorial coordinates. The \emph{SDSS stripe} is about $2.5^\circ$ wide and $120^\circ$ long, while the SNLS samples 4 regions of low galactic extinction with area 1 square degree each. This distribution on the sky makes the high redshift surveys particularly bad for dipole searches \` a la Sec.~\ref{sec:cos:darkflow}, since we have information only in very limited sections of the sky --- any multipole expansion of the velocity field of the far away universe will be wildly unstable. The redshift coverage of the different surveys is shown in Fig.~\ref{fig:ev:zcov}. Notice in particular how the SDSS has filled in a gap around redshift $0.2$. This gives some constraining power over the most naive implementation of void models, where the Hubble parameter would 'jump' between the datasets.

\begin{figure}[htbp]
\begin{center}
\includegraphics[width=0.49\textwidth]{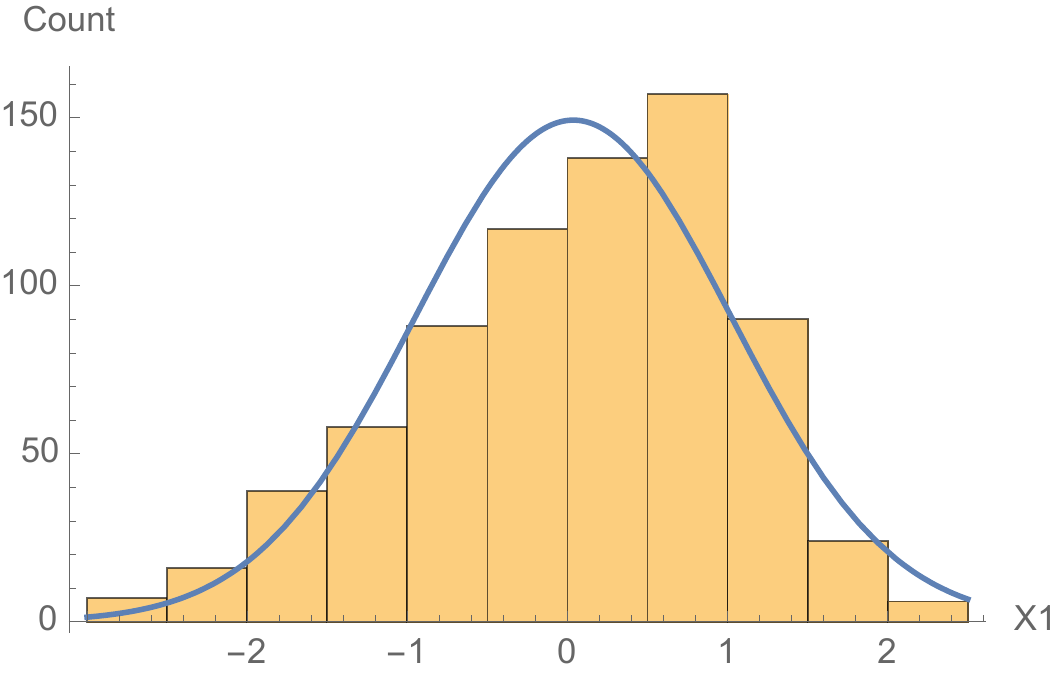}
\includegraphics[width=0.49\textwidth]{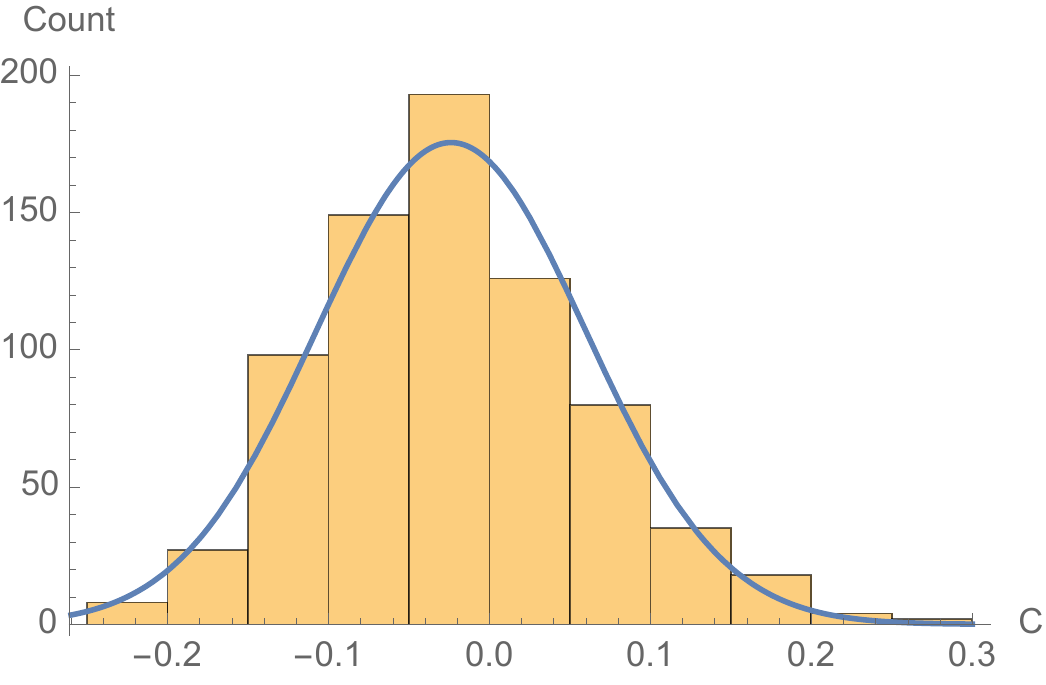}
\caption[Distribution of measured SN correction parameters]{Distribution of the measured $\hat x_1,\hat c$. A gaussian with matching mean and variance is superimposed. The approximate standard deviation of these distributions are $\sigma_{x1}\approx 0.99$ and $\sigma_c\approx 0.084$. These numbers are guiding, and play absolutely no role in the fitting procedure.}
\label{fig:ev:corrpar}
\end{center}
\end{figure}

\begin{figure}[htbp]
\begin{center}
\includegraphics[width=0.49\textwidth]{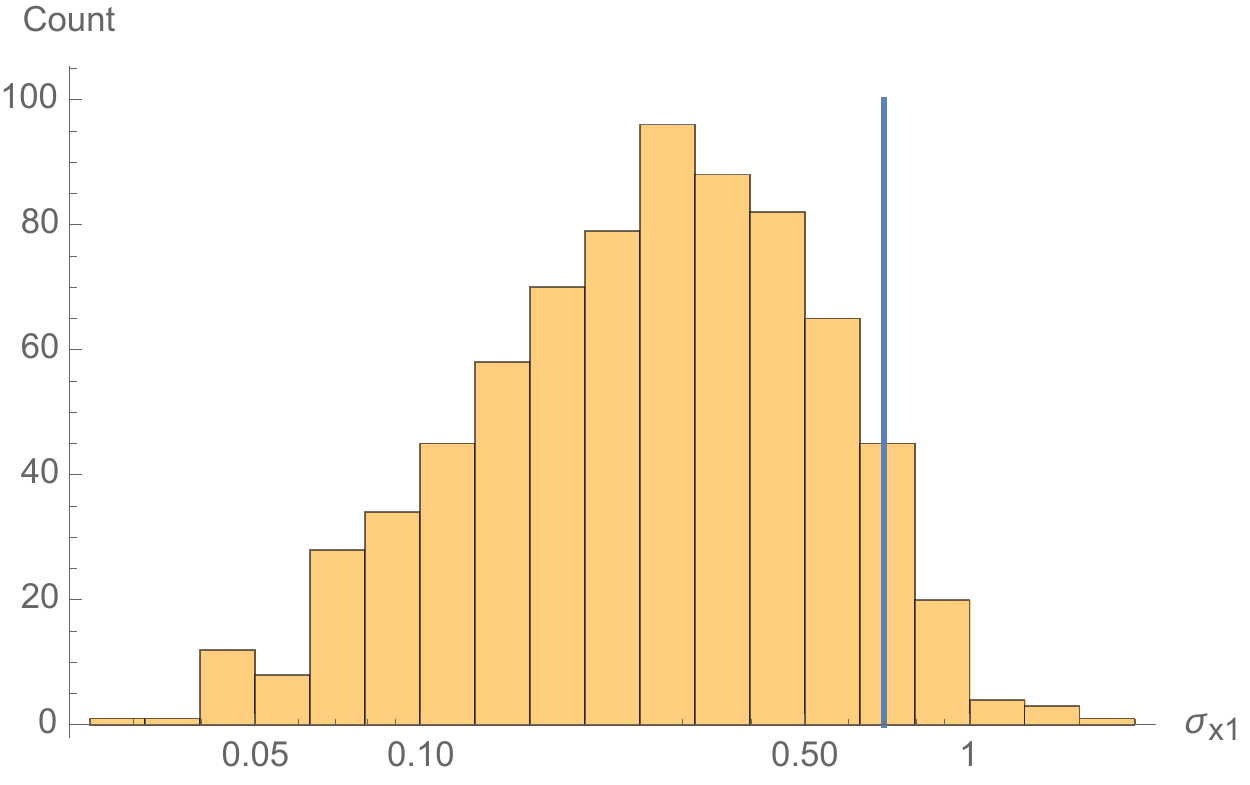}
\includegraphics[width=0.49\textwidth]{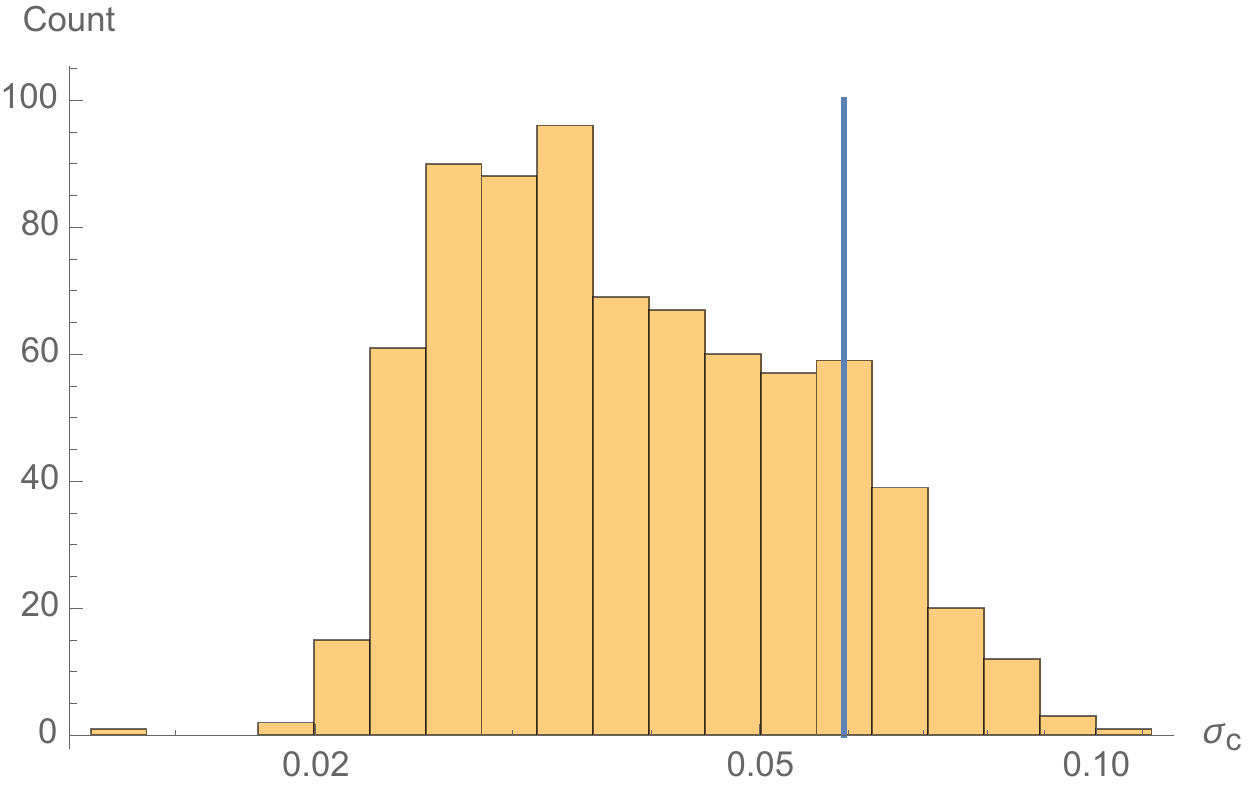}
\caption[Distribution of estimated errors on correction parameters]{Distributions in logarithmic bins, of the square root of the diagonal elements of the covariance matrix for the correction parameters $x_1,c$. These errors are to be compared with the variance of the distribution in Fig.~\ref{fig:ev:corrpar}. It is immediately obvious that we cannot neglect these errors when constructing the likelihood function. Indeed this looks a bit like the example \emph{Unequal errors on measurements} in Sec.~\ref{sec:stat:MC} --- we have some intrinsic distribution which we contaminate with experimental uncertainties, so the total error is the intrinsic error and the experimental error added in quadrature. Where the two errors would be approximately equal (taken as simply the numbers quoted below Fig.~\ref{fig:ev:corrpar} divided by $\sqrt 2$) is marked by the blue line. }
\label{fig:ev:corrunc}
\end{center}
\end{figure}

Next, let's have a look at the correction parameters $x_1,c$ of the SALT fits. These are presented in Fig.~\ref{fig:ev:corrpar}. The superimposed gaussians are simply put there to guide the eye. As of writing this, I know of no theory of the distribution of these parameters. That is why we assume the theoreticians dream, that they come from gaussians. This is almost certainly not true, but will be our first step towards finding out more about these distributions. As we will see, we absolutely need to deal with these distributions. To see why this is the case, let's look at the distribution of the uncertainties of the measurements. The diagonal elements of the covariance matrix are presented in Fig.~\ref{fig:ev:corrunc}. For some of these measurements, not only does the intrinsic error and the experimental one have similar size --- the experimental noise may completely dominate for the most poorly sampled lightcurves. For most of them though, this is not the case, and the measurement is quite good. The point still remains in principle though, and we lose nothing --- except computing time --- by doing a more thorough analysis.

Naturally, one could consider more intricate distributions for these two parameters. This is simply a matter of introducing more and more parameters to describe them. However, without any physical motivation to do so, we simply put in gaussians. Even if they are not optimal, they capture the most prominent feature --- they have some spread, which we wish to quantify.

\section{A statistical model of calibrated standard candles}
As stated in Sec.~\ref{sec:sn:obs}, the corrected SN distance modulus is taken to be\footnote{Note that we do not include the newer \emph{host galaxy mass correction} employed in \cite{Betoule:2014frx}, as it is not of immediate relevance to the problem we are addressing. The change due to the exclusion of this parameter is negligible.}
\al{
\mu_\text{SN} = m_B^* - (M-\alpha x_1 + \beta c),
}
Now we take it to be the true values that obey this relation. Writing down the likelihood for the dataset $(\hat m^*_{B1},\hat x_{11},\hat c_1,\dots)$ is then straight forward. With a hat on observed values, we have
\al{
\mathcal L = p[(\hat m^*_{B},\hat x_{1},\hat c)|\theta]
 = \int p [(\hat{m}^*_B, \hat{x}_1, \hat{c}) | (M, x_1, c), \theta]
  \ p [(M, x_1, c) | \theta] \text{d}M \text{d}x_1 \text{d}c \label{eq:ev:conlike}
}
Here I have simply used Eq.~\eqref{eq:stat:intout} to integrate out my ignorance of the true values of the observables. I stress that $p [(M, x_1, c) | \theta]$ is my simple model of the distribution of the correction parameters --- \emph{not} a (Bayesian) prior. What we need now is just the two expressions for the pdfs in this equation.

Inspired by Fig.~\ref{fig:ev:corrpar} we will take the intrinsic distribution of all the parameters $M,x_1,c$ to be independent,\footnote{This is easily extended to correlated distributions if one wants to do so --- here we simply use the minimal reasonable amount of parameters} gaussian, and redshift independent, giving us the following pdf,
\al{
&p [(M, x_1, c)|\theta] = p (M|\theta) p(x_1|\theta) p(c| \theta), 
 \quad \text{where:} \\
&p (M|\theta) = (2\pi\sigma_{M_0}^2)^{-1/2} 
 \exp\left\{-\left[\left({M - M_0}\right)/\sigma_{M_0}\right]^2/2\right\},
 \nonumber \\
 &p (x_1|\theta) = (2\pi\sigma_{x_0}^2)^{-1/2} 
 \exp\left\{-\left[\left({x_1 - x_0}\right)/\sigma_{x_0}\right]^2/2\right\}, 
 \nonumber \\
 &p (c|\theta) = (2\pi\sigma_{c_0}^2)^{-1/2} 
 \exp\left\{-\left[\left({c - c_0}\right)/\sigma_{c_0}\right]^2/2\right\}. \nonumber
}
All the 6 parameters $\{M_0,\sigma_{M_0}, x_0, \sigma_{x_0}, c_0, \sigma_{c_0}\}$ are fitted along with the cosmological parameters, since we have no theoretical model for what they should be. To simplify our calculations, we write this in terms of a vector $Y= \{M_1, x_{11}, c_1, \dots M_N, x_{1N}, c_N \}$, the corresponding zero-points $Y_0$, and the covariance matrix $\Sigma_l = \text{diag}(\sigma_{M_0}^2,\sigma_{x0}^2,\sigma_{c0}^2,\dots)$,
\al{
p[(M, x_1, c)|\theta] = p[Y|Y_0,\theta] =  {|2\pi\Sigma_l|}^{-1/2} 
 \exp\left[-(Y - Y_0) \Sigma_l^{-1} (Y - Y_0)^\text{T}/2\right].
}
Note that this gaussian approximation is just the simplest reasonable model for these data. Introducing skewness or other such distributions may obviously lead to higher likelihoods. The main point here is that \emph{we desperately need some model}, and we have no theoretical motivation for any one over another --- we merely pick the simplest one.

Taking the experimental errors, statistical as well as systematic, to be described by gaussians as well, gives us the following pdf
\al{
p (\hat X|X, \theta) = {|2\pi \Sigma_\text{d}|}^{-1/2}
 \exp \left[-(\hat{X} - X) \Sigma_\text{d}^{-1} (\hat{X} - X)^\text{T}/2 \right],
}
where $X = \{m^*_{B1}, x_{11}, c_1,\dots\}$, $\hat X$ is the observed $X$, and $\Sigma_\text{d}$ is the estimated experimental covariance matrix. To combine the two, we write
\al{
\hat X& - X = (\hat{Z}A^{-1} - Y) A \hspace{1cm} \text{where} \\
\hat Z &= \{ \hat m^*_{B1} - \mu_{C1}, \hat x_{11}, \hat c_1, \dots \}, \nonumber \\
A &= \begin{pmatrix}
1 & 0 & 0 & \\
-\alpha & 1 & 0 & 0 \\
\beta & 0 & 1&  \\
 & 0 & & \ddots
\end{pmatrix}, \nonumber
}
where the 3-by-3 block repeats all the way down, and all other elements are zero. Hence $p[\hat{X} | X, \theta] = p[\hat{Z} | Y, \theta]$ and we get for the likelihood
\al{
\mathcal L  
&= \int p[\hat{Z} | Y, \theta] \ p[Y | Y_0, \theta] \text{d}Y  \\
&= {|2\pi \Sigma_\text{d}|}^{-1/2} {|2\pi\Sigma_l|}^{-1/2}
	 \int\text{d}Y  \nonumber \\	
&\phantom{=}\times\exp\left(-(Y-Y_0) \Sigma_l^{-1} (Y-Y_0)^\text{T}/2 \right) \nonumber \\
&\phantom{=}\times\exp\left(-(Y-\hat{Z} A^{-1}) A \Sigma_d^{-1} 
  A^\text{T}(Y - \hat{Z} A^{-1})^\text{T}/2 \right) \nonumber \\
 & = {|2\pi(\Sigma_\text{d} + A^\text{T}\Sigma_l A)|}^{-1/2} \label{eq:ev:reallike}\\
&\phantom{=} \times \exp\left[-(\hat{Z} - Y_0 A)(\Sigma_\text{d} + A^\text{T}
 \Sigma_l A)^{-1}(\hat{Z} - Y_0 A)^\text{T}/2 \right].\nonumber 
}
The gaussian form of the integrand makes this integral very simple. Remember that this likelihood reflects not just our calibration of the SNe, but also the modelling of the correction parameters. From this likelihood we will find the MLE and derive confidence limits on both cosmological quantities, $\Omega_m,\Omega_\Lambda$, but are also able to place constraints on the absolute  magnitude scatter, the correction coefficients and the distributions of correction parameters. All these tell us about how good our model is and how well our candles are calibrated --- how \emph{standard} they really are.

\section{Results of the main fit}\label{sec:ev:mainfit}
In this section I will present the main result of the work --- the MLE and confidence regions of our fit to the latest, greatest catalogue of SNe to date.\footnote{The code and data used in the analysis is available for the interested at \url{http://nbia.dk/astroparticle/SNMLE/}. It uses Python 2.7 and the scientific library SciPy, both of which are open source.} Tab.~\ref{tab:ev:results} presents the best fits under specific constraints. The parameters are described in the previous section. In particular we see that the calibration brings the intrinsic (or at least unaccounted-for) variation to $0.108$ mag. We also get out the variances of the correction parameter distribution, which are relatively independent of the other parameters. Compared to the numbers quoted in Fig.~\ref{fig:ev:corrpar} we see that the effect of experimental uncertainties is most pronounced in the colour correction, $c$, which might have been anticipated from Fig.~\ref{fig:ev:corrunc}.

\begin{figure}[htb]
\begin{center}
\includegraphics[width=0.8\textwidth]{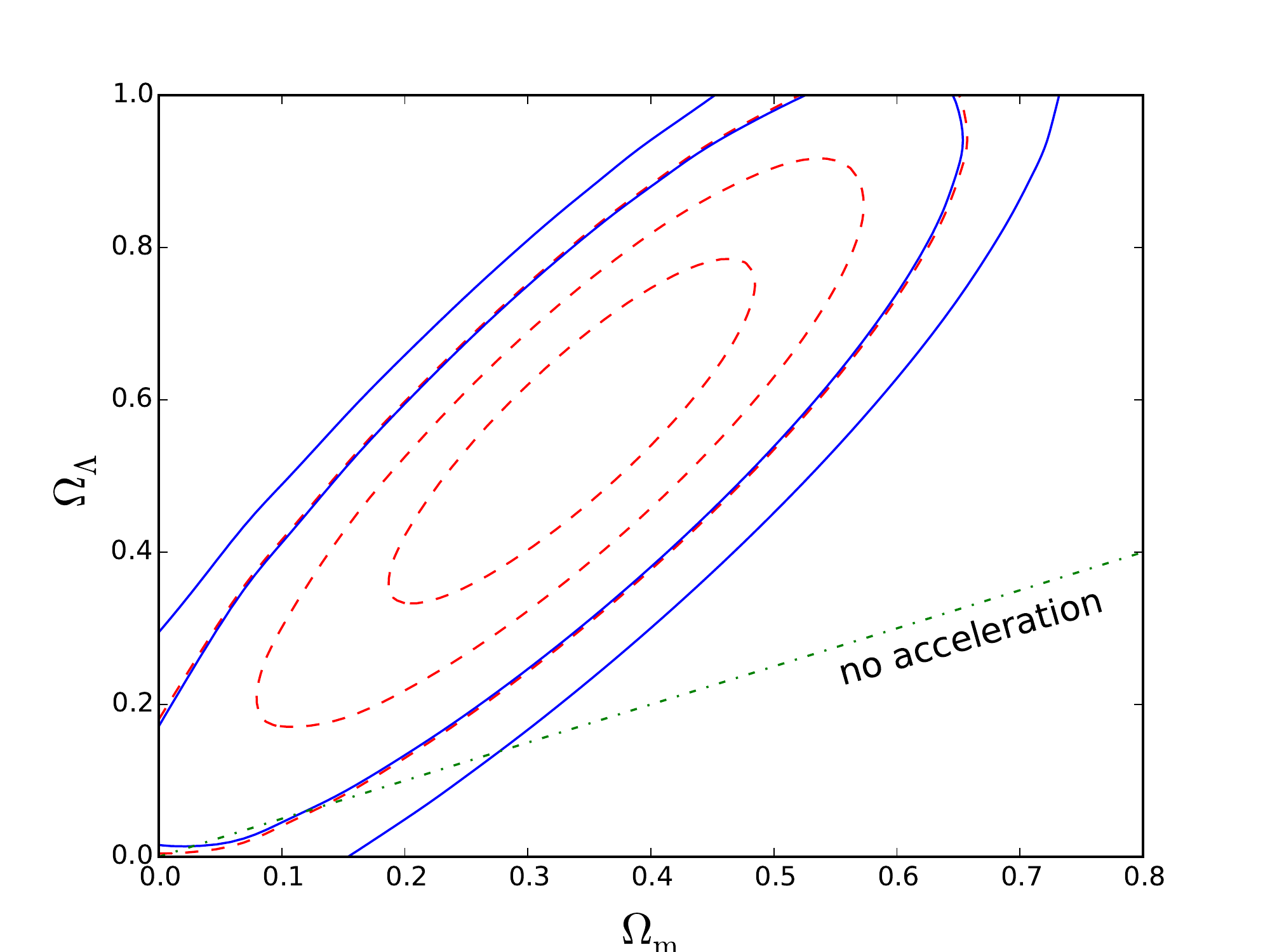}
\caption[Contour plot of fit to cosmological parameters]{Contour plot of the profile likelihood in the $\Omega_\text{m},\Omega_\Lambda$ plane. 1, 2 and 3$\sigma$ contours, regarding \emph{all other parameters} as nuisance parameters, are shown as red dashed lines. Blue lines mark the $1$ and $2\sigma$ 10D contours projected on to the plane (see Fig.~\ref{fig:stat:proflike} --- the 2D contours describe the confidence in only $a_\omega$, while the 10D are the joint contour of $a_\bot$ and $a_\omega$, but only shown in $a_\omega$ space because of the dimensional limitations of paper).}
\label{fig:ev:contours}
\end{center}
\end{figure}

\begin{figure}[hp]
\begin{center}
\includegraphics[width=0.6\textwidth]{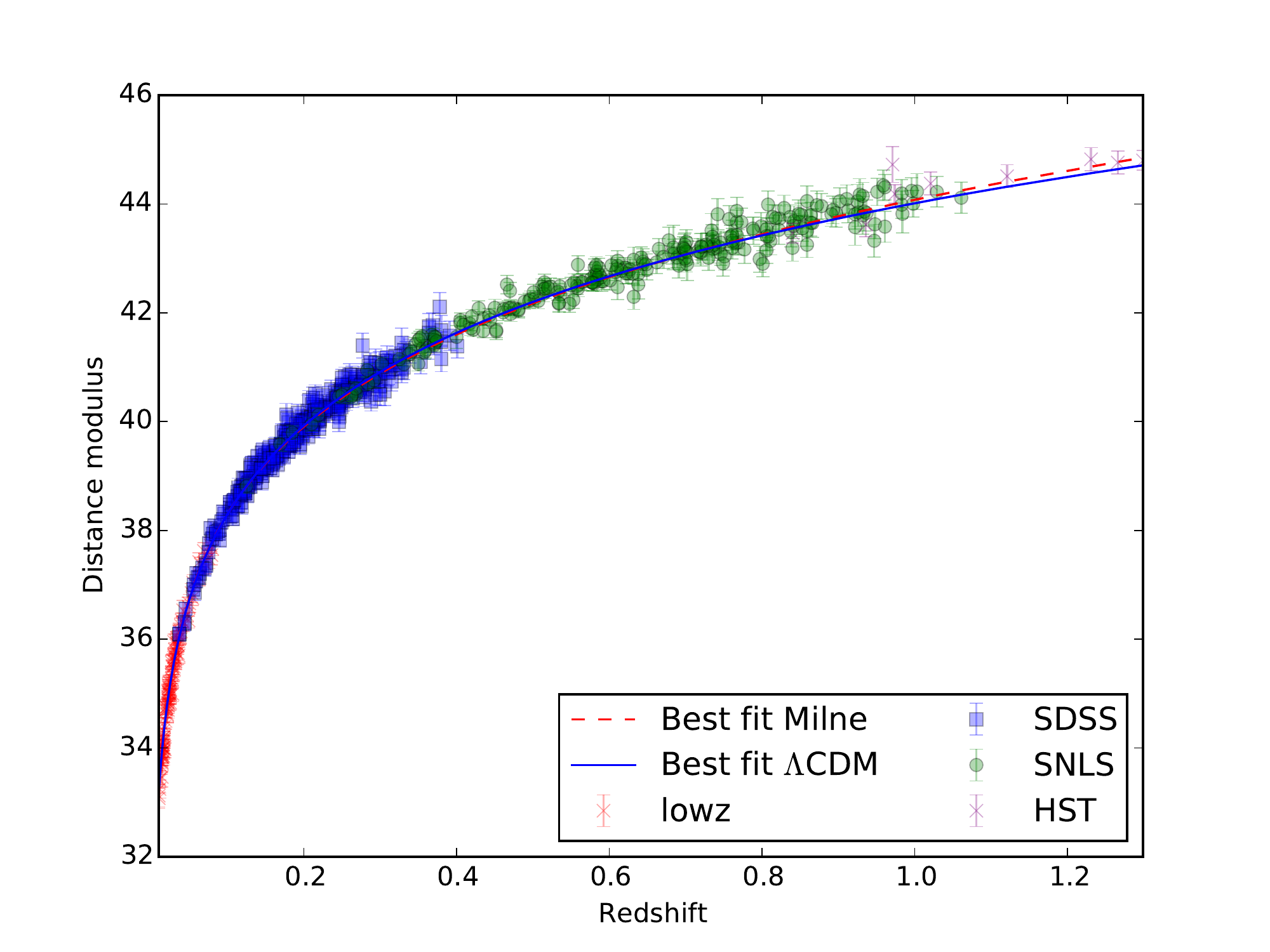}
\caption[Hubble plot]{Comparison of data and model. The measured distance magnitudes, $\hat{\mu_\text{SN}} = \hat{m}^*_B - M_0 + \alpha \hat {x}_1 -\beta\hat{c}$ with different markers depending on the survey. The expected value in two cosmological models are also plotted. $\Lambda$CDM is the best fit accelerating universe, and Milne is an empty universe expanding with constant velocity. The error bars are the square root of the diagonal elements of $\Sigma_l + A^{\text{T}-1}\Sigma_\text{d} A^{-1}$, and include both experimental uncertainties and intrinsic dispersion. These error bars are therefore mildly correlated.}
\label{fig:ev:hubble}
\end{center}
\end{figure}

\begin{figure}[hp]
\begin{center}
\includegraphics[width=0.6\textwidth]{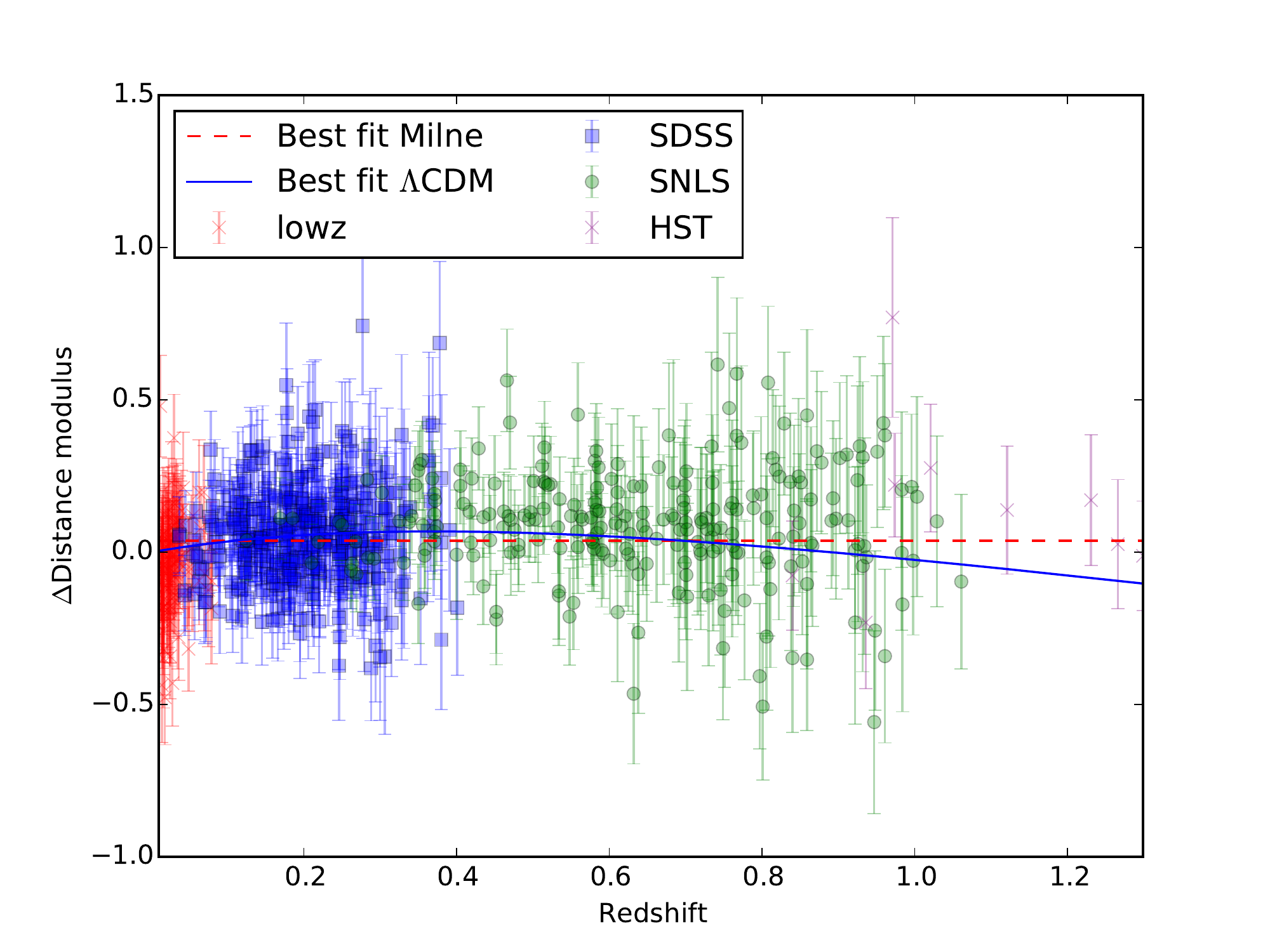}
\caption[Hubble residual plot]{As Fig.~\ref{fig:ev:hubble}, but with the Milne model subtracted. The Milne model plotted has had its Hubble constant corrected by changing the zero point slightly to correct for the change in $M_0$ --- see the discussion following Eq.~\eqref{eq:sn:dismod2}.}
\label{fig:ev:hubbleres}
\end{center}
\end{figure}

\begin{table}
\caption{MLE under specific constraints, marked in boldface. $\Delta\chi^2$ here is short for $-2\log\mathcal L/\mathcal{L}_\text{max}$. ($-2\log\mathcal L_\text{max} = -214.97$) } 
 \label{tab:ev:results}
\makebox[350pt][c]{
\begin{minipage}[b]{\linewidth}
 \begin{tabular}{llcccccccccc}
Constraint &
$\Delta\chi^2$
& $\Omega_\text{m}$ 
 & $\Omega_\Lambda$ & $\alpha$ & $x_0$ & $\sigma_{x_0}$ & $\beta$ 
 & $c_0$ & $\sigma_{c_0}$ & $M_0$ & $\sigma_{M_0}$ \\
	  \hline
None (best fit) & $\mathbf{0}$ & 0.341 & 0.569 & 0.134 & 0.038 & 0.932 
 & 3.059 & -0.016 & 0.071 & -19.052 & 0.108\\
Flat geometry & 0.147 & 0.376 & $\mathbf{0.624}$ & 0.135 & 0.039 & 0.932 
 & 3.060 &-0.016 & 0.071 & -19.055 & 0.108\\
Empty universe & 11.9 & $\mathbf{0.000}$ & $\mathbf{0.000}$ & 0.133 & 0.034 
 & 0.932 & 3.051 & -0.015 & 0.071 & -19.014 & 0.109 \\ 
Non-accelerating & 11.0 & 0.068 & $\mathbf{0.034}$ & 0.132 & 0.033 & 0.931 
 & 3.045 & -0.013 & 0.071 &-19.006 & 0.109 \\
Matter-less universe & 10.4 & $\mathbf{0.000}$ & 0.094 & 0.134 & 0.036 & 0.932 
 & 3.059 & -0.017 & 0.071 & -19.032 & 0.109 \\
Einstein-deSitter & 221.97 & $\mathbf{1.000}$ & $\mathbf{0.000}$ & 0.123 & 0.014 
 & 0.927 & 3.039 & 0.009 & 0.072 & -18.839 & 0.125
 \end{tabular}
\end{minipage}
} 
\end{table}
In the spirit of cosmology, Fig.~\ref{fig:ev:contours} presents the $1,2,$ and $3\sigma$ contours of the $\Omega_m,\Omega_\Lambda$ profile likelihood and the $1$ and $2\sigma$ contours of the full likelihood, projected to the $\Omega_m,\Omega_\Lambda$-plane (see Fig.~\ref{fig:stat:proflike} for how the two differ).
From this figure, and the tabulated numbers, we see in particular that the non-accelerating universe is excluded at about $3\sigma$ --- not overwhelming evidence.\footnote{The usual criterion in particle physics is $5\sigma$, or a \emph{local} p-value of about $5.7\cdot 10^{-7}$, for a rigorous discovery, however see \citep{Lyons:2013yja,Richard2015} for a discussion of this convention.} The Hubble plot we find is presented in Figs.~\ref{fig:ev:hubble} and \ref{fig:ev:hubbleres}.

\begin{figure}[htbp]
\begin{center}
\includegraphics[width=0.6\textwidth]{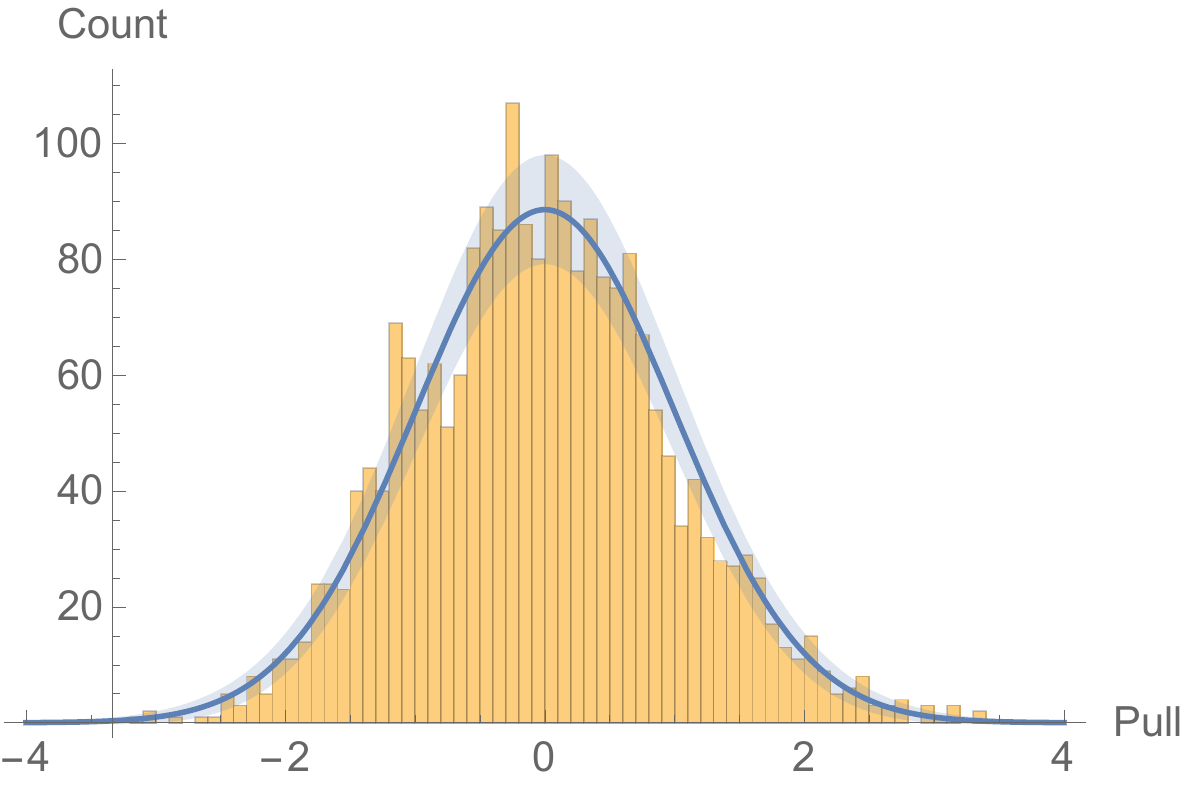}
\caption[Pull distribution]{Pull distribution of the best fit model. Pulls are defined as described in Eq.~\eqref{eq:ev:pulls}. According to our very simple model, everything should be gaussian. Therefore we superimpose a normalised gaussian with the expected Poisson noise ($1/\sqrt N$).}
\label{fig:ev:pulls}
\end{center}
\end{figure}

Now we want to assess how good the fit is. We simply put in gaussians as our model before, now we want to see how well this actually describes the data, along with a determination of how well the statistical approximations we do are --- this is \emph{not} a linear model, yet we use Wilks' theorem to find confidence regions. In Fig.~\ref{fig:ev:pulls} are the pulls, defined as the residual, normalised to the combined expected error,
\al{\label{eq:ev:pulls}
\text{pull} = (\hat{Z} - Y_0 A)U^{-1},
}
where $U$ is the upper triangular Cholesky factor of the estimated covariance matrix, ie. $ U^TU = \Sigma_\text{d} + A^\text{T}\Sigma_l A$. We see from the figure that while it is not a perfect description, neither is it obviously invalid. Performing a series of \emph{goodness-of-fit} tests of the pull distribution to a normal distribution, we get the p-values in Tab.~\ref{tab:ev:pvals}. All four tests here are looking at the cumulative distribution function in different ways. Looking at more targeted tests may give radically different answers. In particular the \emph{skewness} (the third moment) is off, which might have been anticipated already from the distribution of $x_1$.
\begin{table}[h]
\caption{p-values from testing the hypothesis that the pull distribution follows a normal distribution, $\mathcal N(0,1)$.}
\begin{center}
\begin{tabular}{lcc}
Test & statistic & p-value \\
\hline
 \text{Anderson-Darling} & 2.528 & 0.0479 \\
 \text{Cram{\' e}r-von Mises} & 0.454 & 0.0522 \\
 \text{Kolmogorov-Smirnov} & 0.0244 & 0.1389 \\
 \text{Kuiper} & 0.0329 & 0.1231
\end{tabular}
\end{center}
\label{tab:ev:pvals}
\end{table}

\begin{figure}[htpb]
\begin{center}
    \includegraphics[width=0.6\columnwidth]{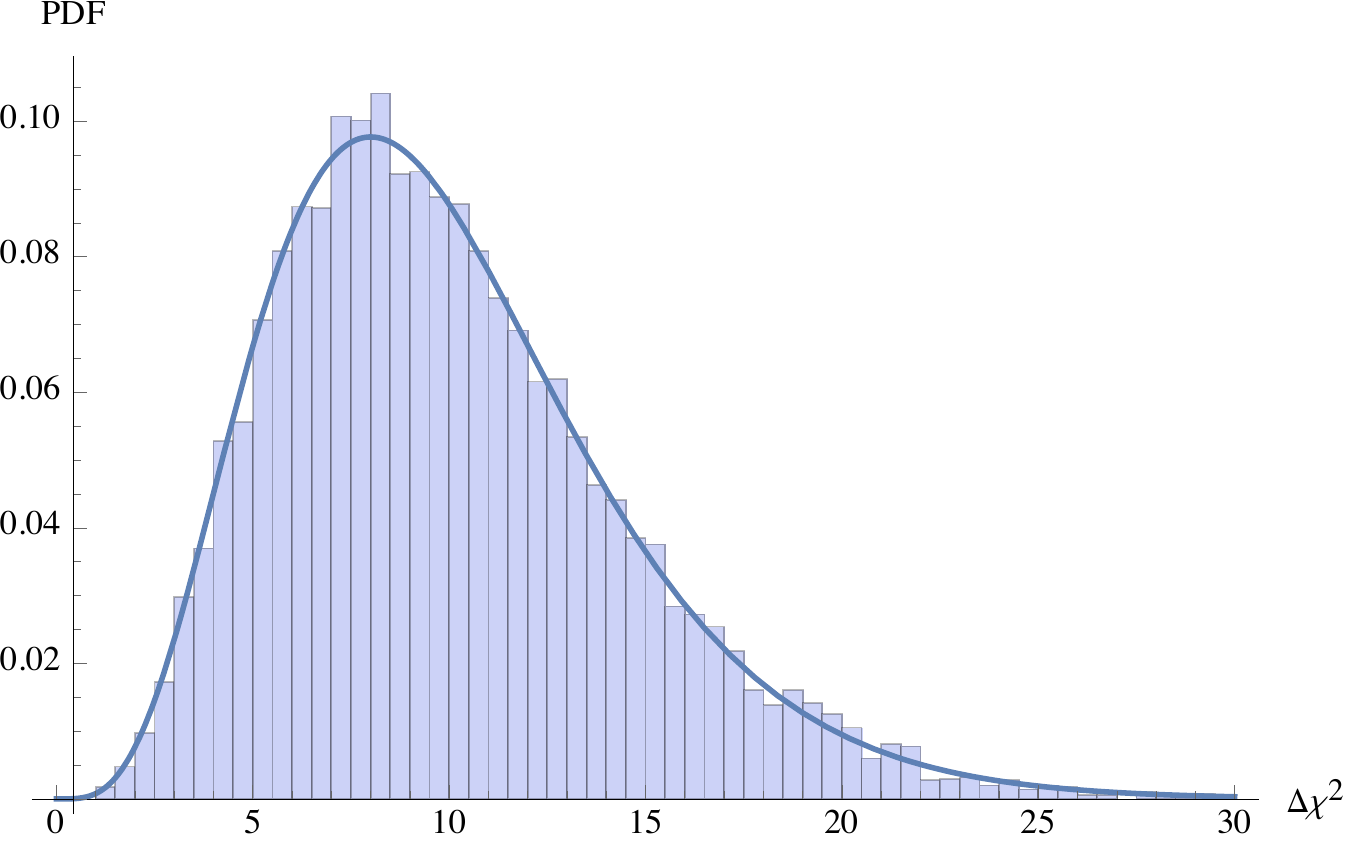}
\caption[Likelihood ratio distribution from MC]{Distribution of the MC likelihood ratio as defined in the text. The expected $\chi^2$ distribution is superimposed. The excellent agreement between the two reinforces our trust in Wilks' theorem.}
\label{fig:ev:wilkscheck}
\end{center}
\end{figure}

We also perform a simple MC test to check that the confidence levels we set by Wilks' theorem are good. To do so, we simulate $10^4$ datasets, assuming the model to be correct and taking the best fit parameters of the actual data as the model.\footnote{This choice is the most relevant, but is not important --- we could in principle choose any value.} We keep the z-values, but draw new $M,x_1,c$ values from the predicted distributions. For every one of these datasets we find the best fit parameters, and in particular the maximum likelihood. Wilks' theorem now states that the distribution of the quantity $-2\log\mathcal L_\text{true} / \mathcal L_\text{max}$ is a $\chi^2$ with $10$ degrees of freedom, where $\mathcal L_\text{true}$ is the likelihood of the true parameters. The distribution from MC and the analytic curve are plotted in Fig.~\ref{fig:ev:wilkscheck}. We see that Wilks' theorem holds true to very high precision, and so we can indeed trust the confidence levels set by the likelihood ratio.

\section{Older analyses}
To emphasise the strength of the present analysis, it is instructive to look at the exact differences between this and previous analyses. The previous analyses are mainly in two categories: a likelihood based one, which I argue is not a good fit, and a residual based one, which in particular is not a likelihood maximisation, and as such, \emph{what we learned in Chap.~\ref{cha:sta} no longer applies}. Let us first take a look at these other methods and then return to a brief comparison with the proposed approach.


\subsection{Residual based method}
This method is the most prominent method used in analyses using the SALT method (or other methods like it) of lightcurve fitting, see eg. \citep{Betoule:2014frx,Conley:2011ku,Astier:2005qq,Kowalski:2008ez,Amanullah:2010vv,Perlmutter:1998np}. The exact procedure varies slightly, but the main point is that one considers the quantity, which is sometimes called a $\chi^2$, but for pedagogical purposes I will simply call it $f$,
\al{
f =& ( m_B^* - M +\alpha x_1 -\beta c - \mu ) [\text{diag}(\sigma_{int}^2) + C(\alpha,\beta)]^{-1}  ( m_B^* - M +\alpha x_1 -\beta c - \mu ) \nonumber \\
\approx& \sum \frac{\Delta\mu^2}{\sigma_\mu^2 + \sigma_{int}^2} \label{eq:ev:fchi}
}
where I have written out explicitly the term $\sigma_{int}$ in the covariance matrix $C$, which I write schematically as $\sigma_{\mu,i}^2 = \sum_{ab} (\Sigma_{d,i})_{ab}r_a r_b$, where $r=(1,\alpha,-\beta)$ mixes in the relevant covariances of $x_1,c$, and $\Sigma_d$ is the covariance matrix as used in the previous section. The newest version of this type of analysis tries to determine independently the $\sigma_{int}$ (\citep{Betoule:2014frx}), while most other analyses have done something curious. First the $f$ is minimised with some plausible value of $\sigma_{int}$ inserted --- typically a guess based on previous analyses. When the minimum is found, one then \emph{adjusts the $\sigma_{int}$ such that $f$ is equal to the number of degrees of freedom}, say $N$. As we saw in Sec.~\ref{stat:sec:nonlin} this is indeed reminiscent of fitting the unknown intrinsic error (the degree to which SNe are standard candles). But, as I stress in Sec.~\ref{sec:stat:MC}, not even the maximum likelihood estimate satisfies this exactly, when the errors on the datapoints are unequal. When the $f$ has now been fixed, the minimum might have moved slightly, so the procedure is iterated until convergence.

The most alarming thing is now, that confidence regions are put in place by Wilks' theorem --- even though as we just saw, \emph{this method is not derived from a likelihood, and as such using Wilks' theorem is manifestly nonsense}. However, as we will see, since the guess $f$ is an educated one, the limits one sets this way are not completely off.


\subsection{Simple likelihood}
It has been realised, that the previous method was indeed not derived from a likelihood, see \cite{March:2011xa,Kim:2011hg,Vishwakarma:2010nc,Wei:2015xca}. If $f$ is to be interpreted as something like $-2\log\mathcal L$, then we have to impose a normalisation, namely $\int \mathcal L\ d\text{(data)}=1$. Taking just the $m_B^*$ as the data, we see that the previous expression \eqref{eq:ev:fchi} should come from a likelihood, which takes the form
\al{
\mathcal L_w = \prod (2\pi[\sigma_\mu^2 + \sigma_{int}^2])^{-1/2} \exp\left(-\frac{1}{2} \frac{\Delta\mu^2}{\sigma_\mu^2 + \sigma_{int}^2} \right)
\label{eq:ev:wronglike}
}
Now we can perform a rigorous likelihood maximisation, and construct confidence regions using Wilks' theorem. But let's first look closer at the likelihood we have constructed. As just stated, the way I constructed this was by regarding the integral as going over \emph{only} the $m_B^*$. But there is no reason why $m_B^*$ should enjoy a privileged status compared to $x_1,c$. It is immediate that trying to integrate over these datapoints will give infinity. So let's see where we went wrong. To do so, we need to go back to the basics of constructing the likelihood, Eq.~\eqref{eq:ev:conlike}. Now put in \emph{flat distributions of $x_1,c$}, ie. instead of all gaussians, we now take
\al{
p(x_1 | \theta ) \propto 1 \nonumber \\
 p(c | \theta ) \propto 1
}
where in principle we need some compact support for these distributions to have a finite (unit) normalisation of the likelihood. Putting these distributions into Eq.~\eqref{eq:ev:conlike}, with the gaussian measurement errors and $p(M|\theta)$, we have, using the notation of the last section
\al{
\mathcal L_w &\propto \frac{1}{\sqrt{|2\pi\Sigma_d| } } 
\int \exp \left[ -\frac{1}{2}(\hat X - X)\Sigma_d^{-1}(\hat X - X)^{\text T} \right]\ dx_1\ dc \nonumber \\ 
&\times (2\pi\sigma_{M_0}^2)^{-1/2} 
 \exp\left\{-\left[\left({M - M_0}\right)/\sigma_{M_0}\right]^2/2\right\}\ dM 
}
Performing first the $x_1,c$ integrals simply mixes in the uncertainties and swaps $\hat x_1,\hat c$ for $x_1,c$ in $\hat X$. We end with, writing for simplicity just a diagonal covariance matrix,
\al{
\mathcal L_w &\propto\int \prod_i \frac{1}{\sqrt{2\pi \sum_{ab}(\Sigma_i)_{ab}r_a r_b} } 
	\exp\left[ -\frac{1}{2} \frac{[\hat m_{Bi}^*- (\mu(z_i) + M -\alpha \hat x_{1i} + \beta \hat c_i)]^2}{\sum_{ab}(\Sigma_i)_{ab}r_a r_b} \right] \nonumber \\ 
	&\times (2\pi\sigma_{int}^2)^{-1/2} \exp\left\{-\left[\left({M - M_0}\right)/\sigma_{int}\right]^2/2\right\}\ dM 
}
The error $\sum_{ab}(\Sigma_i)_{ab}r_a r_b$ here is what we schematically called $\sigma_\mu^2$ before. Performing the $M$ integral now simply adds the $\sigma_{int}$ error and swaps $M$ for $M_0$ in the residual, which gives us the expected expression, given in Eq.~\eqref{eq:ev:wronglike} --- up to a constant, which in principle is infinite. Putting in compact support of the $x_1,c$ distributions complicates the integrals, but provided the limits are far away from the interesting region, the results will be almost the same. I won't go into details about fitting these limits. The main point is that now we know exactly how this likelihood comes about, and so can say confidently that \emph{it is wrong}, as we see quite clearly from Fig.~\ref{fig:ev:corrpar}, that we have no reason to believe the distributions of $x_1,c$ are flat in the full range of the variables. One can of course impose narrow uniform distributions, but this will still include two extra parameters in the fit, just as the gaussian distributions do --- there is no such thing as a free distribution.

\subsection{Comparison with the present analysis}
We have already seen how the naive likelihood based method compares to the present analysis. For the present purposes, let us rewrite the likelihood of Eq.~\eqref{eq:ev:reallike} as
\al{
\mathcal L &= {|2\pi(\Sigma_\text{d} + A^\text{T}\Sigma_l A)|}^{-1/2} \\
&\phantom{=} \times \exp\left[-(\hat{Z}A^{-1} - Y_0)(A^{T-1}\Sigma_\text{d} A^{-1} + \Sigma_l )^{-1}(\hat{Z}A^{-1} - Y_0)^\text{T}/2 \right] \nonumber 
}
Taking for simplicity $\Sigma_\text d$ to be diagonal, we see that the combination of every third term in the expansion of the sum looks just like Eq.~\eqref{eq:ev:fchi}, notice the form of the residual here is
\al{
\hat{Z}A^{-1} - Y_0 = (m_B^* - M_0 +\alpha \hat x_1 - \beta \hat c - \mu, \hat x_1 - x_{10}, \hat c - c_0,\dots)
}
This means we are more or less fitting the same expressions we were before. The important thing, which has been left out of the other analyses are the terms linking residuals in $\hat x_1,\hat c$ to those of $\hat m_B^*$ --- the off-diagonal terms of $A^{T-1}\Sigma_\text{d} A^{-1} + \Sigma_l$. Since these analyses did not consider residuals of these terms, they obviously cannot include these corrections.

\begin{figure}[htbp]
\begin{center}
\includegraphics[width=\textwidth]{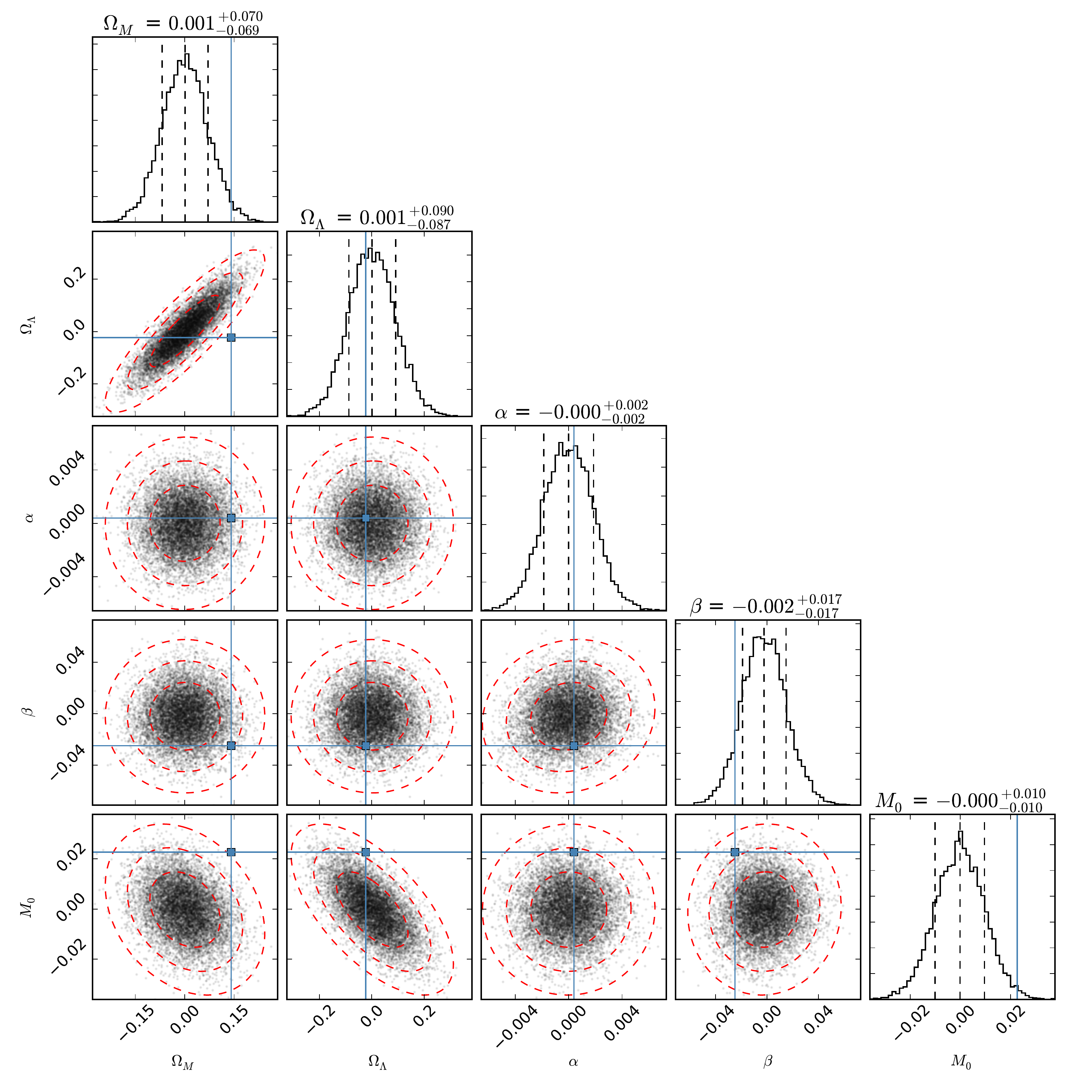}
\caption[MLE-residual method correlation]{Correlations between MLE and residual method parameters, for the relevant parameters. Plotted is MLE$-$residual estimate. The dashed ellipses show approximate $1,2,3\sigma$ 2d profile contours. The blue markers show the value obtained from the real data. We see that in particular the $\Omega_m,\Omega_\Lambda$ point is off. In total, there are $9945$ simulated datasets plotted.}
\label{fig:ev:parcorr}
\end{center}
\end{figure}

Now, that is not to say that the residual based method will give a wrong result. We just need to show it by some other means than for the MLE. In particular, since we cannot do the calculation analytically, we will have to show it by simulations. To do such a simulation though, \emph{we have to assume a distribution} from which we draw the values of $x_1$ and $c$. To show how well the two methods should agree, I reuse the MC from Sec.~\ref{sec:ev:mainfit} and now do both a fit with our MLE method and the residual based method. As a slight correction to the method, I use for the $\sigma_\text{int}$ term in the residual method the value I find from the MLE. Now, for every simulated dataset, I plot the difference in the obtained parameters in Fig.~\ref{fig:ev:parcorr}. This MC study is then compared to the obtained values of the actual dataset. Fitting the original dataset by the residual method, our best fit is 
\al{
\{ \Omega_m, \Omega_\Lambda,\alpha,\beta,M_0 \} = \{0.200,   0.591,   0.134,   3.08, -19.07\}
}
First of all, notice that the two methods actually agree on average! This is somewhat surprising, but certainly possible. This just means that within this model the two methods agree, more or less --- to the degree of spread in the figure. However, looking at the value of the real dataset, we see that it doesn't quite agree. To put this into numbers, I first calculate the sample covariance of the MC values, $\hat{\mathcal I}$. Seeing the distribution of MC points as an estimate of the pdf, $\hat{\mathcal I}$ is the covariance of the 5d approximate gaussian distribution. We can now calculate a $\chi^2$ of the difference we see in the real data, as
\al{
\Delta\chi^2 = \Delta\text{parameters}\cdot\hat{\mathcal I}^{-1} \cdot \Delta\text{parameters} \approx 22.73,
}
which for a $\chi^2$ distribution with 5 degrees of freedom is about $3.6\sigma$. That means, taking into consideration both fits, it is rather unlikely that they would differ by this much, \emph{if the gaussian model is correct}. This fit is explicitly carried out with gaussian distributions --- even for the residual method, which superficially completely disregards this information.

It is important to realise that any result obtained by the residual based method \emph{can only be validated by a MC study}. This does by no means forbid it be a good estimator. What is dangerous though, is that one cannot extract from this the parameters of the underlying distributions, and so we lose the ability to eg. plot Fig.~\ref{fig:ev:pulls} in this framework. We lose the ability to assess the model of correction parameter which I emphasise, \emph{we must assume to validate the method}, whether we like it or not. In particular, doing the MLE method, we have made only completely standard assumptions.

\chapter{Conclusion}\label{cha:con}
In this thesis I have presented in detail my knowledge of fitting cosmological parameters with supernovae. I analysed the latest large compilation of supernovae and found a result in significant contrast to the canonical result. In particular, a non-accelerating universe is only excluded at $3\sigma$ by supernovae alone. This is not to say that supernovae prefer this universe --- the best fit point of this \emph{very simple model} is still with a significant dark energy contribution! What we have done is to state explicitly all assumptions that are usually made. These assumptions will surely change, even in the near future. Preliminary work such as \cite{Marriner:2011mf,Kessler:2012gn} suggests using more and more contrived versions of the residuals based method due to selection effects, and a very recent study proposes two populations of supernovae \cite{Milne:2014rfa}. I have in this analysis not made any such assumptions or speculations. The result I present has standard assumptions and models --- only we need to state this explicitly, because we \emph{have to write down a likelihood.} 

This effort, I think, has two obvious products. First of all, we now have well defined confidence regions, readily compared to other analyses. We know at all times exactly what we are fitting and modelling, because we are forced to write all this down. Secondly, it gives an avenue to further exploration of the correction parameter distributions. As stated before, it is hardly believable that the distributions should be exactly gaussian, or that there is no evolution of supernovae during the evolution of the universe. The method presented here is very easily capable of dealing with this issue.

\listoffigures
\clearpage
\begingroup
\let\cleardoublepage\relax

\begin{thebibliography}{}

\bibitem[\protect\citeauthoryear{Nielsen, Guffanti, and Sarkar}{Nielsen
  et~al\mbox{.}}{2015}]{Nielsen:2015pga}
{\sc Nielsen, J.~T.}, {\sc Guffanti, A.}, {\sc and} {\sc Sarkar, S.} 2015.
\newblock {Marginal evidence for cosmic acceleration from Type Ia supernovae},
  \href{http://www.arxiv.org/abs/1506.01354}{arXiv:[1506.01354]}.

\bibitem[\protect\citeauthoryear{Perlmutter et~al\mbox{.}}{Perlmutter
  et~al\mbox{.}}{1999}]{Perlmutter:1998np}
{\sc Perlmutter, S.} {\sc et~al\mbox{.}} 1999.
\newblock {Measurements of Omega and Lambda from 42 high redshift supernovae}.
\newblock \href{http://dx.doi.org/10.1086/307221}{{\em Astrophys.J.\/}~{\bf
  517}, 565--586},
  \href{http://www.arxiv.org/abs/astro-ph/9812133}{arXiv:[astro-ph/9812133]}.

\bibitem[\protect\citeauthoryear{Riess et~al\mbox{.}}{Riess
  et~al\mbox{.}}{1998}]{Riess:1998cb}
{\sc Riess, A.~G.} {\sc et~al\mbox{.}} 1998.
\newblock {Observational evidence from supernovae for an accelerating universe
  and a cosmological constant}.
\newblock \href{http://dx.doi.org/10.1086/300499}{{\em Astron.J.\/}~{\bf 116},
  1009--1038},
  \href{http://www.arxiv.org/abs/astro-ph/9805201}{arXiv:[astro-ph/9805201]}.

\bibitem[\protect\citeauthoryear{Cyburt, Fields, and Olive}{Cyburt
  et~al\mbox{.}}{2008}]{Cyburt:2008kw}
{\sc Cyburt, R.~H.}, {\sc Fields, B.~D.}, {\sc and} {\sc Olive, K.~A.} 2008.
\newblock {An Update on the big bang nucleosynthesis prediction for Li-7: The
  problem worsens}.
\newblock \href{http://dx.doi.org/10.1088/1475-7516/2008/11/012}{{\em
  JCAP\/}~{\bf 0811}, 012},
  \href{http://www.arxiv.org/abs/0808.2818}{arXiv:[0808.2818]}.

\bibitem[\protect\citeauthoryear{Fisher}{Fisher}{1930}]{fisher1930inverse}
{\sc Fisher, R.~A.} 1930.
\newblock Inverse probability.
\newblock
  \href{https://www.nada.kth.se/kurser/kth/2D5342/kurspaket/fiducial.pdf}{{\em
  Mathematical Proceedings of the Cambridge Philosophical Society\/}~{\bf 26},
  04, 528--535}.

\bibitem[\protect\citeauthoryear{Hankel}{Hankel}{1864}]{hankel1864}
{\sc Hankel, H.} 1864.
\newblock Die euler'schen integrale bei unbeschr{\"a}nkter variabilit{\"a}t des
  arguments.
\newblock
  \href{https://ia902205.us.archive.org/28/items/dieeulerschenin00hankgoog/dieeulerschenin00hankgoog.pdf}{{\em
  Z. Math. Phys\/}~{\bf 9}, 1--21}.

\bibitem[\protect\citeauthoryear{Rao}{Rao}{2009}]{rao2009linear}
{\sc Rao, C.~R.} 2009.
\newblock {\em Linear statistical inference and its applications}. Vol.~22.
\newblock John Wiley \& Sons.

\bibitem[\protect\citeauthoryear{Wilks}{Wilks}{1938}]{wilks1938large}
{\sc Wilks, S.~S.} 1938.
\newblock The large-sample distribution of the likelihood ratio for testing
  composite hypotheses.
\newblock
  \href{http://projecteuclid.org/download/pdf_1/euclid.aoms/1177732360}{{\em
  The Annals of Mathematical Statistics\/}~{\bf 9}, 1, 60--62}.

\bibitem[\protect\citeauthoryear{Peebles}{Peebles}{1993}]{peebles1993principles}
{\sc Peebles, P. J.~E.} 1993.
\newblock {\em Principles of physical cosmology}.
\newblock Princeton University Press.

\bibitem[\protect\citeauthoryear{Carroll}{Carroll}{2004}]{carroll2004spacetime}
{\sc Carroll, S.~M.} 2004.
\newblock {\em Spacetime and geometry. An introduction to general relativity}.
  Vol.~1.

\bibitem[\protect\citeauthoryear{Kolb and Turner}{Kolb and
  Turner}{1990}]{kolb1990early}
{\sc Kolb, E.~W.} {\sc and} {\sc Turner, M.~S.} 1990.
\newblock {\em The early universe.} Vol.~1.

\bibitem[\protect\citeauthoryear{Weinberg}{Weinberg}{1972}]{Weinberg:1972aa}
{\sc Weinberg, S.} 1972.
\newblock {\em Gravitation and cosmology : principles and applications of the
  general theory of relativity}.
\newblock Wiley, New York.

\bibitem[\protect\citeauthoryear{Ade et~al\mbox{.}}{Ade
  et~al\mbox{.}}{2014}]{Ade:2013zuv}
{\sc Ade, P.} {\sc et~al\mbox{.}} 2014.
\newblock {Planck 2013 results. XVI. Cosmological parameters}.
\newblock \href{http://dx.doi.org/10.1051/0004-6361/201321591}{{\em
  Astron.Astrophys.\/}~{\bf 571}, A16},
  \href{http://www.arxiv.org/abs/1303.5076}{arXiv:[1303.5076]}.

\bibitem[\protect\citeauthoryear{Hui and Greene}{Hui and
  Greene}{2006}]{Hui:2005nm}
{\sc Hui, L.} {\sc and} {\sc Greene, P.~B.} 2006.
\newblock {Correlated Fluctuations in Luminosity Distance and the (Surprising)
  Importance of Peculiar Motion in Supernova Surveys}.
\newblock \href{http://dx.doi.org/10.1103/PhysRevD.73.123526}{{\em
  Phys.Rev.\/}~{\bf D73}, 123526},
  \href{http://www.arxiv.org/abs/astro-ph/0512159}{arXiv:[astro-ph/0512159]}.

\bibitem[\protect\citeauthoryear{Kogut, Lineweaver, Smoot, Bennett, Banday,
  et~al\mbox{.}}{Kogut et~al\mbox{.}}{1993}]{Kogut:1993ag}
{\sc Kogut, A.}, {\sc Lineweaver, C.}, {\sc Smoot, G.~F.}, {\sc Bennett, C.},
  {\sc Banday, A.}, {\sc et~al\mbox{.}} 1993.
\newblock {Dipole anisotropy in the COBE DMR first year sky maps}.
\newblock \href{http://dx.doi.org/10.1086/173453}{{\em Astrophys.J.\/}~{\bf
  419}, 1},
  \href{http://www.arxiv.org/abs/astro-ph/9312056}{arXiv:[astro-ph/9312056]}.

\bibitem[\protect\citeauthoryear{Aghanim et~al\mbox{.}}{Aghanim
  et~al\mbox{.}}{2014}]{Aghanim:2013suk}
{\sc Aghanim, N.} {\sc et~al\mbox{.}} 2014.
\newblock {Planck 2013 results. XXVII. Doppler boosting of the CMB: Eppur si
  muove}.
\newblock \href{http://dx.doi.org/10.1051/0004-6361/201321556}{{\em
  Astron.Astrophys.\/}~{\bf 571}, A27},
  \href{http://www.arxiv.org/abs/1303.5087}{arXiv:[1303.5087]}.

\bibitem[\protect\citeauthoryear{Colin, Mohayaee, Sarkar, and Shafieloo}{Colin
  et~al\mbox{.}}{2011}]{Colin:2010ds}
{\sc Colin, J.}, {\sc Mohayaee, R.}, {\sc Sarkar, S.}, {\sc and} {\sc
  Shafieloo, A.} 2011.
\newblock {Probing the anisotropic local universe and beyond with SNe Ia data}.
\newblock \href{http://dx.doi.org/10.1111/j.1365-2966.2011.18402.x}{{\em
  Mon.Not.Roy.Astron.Soc.\/}~{\bf 414}, 264--271},
  \href{http://www.arxiv.org/abs/1011.6292}{arXiv:[1011.6292]}.

\bibitem[\protect\citeauthoryear{Bonvin, Durrer, and Kunz}{Bonvin
  et~al\mbox{.}}{2006}]{Bonvin:2006en}
{\sc Bonvin, C.}, {\sc Durrer, R.}, {\sc and} {\sc Kunz, M.} 2006.
\newblock {The dipole of the luminosity distance: a direct measure of h(z)}.
\newblock \href{http://dx.doi.org/10.1103/PhysRevLett.96.191302}{{\em
  Phys.Rev.Lett.\/}~{\bf 96}, 191302},
  \href{http://www.arxiv.org/abs/astro-ph/0603240}{arXiv:[astro-ph/0603240]}.

\bibitem[\protect\citeauthoryear{Feindt, Kerschhaggl, Kowalski, Aldering,
  Antilogus, et~al\mbox{.}}{Feindt et~al\mbox{.}}{2013}]{Feindt:2013pma}
{\sc Feindt, U.}, {\sc Kerschhaggl, M.}, {\sc Kowalski, M.}, {\sc Aldering,
  G.}, {\sc Antilogus, P.}, {\sc et~al\mbox{.}} 2013.
\newblock {Measuring cosmic bulk flows with Type Ia Supernovae from the Nearby
  Supernova Factory}.
\newblock \href{http://dx.doi.org/10.1051/0004-6361/201321880}{{\em
  Astron.Astrophys.\/}~{\bf 560}, A90},
  \href{http://www.arxiv.org/abs/1310.4184}{arXiv:[1310.4184]}.

\bibitem[\protect\citeauthoryear{Davis, Hui, Frieman, Haugbolle, Kessler,
  et~al\mbox{.}}{Davis et~al\mbox{.}}{2011}]{Davis:2010jq}
{\sc Davis, T.~M.}, {\sc Hui, L.}, {\sc Frieman, J.~A.}, {\sc Haugbolle, T.},
  {\sc Kessler, R.}, {\sc et~al\mbox{.}} 2011.
\newblock {The effect of peculiar velocities on supernova cosmology}.
\newblock \href{http://dx.doi.org/10.1088/0004-637X/741/1/67}{{\em
  Astrophys.J.\/}~{\bf 741}, 67},
  \href{http://www.arxiv.org/abs/1012.2912}{arXiv:[1012.2912]}.

\bibitem[\protect\citeauthoryear{Weinberg}{Weinberg}{1989}]{Weinberg:1988cp}
{\sc Weinberg, S.} 1989.
\newblock {The Cosmological Constant Problem}.
\newblock \href{http://dx.doi.org/10.1103/RevModPhys.61.1}{{\em
  Rev.Mod.Phys.\/}~{\bf 61}, 1--23}.

\bibitem[\protect\citeauthoryear{Carroll, Press, and Turner}{Carroll
  et~al\mbox{.}}{1992}]{Carroll:1991mt}
{\sc Carroll, S.~M.}, {\sc Press, W.~H.}, {\sc and} {\sc Turner, E.~L.} 1992.
\newblock {The Cosmological constant}.
\newblock \href{http://dx.doi.org/10.1146/annurev.aa.30.090192.002435}{{\em
  Ann.Rev.Astron.Astrophys.\/}~{\bf 30}, 499--542}.

\bibitem[\protect\citeauthoryear{Martin}{Martin}{2012}]{Martin:2012bt}
{\sc Martin, J.} 2012.
\newblock {Everything You Always Wanted To Know About The Cosmological Constant
  Problem (But Were Afraid To Ask)}.
\newblock \href{http://dx.doi.org/10.1016/j.crhy.2012.04.008}{{\em Comptes
  Rendus Physique\/}~{\bf 13}, 566--665},
  \href{http://www.arxiv.org/abs/1205.3365}{arXiv:[1205.3365]}.

\bibitem[\protect\citeauthoryear{Velten, vom Marttens, and Zimdahl}{Velten
  et~al\mbox{.}}{2014}]{Velten:2014nra}
{\sc Velten, H.}, {\sc vom Marttens, R.}, {\sc and} {\sc Zimdahl, W.} 2014.
\newblock {Aspects of the cosmological ``coincidence problem''}.
\newblock \href{http://dx.doi.org/10.1140/epjc/s10052-014-3160-4}{{\em
  Eur.Phys.J.\/}~{\bf C74}, 11, 3160},
  \href{http://www.arxiv.org/abs/1410.2509}{arXiv:[1410.2509]}.

\bibitem[\protect\citeauthoryear{Rugh and Zinkernagel}{Rugh and
  Zinkernagel}{2002}]{Rugh:2000ji}
{\sc Rugh, S.~E.} {\sc and} {\sc Zinkernagel, H.} 2002.
\newblock {The Quantum vacuum and the cosmological constant problem}.
\newblock \href{http://dx.doi.org/10.1016/S1355-2198(02)00033-3}{{\em
  Stud.Hist.Philos.Mod.Phys.\/}~{\bf 33}, 663--705},
  \href{http://www.arxiv.org/abs/hep-th/0012253}{arXiv:[hep-th/0012253]}.

\bibitem[\protect\citeauthoryear{Koksma and Prokopec}{Koksma and
  Prokopec}{2011}]{Koksma:2011cq}
{\sc Koksma, J.~F.} {\sc and} {\sc Prokopec, T.} 2011.
\newblock {The Cosmological Constant and Lorentz Invariance of the Vacuum
  State}, \href{http://www.arxiv.org/abs/1105.6296}{arXiv:[1105.6296]}.

\bibitem[\protect\citeauthoryear{Starobinsky}{Starobinsky}{1980}]{Starobinsky:1980te}
{\sc Starobinsky, A.~A.} 1980.
\newblock {A New Type of Isotropic Cosmological Models Without Singularity}.
\newblock \href{http://dx.doi.org/10.1016/0370-2693(80)90670-X}{{\em
  Phys.Lett.\/}~{\bf B91}, 99--102}.

\bibitem[\protect\citeauthoryear{Bertotti, Iess, and Tortora}{Bertotti
  et~al\mbox{.}}{2003}]{Bertotti:2003rm}
{\sc Bertotti, B.}, {\sc Iess, L.}, {\sc and} {\sc Tortora, P.} 2003.
\newblock {A test of general relativity using radio links with the Cassini
  spacecraft}.
\newblock \href{http://dx.doi.org/10.1038/nature01997}{{\em Nature\/}~{\bf
  425}, 374}.

\bibitem[\protect\citeauthoryear{Sotiriou and Faraoni}{Sotiriou and
  Faraoni}{2010}]{Sotiriou:2008rp}
{\sc Sotiriou, T.~P.} {\sc and} {\sc Faraoni, V.} 2010.
\newblock {f(R) Theories Of Gravity}.
\newblock \href{http://dx.doi.org/10.1103/RevModPhys.82.451}{{\em
  Rev.Mod.Phys.\/}~{\bf 82}, 451--497},
  \href{http://www.arxiv.org/abs/0805.1726}{arXiv:[0805.1726]}.

\bibitem[\protect\citeauthoryear{Buchert}{Buchert}{2000}]{Buchert:1999er}
{\sc Buchert, T.} 2000.
\newblock {On average properties of inhomogeneous fluids in general relativity.
  1. Dust cosmologies}.
\newblock \href{http://dx.doi.org/10.1023/A:1001800617177}{{\em
  Gen.Rel.Grav.\/}~{\bf 32}, 105--125},
  \href{http://www.arxiv.org/abs/gr-qc/9906015}{arXiv:[gr-qc/9906015]}.

\bibitem[\protect\citeauthoryear{Buchert, Larena, and Alimi}{Buchert
  et~al\mbox{.}}{2006}]{Buchert:2006ya}
{\sc Buchert, T.}, {\sc Larena, J.}, {\sc and} {\sc Alimi, J.-M.} 2006.
\newblock {Correspondence between kinematical backreaction and scalar field
  cosmologies: The `Morphon field'}.
\newblock \href{http://dx.doi.org/10.1088/0264-9381/23/22/018}{{\em
  Class.Quant.Grav.\/}~{\bf 23}, 6379--6408},
  \href{http://www.arxiv.org/abs/gr-qc/0606020}{arXiv:[gr-qc/0606020]}.

\bibitem[\protect\citeauthoryear{Buchert}{Buchert}{2008}]{Buchert:2007ik}
{\sc Buchert, T.} 2008.
\newblock {Dark Energy from Structure: A Status Report}.
\newblock \href{http://dx.doi.org/10.1007/s10714-007-0554-8}{{\em
  Gen.Rel.Grav.\/}~{\bf 40}, 467--527},
  \href{http://www.arxiv.org/abs/0707.2153}{arXiv:[0707.2153]}.

\bibitem[\protect\citeauthoryear{Tolman}{Tolman}{1934}]{Tolman:1934aa}
{\sc Tolman, R.~C.} 1934.
\newblock Effect of inhomogeneity on cosmological models.
\newblock \href{http://www.ncbi.nlm.nih.gov/pmc/articles/PMC1076370/}{{\em
  Proceedings of the National Academy of Sciences of the United States of
  America\/}~{\bf 20}, 3 (03), 169--176}.

\bibitem[\protect\citeauthoryear{Bondi}{Bondi}{1947}]{Bondi01121947}
{\sc Bondi, H.} 1947.
\newblock Spherically symmetrical models in general relativity.
\newblock \href{http://dx.doi.org/10.1093/mnras/107.5-6.410}{{\em Monthly
  Notices of the Royal Astronomical Society\/}~{\bf 107}, 5-6, 410--425}.

\bibitem[\protect\citeauthoryear{Celerier}{Celerier}{2000}]{Celerier:1999hp}
{\sc Celerier, M.-N.} 2000.
\newblock {Do we really see a cosmological constant in the supernovae data?}
\newblock \href{http://adsabs.harvard.edu/abs/2000A&A...353...63C}{{\em
  Astron.Astrophys.\/}~{\bf 353}, 63--71},
  \href{http://www.arxiv.org/abs/astro-ph/9907206}{arXiv:[astro-ph/9907206]}.

\bibitem[\protect\citeauthoryear{Nadathur and Sarkar}{Nadathur and
  Sarkar}{2011}]{Nadathur:2010zm}
{\sc Nadathur, S.} {\sc and} {\sc Sarkar, S.} 2011.
\newblock {Reconciling the local void with the CMB}.
\newblock \href{http://dx.doi.org/10.1103/PhysRevD.83.063506}{{\em
  Phys.Rev.\/}~{\bf D83}, 063506},
  \href{http://www.arxiv.org/abs/1012.3460}{arXiv:[1012.3460]}.

\bibitem[\protect\citeauthoryear{Garcia-Bellido and Haugboelle}{Garcia-Bellido
  and Haugboelle}{2008}]{GarciaBellido:2008gd}
{\sc Garcia-Bellido, J.} {\sc and} {\sc Haugboelle, T.} 2008.
\newblock {Looking the void in the eyes - the kSZ effect in LTB models}.
\newblock \href{http://dx.doi.org/10.1088/1475-7516/2008/09/016}{{\em
  JCAP\/}~{\bf 0809}, 016},
  \href{http://www.arxiv.org/abs/0807.1326}{arXiv:[0807.1326]}.

\bibitem[\protect\citeauthoryear{Bull, Clifton, and Ferreira}{Bull
  et~al\mbox{.}}{2012}]{Bull:2011wi}
{\sc Bull, P.}, {\sc Clifton, T.}, {\sc and} {\sc Ferreira, P.~G.} 2012.
\newblock {The kSZ effect as a test of general radial inhomogeneity in LTB
  cosmology}.
\newblock \href{http://dx.doi.org/10.1103/PhysRevD.85.024002}{{\em
  Phys.Rev.\/}~{\bf D85}, 024002},
  \href{http://www.arxiv.org/abs/1108.2222}{arXiv:[1108.2222]}.

\bibitem[\protect\citeauthoryear{Celerier}{Celerier}{2012}]{Celerier:2011zh}
{\sc Celerier, M.-N.} 2012.
\newblock {Some clarifications about Lema\^itre-Tolman models of the Universe
  used to deal with the dark energy problem}.
\newblock \href{http://dx.doi.org/10.1051/0004-6361/201219104}{{\em
  Astron.Astrophys.\/}~{\bf 543}, A71},
  \href{http://www.arxiv.org/abs/1108.1373}{arXiv:[1108.1373]}.

\bibitem[\protect\citeauthoryear{Krasinski, Hellaby, Bolejko, and
  Celerier}{Krasinski et~al\mbox{.}}{2010}]{Krasinski:2009qq}
{\sc Krasinski, A.}, {\sc Hellaby, C.}, {\sc Bolejko, K.}, {\sc and} {\sc
  Celerier, M.-N.} 2010.
\newblock {Imitating accelerated expansion of the Universe by matter
  inhomogeneities: Corrections of some misunderstandings}.
\newblock \href{http://dx.doi.org/10.1007/s10714-010-0993-5}{{\em
  Gen.Rel.Grav.\/}~{\bf 42}, 2453--2475},
  \href{http://www.arxiv.org/abs/0903.4070}{arXiv:[0903.4070]}.

\bibitem[\protect\citeauthoryear{Bolejko and Sussman}{Bolejko and
  Sussman}{2011}]{Bolejko:2010wc}
{\sc Bolejko, K.} {\sc and} {\sc Sussman, R.~A.} 2011.
\newblock {Cosmic spherical void via coarse-graining and averaging
  non-spherical structures}.
\newblock \href{http://dx.doi.org/10.1016/j.physletb.2011.02.007}{{\em
  Phys.Lett.\/}~{\bf B697}, 265--270},
  \href{http://www.arxiv.org/abs/1008.3420}{arXiv:[1008.3420]}.

\bibitem[\protect\citeauthoryear{Szekeres}{Szekeres}{1975}]{Szekeres:1974ct}
{\sc Szekeres, P.} 1975.
\newblock {A Class of Inhomogeneous Cosmological Models}.
\newblock \href{http://dx.doi.org/10.1007/BF01608547}{{\em
  Commun.Math.Phys.\/}~{\bf 41}, 55}.

\bibitem[\protect\citeauthoryear{Bolejko and Celerier}{Bolejko and
  Celerier}{2010}]{Bolejko:2010eb}
{\sc Bolejko, K.} {\sc and} {\sc Celerier, M.-N.} 2010.
\newblock {Szekeres Swiss-Cheese model and supernova observations}.
\newblock \href{http://dx.doi.org/10.1103/PhysRevD.82.103510}{{\em
  Phys.Rev.\/}~{\bf D82}, 103510},
  \href{http://www.arxiv.org/abs/1005.2584}{arXiv:[1005.2584]}.

\bibitem[\protect\citeauthoryear{Biswas and Notari}{Biswas and
  Notari}{2008}]{Biswas:2007gi}
{\sc Biswas, T.} {\sc and} {\sc Notari, A.} 2008.
\newblock {Swiss-Cheese Inhomogeneous Cosmology and the Dark Energy Problem}.
\newblock \href{http://dx.doi.org/10.1088/1475-7516/2008/06/021}{{\em
  JCAP\/}~{\bf 0806}, 021},
  \href{http://www.arxiv.org/abs/astro-ph/0702555}{arXiv:[astro-ph/0702555]}.

\bibitem[\protect\citeauthoryear{Kashlinsky, Atrio-Barandela, Kocevski, and
  Ebeling}{Kashlinsky et~al\mbox{.}}{2009}]{Kashlinsky:2008ut}
{\sc Kashlinsky, A.}, {\sc Atrio-Barandela, F.}, {\sc Kocevski, D.}, {\sc and}
  {\sc Ebeling, H.} 2009.
\newblock {A measurement of large-scale peculiar velocities of clusters of
  galaxies: results and cosmological implications}.
\newblock \href{http://dx.doi.org/10.1086/592947}{{\em Astrophys.J.\/}~{\bf
  686}, L49--L52},
  \href{http://www.arxiv.org/abs/0809.3734}{arXiv:[0809.3734]}.

\bibitem[\protect\citeauthoryear{Watkins, Feldman, and Hudson}{Watkins
  et~al\mbox{.}}{2009}]{Watkins:2008hf}
{\sc Watkins, R.}, {\sc Feldman, H.~A.}, {\sc and} {\sc Hudson, M.~J.} 2009.
\newblock {Consistently Large Cosmic Flows on Scales of 100 Mpc/h: a Challenge
  for the Standard LCDM Cosmology}.
\newblock \href{http://dx.doi.org/10.1111/j.1365-2966.2008.14089.x}{{\em
  Mon.Not.Roy.Astron.Soc.\/}~{\bf 392}, 743--756},
  \href{http://www.arxiv.org/abs/0809.4041}{arXiv:[0809.4041]}.

\bibitem[\protect\citeauthoryear{Tsagas}{Tsagas}{2010}]{Tsagas11062010}
{\sc Tsagas, C.~G.} 2010.
\newblock Large-scale peculiar motions and cosmic acceleration.
\newblock \href{http://dx.doi.org/10.1111/j.1365-2966.2010.16460.x}{{\em
  Monthly Notices of the Royal Astronomical Society\/}~{\bf 405}, 1, 503--508},
  \href{http://www.arxiv.org/abs/0902.3232}{arXiv:[0902.3232]}.

\bibitem[\protect\citeauthoryear{Tsagas}{Tsagas}{2011}]{Tsagas:2011wq}
{\sc Tsagas, C.~G.} 2011.
\newblock {Peculiar motions, accelerated expansion and the cosmological axis}.
\newblock \href{http://dx.doi.org/10.1103/PhysRevD.84.063503}{{\em
  Phys.Rev.\/}~{\bf D84}, 063503},
  \href{http://www.arxiv.org/abs/1107.4045}{arXiv:[1107.4045]}.

\bibitem[\protect\citeauthoryear{{Minkowski}}{{Minkowski}}{1941}]{Minko1941}
{\sc {Minkowski}, R.} 1941.
\newblock {Spectra of Supernovae}.
\newblock \href{http://dx.doi.org/10.1086/125315}{{\em Publications of the
  Astronomical Society of the Pacific\/}~{\bf 53}, 224}.

\bibitem[\protect\citeauthoryear{Turatto}{Turatto}{2003}]{Turatto:2003np}
{\sc Turatto, M.} 2003.
\newblock {Classification of supernovae}.
\newblock
  \href{http://link.springer.com/chapter/10.1007%2F3-540-45863-8_3}{{\em
  Lect.Notes Phys.\/}~{\bf 598}, 21},
  \href{http://www.arxiv.org/abs/astro-ph/0301107}{arXiv:[astro-ph/0301107]}.

\bibitem[\protect\citeauthoryear{Foley, Challis, Chornock, Ganeshalingam, Li,
  et~al\mbox{.}}{Foley et~al\mbox{.}}{2013}]{Foley:2012tu}
{\sc Foley, R.~J.}, {\sc Challis, P.}, {\sc Chornock, R.}, {\sc Ganeshalingam,
  M.}, {\sc Li, W.}, {\sc et~al\mbox{.}} 2013.
\newblock {Type Iax Supernovae: A New Class of Stellar Explosion}.
\newblock \href{http://dx.doi.org/10.1088/0004-637X/767/1/57}{{\em
  Astrophys.J.\/}~{\bf 767}, 57},
  \href{http://www.arxiv.org/abs/1212.2209}{arXiv:[1212.2209]}.

\bibitem[\protect\citeauthoryear{Clocchiatti}{Clocchiatti}{2011}]{Clocchiatti:2011fw}
{\sc Clocchiatti, A.} 2011.
\newblock {Type Ia Supernovae and the discovery of the Cosmic Acceleration},
  \href{http://www.arxiv.org/abs/1112.0706}{arXiv:[1112.0706]}.

\bibitem[\protect\citeauthoryear{{Koester} and {Chanmugam}}{{Koester} and
  {Chanmugam}}{1990}]{Koester1990}
{\sc {Koester}, D.} {\sc and} {\sc {Chanmugam}, G.} 1990.
\newblock {REVIEW: Physics of white dwarf stars}.
\newblock \href{http://dx.doi.org/10.1088/0034-4885/53/7/001}{{\em Reports on
  Progress in Physics\/}~{\bf 53}, 837--915}.

\bibitem[\protect\citeauthoryear{{Chandrasekhar}}{{Chandrasekhar}}{1931}]{Chandra1931}
{\sc {Chandrasekhar}, S.} 1931.
\newblock {The Maximum Mass of Ideal White Dwarfs}.
\newblock \href{http://dx.doi.org/10.1086/143324}{{\em Astrophysical
  Journal\/}~{\bf 74}, 81}.

\bibitem[\protect\citeauthoryear{Dilday, Howell, Cenko, Silverman, Nugent,
  et~al\mbox{.}}{Dilday et~al\mbox{.}}{2012}]{Dilday:2012cy}
{\sc Dilday, B.}, {\sc Howell, D.}, {\sc Cenko, S.}, {\sc Silverman, J.}, {\sc
  Nugent, P.}, {\sc et~al\mbox{.}} 2012.
\newblock {PTF11kx: A Type-Ia Supernova with a Symbiotic Nova Progenitor}.
\newblock \href{http://dx.doi.org/10.1126/science.1219164}{{\em Science\/}~{\bf
  337}, 942}, \href{http://www.arxiv.org/abs/1207.1306}{arXiv:[1207.1306]}.

\bibitem[\protect\citeauthoryear{Mazzali, Ropke, Benetti, and
  Hillebrandt}{Mazzali et~al\mbox{.}}{2007}]{Mazzali:2007et}
{\sc Mazzali, P.~A.}, {\sc Ropke, F.~K.}, {\sc Benetti, S.}, {\sc and} {\sc
  Hillebrandt, W.} 2007.
\newblock {A Common Explosion Mechanism for Type Ia Supernovae}.
\newblock \href{http://dx.doi.org/10.1126/SCIENCE.1136259}{{\em Science\/}~{\bf
  315}, 825},
  \href{http://www.arxiv.org/abs/astro-ph/0702351}{arXiv:[astro-ph/0702351]}.

\bibitem[\protect\citeauthoryear{Karpenka}{Karpenka}{2015}]{Karpenka:2015vva}
{\sc Karpenka, N.} 2015.
\newblock {The supernova cosmology cookbook: Bayesian numerical recipes},
  \href{http://www.arxiv.org/abs/1503.03844}{arXiv:[1503.03844]}.

\bibitem[\protect\citeauthoryear{Johnson and Morgan}{Johnson and
  Morgan}{1953}]{Johnson:1953zz}
{\sc Johnson, H.} {\sc and} {\sc Morgan, W.} 1953.
\newblock {Fundamental stellar photometry for standards of spectral type on the
  revised system of the Yerkes spectral atlas}.
\newblock \href{http://dx.doi.org/10.1086/145697}{{\em Astrophys.J.\/}~{\bf
  117}, 313}.

\bibitem[\protect\citeauthoryear{Bessell}{Bessell}{2005}]{Bessell:2005zz}
{\sc Bessell, M.~S.} 2005.
\newblock {Standard Photometric Systems}.
\newblock \href{http://dx.doi.org/10.1146/annurev.astro.41.082801.100251}{{\em
  Ann.Rev.Astron.Astrophys.\/}~{\bf 43}, 293--336}.

\bibitem[\protect\citeauthoryear{Goobar and Leibundgut}{Goobar and
  Leibundgut}{2011}]{Goobar:2011iv}
{\sc Goobar, A.} {\sc and} {\sc Leibundgut, B.} 2011.
\newblock {Supernova cosmology: legacy and future}.
\newblock \href{http://dx.doi.org/10.1146/annurev-nucl-102010-130434}{{\em
  Ann.Rev.Nucl.Part.Sci.\/}~{\bf 61}, 251--279},
  \href{http://www.arxiv.org/abs/1102.1431}{arXiv:[1102.1431]}.

\bibitem[\protect\citeauthoryear{Phillips}{Phillips}{1993}]{Phillips:1993ng}
{\sc Phillips, M.} 1993.
\newblock {The absolute magnitudes of Type IA supernovae}.
\newblock \href{http://dx.doi.org/10.1086/186970}{{\em Astrophys.J.\/}~{\bf
  413}, L105--L108}.

\bibitem[\protect\citeauthoryear{Tripp}{Tripp}{1998}]{Tripp:1997wt}
{\sc Tripp, R.} 1998.
\newblock {A Two-parameter luminosity correction for type Ia supernovae}.
\newblock
  \href{http://articles.adsabs.harvard.edu/cgi-bin/nph-iarticle_query?1998A%26A...331..815T&amp;data_type=PDF_HIGH&amp;whole_paper=YES&amp;type=PRINTER&amp;filetype=.pdf}{{\em
  Astron.Astrophys.\/}~{\bf 331}, 815--820}.

\bibitem[\protect\citeauthoryear{Kelly, Hicken, Burke, Mandel, and
  Kirshner}{Kelly et~al\mbox{.}}{2010}]{Kelly:2009iy}
{\sc Kelly, P.~L.}, {\sc Hicken, M.}, {\sc Burke, D.~L.}, {\sc Mandel, K.~S.},
  {\sc and} {\sc Kirshner, R.~P.} 2010.
\newblock {Hubble Residuals of Nearby Type Ia Supernovae Are Correlated with
  Host Galaxy Masses}.
\newblock \href{http://dx.doi.org/10.1088/0004-637X/715/2/743}{{\em
  Astrophys.J.\/}~{\bf 715}, 743--756},
  \href{http://www.arxiv.org/abs/0912.0929}{arXiv:[0912.0929]}.

\bibitem[\protect\citeauthoryear{Hayden, Gupta, Garnavich, Mannucci, Nichol,
  et~al\mbox{.}}{Hayden et~al\mbox{.}}{2013}]{Hayden:2012aa}
{\sc Hayden, B.~T.}, {\sc Gupta, R.~R.}, {\sc Garnavich, P.~M.}, {\sc Mannucci,
  F.}, {\sc Nichol, R.~C.}, {\sc et~al\mbox{.}} 2013.
\newblock {The Fundamental Metallicity Relation Reduces Type Ia SN Hubble
  Residuals More Than Host Mass Alone}.
\newblock \href{http://dx.doi.org/10.1088/0004-637X/764/2/191}{{\em
  Astrophys.J.\/}~{\bf 764}, 191},
  \href{http://www.arxiv.org/abs/1212.4848}{arXiv:[1212.4848]}.

\bibitem[\protect\citeauthoryear{Guy, Astier, Nobili, Regnault, and Pain}{Guy
  et~al\mbox{.}}{2005}]{Guy:2005me}
{\sc Guy, J.}, {\sc Astier, P.}, {\sc Nobili, S.}, {\sc Regnault, N.}, {\sc
  and} {\sc Pain, R.} 2005.
\newblock {SALT: A Spectral adaptive Light curve Template for Type Ia
  supernovae}.
\newblock \href{http://dx.doi.org/10.1051/0004-6361:20053025}{{\em
  Astron.Astrophys.\/}~{\bf 443}, 781--791},
  \href{http://www.arxiv.org/abs/astro-ph/0506583}{arXiv:[astro-ph/0506583]}.

\bibitem[\protect\citeauthoryear{Guy et~al\mbox{.}}{Guy
  et~al\mbox{.}}{2007}]{Guy:2007dv}
{\sc Guy, J.} {\sc et~al\mbox{.}} 2007.
\newblock {SALT2: Using distant supernovae to improve the use of Type Ia
  supernovae as distance indicators}.
\newblock \href{http://dx.doi.org/10.1051/0004-6361:20066930}{{\em
  Astron.Astrophys.\/}~{\bf 466}, 11--21},
  \href{http://www.arxiv.org/abs/astro-ph/0701828}{arXiv:[astro-ph/0701828]}.

\bibitem[\protect\citeauthoryear{Riess, Press, and Kirshner}{Riess
  et~al\mbox{.}}{1996}]{Riess:1996pa}
{\sc Riess, A.~G.}, {\sc Press, W.~H.}, {\sc and} {\sc Kirshner, R.~P.} 1996.
\newblock {A Precise distance indicator: Type Ia supernova multicolor light
  curve shapes}.
\newblock \href{http://dx.doi.org/10.1086/178129}{{\em Astrophys.J.\/}~{\bf
  473}, 88},
  \href{http://www.arxiv.org/abs/astro-ph/9604143}{arXiv:[astro-ph/9604143]}.

\bibitem[\protect\citeauthoryear{Jha, Riess, and Kirshner}{Jha
  et~al\mbox{.}}{2007}]{Jha:2006fm}
{\sc Jha, S.}, {\sc Riess, A.~G.}, {\sc and} {\sc Kirshner, R.~P.} 2007.
\newblock {Improved Distances to Type Ia Supernovae with Multicolor Light Curve
  Shapes: MLCS2k2}.
\newblock \href{http://dx.doi.org/10.1086/512054}{{\em Astrophys.J.\/}~{\bf
  659}, 122--148},
  \href{http://www.arxiv.org/abs/astro-ph/0612666}{arXiv:[astro-ph/0612666]}.

\bibitem[\protect\citeauthoryear{Conley et~al\mbox{.}}{Conley
  et~al\mbox{.}}{2008}]{Conley:2008xx}
{\sc Conley, A.~J.} {\sc et~al\mbox{.}} 2008.
\newblock {SiFTO: An Empirical Method for Fitting SNe Ia Light Curves}.
\newblock \href{http://dx.doi.org/10.1086/588518}{{\em Astrophys.J.\/}~{\bf
  681}, 482--498},
  \href{http://www.arxiv.org/abs/0803.3441}{arXiv:[0803.3441]}.

\bibitem[\protect\citeauthoryear{{Oke} and {Sandage}}{{Oke} and
  {Sandage}}{1968}]{Oke1968}
{\sc {Oke}, J.~B.} {\sc and} {\sc {Sandage}, A.} 1968.
\newblock {Energy Distributions, K Corrections, and the Stebbins-Whitford
  Effect for Giant Elliptical Galaxies}.
\newblock \href{http://dx.doi.org/10.1086/149737}{{\em Astrophysical
  Journal\/}~{\bf 154}, 21}.

\bibitem[\protect\citeauthoryear{Hamuy, Phillips, Wells, and Maza}{Hamuy
  et~al\mbox{.}}{1993}]{Hamuy1993}
{\sc Hamuy, M.}, {\sc Phillips, M.~M.}, {\sc Wells, L.~A.}, {\sc and} {\sc
  Maza, J.} 1993.
\newblock K corrections for type la supernovae.
\newblock \href{http://www.jstor.org/stable/40680108}{{\em Publications of the
  Astronomical Society of the Pacific\/}~{\bf 105}, 689, pp. 787--793}.

\bibitem[\protect\citeauthoryear{Schneider, Kochanek, and Wambsganss}{Schneider
  et~al\mbox{.}}{2006}]{Schneider2006}
{\sc Schneider, P.}, {\sc Kochanek, C.}, {\sc and} {\sc Wambsganss, J.} 2006.
\newblock {\em Gravitational Lensing: Strong, Weak and Micro}.
\newblock Springer-Verlag Berlin Heidelberg.

\bibitem[\protect\citeauthoryear{Kantowski, Vaughan, and Branch}{Kantowski
  et~al\mbox{.}}{1995}]{Kantowski:1995bd}
{\sc Kantowski, R.}, {\sc Vaughan, T.}, {\sc and} {\sc Branch, D.} 1995.
\newblock {The Effects of inhomogeneities on evaluating the deceleration
  parameter q(0)}.
\newblock \href{http://dx.doi.org/10.1086/175854}{{\em Astrophys.J.\/}~{\bf
  447}, 35--42},
  \href{http://www.arxiv.org/abs/astro-ph/9511108}{arXiv:[astro-ph/9511108]}.

\bibitem[\protect\citeauthoryear{Frieman}{Frieman}{1996}]{Frieman:1996xk}
{\sc Frieman, J.~A.} 1996.
\newblock {Weak lensing and the measurement of q(0) from type Ia supernovae}.
\newblock {\em Comments Astrophys.\/}~{\bf 18}, 323,
  \href{http://www.arxiv.org/abs/astro-ph/9608068}{arXiv:[astro-ph/9608068]}.

\bibitem[\protect\citeauthoryear{Gunnarsson, Dahlen, Goobar, Jonsson, and
  Mortsell}{Gunnarsson et~al\mbox{.}}{2006}]{Gunnarsson:2005qu}
{\sc Gunnarsson, C.}, {\sc Dahlen, T.}, {\sc Goobar, A.}, {\sc Jonsson, J.},
  {\sc and} {\sc Mortsell, E.} 2006.
\newblock {Corrections for gravitational lensing of supernovae: Better than
  average?}
\newblock \href{http://dx.doi.org/10.1086/499346}{{\em Astrophys.J.\/}~{\bf
  640}, 417--427},
  \href{http://www.arxiv.org/abs/astro-ph/0506764}{arXiv:[astro-ph/0506764]}.

\bibitem[\protect\citeauthoryear{Jonsson, Sullivan, Hook, Basa, Carlberg,
  et~al\mbox{.}}{Jonsson et~al\mbox{.}}{2010}]{Jonsson:2010wx}
{\sc Jonsson, J.}, {\sc Sullivan, M.}, {\sc Hook, I.}, {\sc Basa, S.}, {\sc
  Carlberg, R.}, {\sc et~al\mbox{.}} 2010.
\newblock {Constraining dark matter halo properties using lensed SNLS
  supernovae}.
\newblock \href{http://dx.doi.org/10.1111/j.1365-2966.2010.16467.x}{{\em
  Mon.Not.Roy.Astron.Soc.\/}~{\bf 405}, 535},
  \href{http://www.arxiv.org/abs/1002.1374}{arXiv:[1002.1374]}.

\bibitem[\protect\citeauthoryear{Holz and Linder}{Holz and
  Linder}{2005}]{Holz:2004xx}
{\sc Holz, D.~E.} {\sc and} {\sc Linder, E.~V.} 2005.
\newblock {Safety in numbers: Gravitational lensing degradation of the
  luminosity distance-redshift relation}.
\newblock \href{http://dx.doi.org/10.1086/432085}{{\em Astrophys.J.\/}~{\bf
  631}, 678--688},
  \href{http://www.arxiv.org/abs/astro-ph/0412173}{arXiv:[astro-ph/0412173]}.

\bibitem[\protect\citeauthoryear{Betoule et~al\mbox{.}}{Betoule
  et~al\mbox{.}}{2014}]{Betoule:2014frx}
{\sc Betoule, M.} {\sc et~al\mbox{.}} 2014.
\newblock {Improved cosmological constraints from a joint analysis of the
  SDSS-II and SNLS supernova samples}.
\newblock \href{http://dx.doi.org/10.1051/0004-6361/201423413}{{\em
  Astron.Astrophys.\/}~{\bf 568}, A22},
  \href{http://www.arxiv.org/abs/1401.4064}{arXiv:[1401.4064]}.

\bibitem[\protect\citeauthoryear{Conley et~al\mbox{.}}{Conley
  et~al\mbox{.}}{2011}]{Conley:2011ku}
{\sc Conley, A.} {\sc et~al\mbox{.}} 2011.
\newblock {Supernova Constraints and Systematic Uncertainties from the First 3
  Years of the Supernova Legacy Survey}.
\newblock \href{http://dx.doi.org/10.1088/0067-0049/192/1/1}{{\em
  Astrophys.J.Suppl.\/}~{\bf 192}, 1},
  \href{http://www.arxiv.org/abs/1104.1443}{arXiv:[1104.1443]}.

\bibitem[\protect\citeauthoryear{Lyons}{Lyons}{2013}]{Lyons:2013yja}
{\sc Lyons, L.} 2013.
\newblock {Discovering the Significance of 5 sigma},
  \href{http://www.arxiv.org/abs/1310.1284}{arXiv:[1310.1284]}.

\bibitem[\protect\citeauthoryear{Dawid}{Dawid}{2015}]{Richard2015}
{\sc Dawid, R.} 2015.
\newblock Higgs discovery and the look elsewhere effect.
\newblock \href{http://www.jstor.org/stable/10.1086/679179}{{\em Philosophy of
  Science\/}~{\bf 82}, 1, pp. 76--96}.

\bibitem[\protect\citeauthoryear{Astier et~al\mbox{.}}{Astier
  et~al\mbox{.}}{2006}]{Astier:2005qq}
{\sc Astier, P.} {\sc et~al\mbox{.}} 2006.
\newblock {The Supernova legacy survey: Measurement of $\Omega_M$,
  $\Omega_\Lambda$ and $w$ from the first year data set}.
\newblock \href{http://dx.doi.org/10.1051/0004-6361:20054185}{{\em
  Astron.Astrophys.\/}~{\bf 447}, 31--48},
  \href{http://www.arxiv.org/abs/astro-ph/0510447}{arXiv:[astro-ph/0510447]}.

\bibitem[\protect\citeauthoryear{Kowalski et~al\mbox{.}}{Kowalski
  et~al\mbox{.}}{2008}]{Kowalski:2008ez}
{\sc Kowalski, M.} {\sc et~al\mbox{.}} 2008.
\newblock {Improved Cosmological Constraints from New, Old and Combined
  Supernova Datasets}.
\newblock \href{http://dx.doi.org/10.1086/589937}{{\em Astrophys.J.\/}~{\bf
  686}, 749--778},
  \href{http://www.arxiv.org/abs/0804.4142}{arXiv:[0804.4142]}.

\bibitem[\protect\citeauthoryear{Amanullah, Lidman, Rubin, Aldering, Astier,
  et~al\mbox{.}}{Amanullah et~al\mbox{.}}{2010}]{Amanullah:2010vv}
{\sc Amanullah, R.}, {\sc Lidman, C.}, {\sc Rubin, D.}, {\sc Aldering, G.},
  {\sc Astier, P.}, {\sc et~al\mbox{.}} 2010.
\newblock {Spectra and Light Curves of Six Type Ia Supernovae at 0.511 $<$ z
  $<$ 1.12 and the Union2 Compilation}.
\newblock \href{http://dx.doi.org/10.1088/0004-637X/716/1/712}{{\em
  Astrophys.J.\/}~{\bf 716}, 712--738},
  \href{http://www.arxiv.org/abs/1004.1711}{arXiv:[1004.1711]}.

\bibitem[\protect\citeauthoryear{March, Trotta, Berkes, Starkman, and
  Vaudrevange}{March et~al\mbox{.}}{2011}]{March:2011xa}
{\sc March, M.}, {\sc Trotta, R.}, {\sc Berkes, P.}, {\sc Starkman, G.}, {\sc
  and} {\sc Vaudrevange, P.} 2011.
\newblock {Improved constraints on cosmological parameters from SNIa data}.
\newblock \href{http://dx.doi.org/10.1111/j.1365-2966.2011.19584.x}{{\em
  Mon.Not.Roy.Astron.Soc.\/}~{\bf 418}, 2308--2329},
  \href{http://www.arxiv.org/abs/1102.3237}{arXiv:[1102.3237]}.

\bibitem[\protect\citeauthoryear{Kim}{Kim}{2011}]{Kim:2011hg}
{\sc Kim, A.} 2011.
\newblock {Type Ia Supernova Intrinsic Magnitude Dispersion and the Fitting of
  Cosmological Parameters}.
\newblock \href{http://dx.doi.org/10.1086/658498}{{\em
  Publ.Astron.Soc.Pac.\/}~{\bf 123}, 230},
  \href{http://www.arxiv.org/abs/1101.3513}{arXiv:[1101.3513]}.

\bibitem[\protect\citeauthoryear{Vishwakarma and Narlikar}{Vishwakarma and
  Narlikar}{2010}]{Vishwakarma:2010nc}
{\sc Vishwakarma, R.~G.} {\sc and} {\sc Narlikar, J.~V.} 2010.
\newblock {A Critique of Supernova Data Analysis in Cosmology}.
\newblock \href{http://dx.doi.org/10.1088/1674-4527/10/12/001}{{\em
  Res.Astron.Astrophys.\/}~{\bf 10}, 1195--1198},
  \href{http://www.arxiv.org/abs/1010.5272}{arXiv:[1010.5272]}.

\bibitem[\protect\citeauthoryear{Wei, Wu, Melia, and Maier}{Wei
  et~al\mbox{.}}{2015}]{Wei:2015xca}
{\sc Wei, J.-J.}, {\sc Wu, X.-F.}, {\sc Melia, F.}, {\sc and} {\sc Maier,
  R.~S.} 2015.
\newblock {A Comparative Analysis of the Supernova Legacy Survey Sample with
  $\Lambda$CDM and the $R_h = ct$ Universe}.
\newblock \href{http://dx.doi.org/10.1088/0004-6256/149/3/102}{{\em
  Astron.J.\/}~{\bf 149}, 102},
  \href{http://www.arxiv.org/abs/1501.02838}{arXiv:[1501.02838]}.

\bibitem[\protect\citeauthoryear{Marriner, Bernstein, Kessler, Lampeitl,
  Miquel, Mosher, Nichol, Sako, and Smith}{Marriner
  et~al\mbox{.}}{2011}]{Marriner:2011mf}
{\sc Marriner, J.}, {\sc Bernstein, J.~P.}, {\sc Kessler, R.}, {\sc Lampeitl,
  H.}, {\sc Miquel, R.}, {\sc Mosher, J.}, {\sc Nichol, R.~C.}, {\sc Sako, M.},
  {\sc and} {\sc Smith, M.} 2011.
\newblock {A More General Model for the Intrinsic Scatter in Type Ia Supernova
  Distance Moduli}.
\newblock \href{http://dx.doi.org/10.1088/0004-637X/740/2/72}{{\em Astrophys.
  J.\/}~{\bf 740}, 72},
  \href{http://www.arxiv.org/abs/1107.4631}{arXiv:[1107.4631]}.

\bibitem[\protect\citeauthoryear{Kessler et~al\mbox{.}}{Kessler
  et~al\mbox{.}}{2013}]{Kessler:2012gn}
{\sc Kessler, R.} {\sc et~al\mbox{.}} 2013.
\newblock {Testing Models of Intrinsic Brightness Variations in Type Ia
  Supernovae, and their Impact on Measuring Cosmological Parameters}.
\newblock \href{http://dx.doi.org/10.1088/0004-637X/764/1/48}{{\em Astrophys.
  J.\/}~{\bf 764}, 48},
  \href{http://www.arxiv.org/abs/1209.2482}{arXiv:[1209.2482]}.

\bibitem[\protect\citeauthoryear{Milne, Foley, Brown, and Narayan}{Milne
  et~al\mbox{.}}{2015}]{Milne:2014rfa}
{\sc Milne, P.~A.}, {\sc Foley, R.~J.}, {\sc Brown, P.~J.}, {\sc and} {\sc
  Narayan, G.} 2015.
\newblock {The Changing Fractions of Type ia Supernova Nuv--optical Subclasses
  With Redshift}.
\newblock \href{http://dx.doi.org/10.1088/0004-637X/803/1/20}{{\em
  Astrophys.J.\/}~{\bf 803}, 1, 20},
  \href{http://www.arxiv.org/abs/1408.1706}{arXiv:[1408.1706]}.

\end{thebibliography}

\endgroup

%

\end{document}